\documentclass[
twocolumn,
showpacs, 
superscriptaddress,
nofootinbib,
floatfix,   
11pt]{revtex4-1}
\usepackage{setspace} 
\usepackage[utf8]{inputenc}
\usepackage{float} 
\usepackage{amsmath,amssymb,amsfonts,bm,dsfont}
\usepackage{graphicx}
\usepackage{multirow}
\usepackage[normalem]{ulem}
\usepackage[usenames,dvipsnames]{color}
\allowdisplaybreaks
\usepackage[
colorlinks=true,
linkcolor=blue,
breaklinks=true,
urlcolor=blue,
citecolor=blue]{hyperref}
\graphicspath{{figs.dir/}}
\usepackage{soul}

\definecolor{green}{rgb}{0,0.6,0}

\newcommand{\D}{{\rm d}}

\newcommand{\be}{\begin{equation}} 
\newcommand{\ee}{\end{equation}}
\newcommand{\bea}{\begin{eqnarray}} 
\newcommand{\eea}{\end{eqnarray}}
\newcommand{\beas}{\begin{eqnarray*}} 
\newcommand{\eeas}{\end{eqnarray*}}

\newcommand{\ba}{\begin{align}} 
\newcommand{\ea}{\end{align}}

\renewcommand{\vec}{\bm}

\usepackage{appendix}

\usepackage{slashed}
\usepackage[compat=1.1.0]{tikz-feynman}
\usepackage[compat=1.1.0]{tikz-feynhand}

\newcommand{\s}[1]{\slashed{#1}}

\newcommand{\sss}{\!\!\! / }
\newcommand{\ssss}{\!\!\!\! / }

\newcommand{\Ds}{{D^*}}

\newcommand{\fpi}{{f_\pi}}
\newcommand{\jpsi}{{J/\psi}}
\newcommand{\hc}{{h_c}}

\newcommand{\mev}{{\,\text{MeV}}}
\newcommand{\gev}{{\,\text{GeV}}}

\newcommand{\rub}{\affiliation{Institut f\"ur Theoretische Physik II, Ruhr-Universit\"at Bochum, D-44780 Bochum, Germany }}

\newcommand{\bonn}{\affiliation{Helmholtz-Institut f\"ur Strahlen- und Kernphysik and Bethe Center for Theoretical Physics,\\ Universit\"at Bonn, D-53115 Bonn, Germany}}

\newcommand{\fzj}{\affiliation{Institute for Advanced Simulation, Institut f\"ur Kernphysik and J\"ulich Center for Hadron Physics, Forschungszentrum J\"ulich, D-52425 J\"ulich, Germany}}

\newcommand{\ucas}{\affiliation{School of Physical Sciences, University of Chinese Academy of Sciences, Beijing 100049, China}}

\newcommand{\ihep}{\affiliation{Institute of High Energy Physics, Chinese Academy of Sciences, Beijing 100049, China}}

\newcommand{\moe}{\affiliation{Key Laboratory of Atomic and Subatomic Structure and Quantum Control (MOE), Guangdong Basic Research Center of Excellence for Structure and Fundamental Interactions of Matter, Institute of Quantum Matter, South China Normal University, Guangzhou 510006, China
}}

\newcommand{\iqm}{\affiliation{Guangdong-Hong Kong Joint Laboratory of Quantum Matter, Guangdong Provincial Key Laboratory of Nuclear Science, Southern Nuclear Science Computing Center, South China Normal University, Guangzhou 510006, China}}

\begin{document}
\title{ How many vector charmoniumlike states lie in the mass range from $4.2$~to~$4.35$~GeV?}

\begin{abstract}
In recent years many vector charmonium(-like) states were reported by 
different electron-positron collider experiments above $4.2$ GeV. However,
so far, there not only exists sizable tension in the parameters of those states,
but  there is also no consensus on the number of the vector states
in this energy range. To some extend, this
might be caused by the fact that the experimental data
were typically analyzed in single channel analyses employing overlapping Breit-Wigner
functions, in particular ignoring the effect of opening thresholds. 
In this study, we focus on the mass range between $4.2$ and $4.35$ GeV, conducting a comprehensive analysis of eight different final states in $e^+ e^-$ annihilation. Our findings demonstrate that,  within this mass range, a single vector charmonium-like state, exhibiting properties consistent with a $D_1\bar D$ molecular structure and characterized by a  pole location $\sqrt{s_\text{pole}^{Y(4230)}}=\left( 4227{\pm} 4 {-} \frac{i}{2}(50^{+8}_{-2}) \right) \text{MeV}$, can effectively describe all the collected data.
This is made possible by allowing for an interference with the well-established vector charmonium $\psi(4160)$ along with
the inclusion of the $D_1\bar D$ threshold effect. 
Moreover, in contrast to experimental analyses, our study reveals that the highly asymmetric
total cross sections for $e^+e^-\to J/\psi\pi\pi$ and $e^+e^-\to J/\psi K\bar K$ around 4230 MeV stem from the same physics, 
rooted in the approximate SU(3) flavor symmetry of QCD.
\end{abstract}

\author{Leon~von~Detten}\email{l.von.detten@fz-juelich.de}
\fzj

\author{Vadim Baru}\email{vadim.baru@tp2.rub.de}
\rub

\author{Christoph~Hanhart}\email{c.hanhart@fz-juelich.de}
\fzj 

\author{Qian~Wang}\email{qianwang@m.scnu.edu.cn}
\moe\iqm\ihep

\author{Daniel~Winney}\email{daniel.winney@gmail.com}
\bonn

\author{Qiang~Zhao}\email{zhaoq@ihep.ac.cn}
\ihep\ucas
\maketitle

\newpage
\noindent
\section{Introduction}
\label{sec:intro}
\begin{figure*}[t]
	\centering
	\includegraphics[width=\linewidth,trim={0pt 5pt 0pt 130pt}]{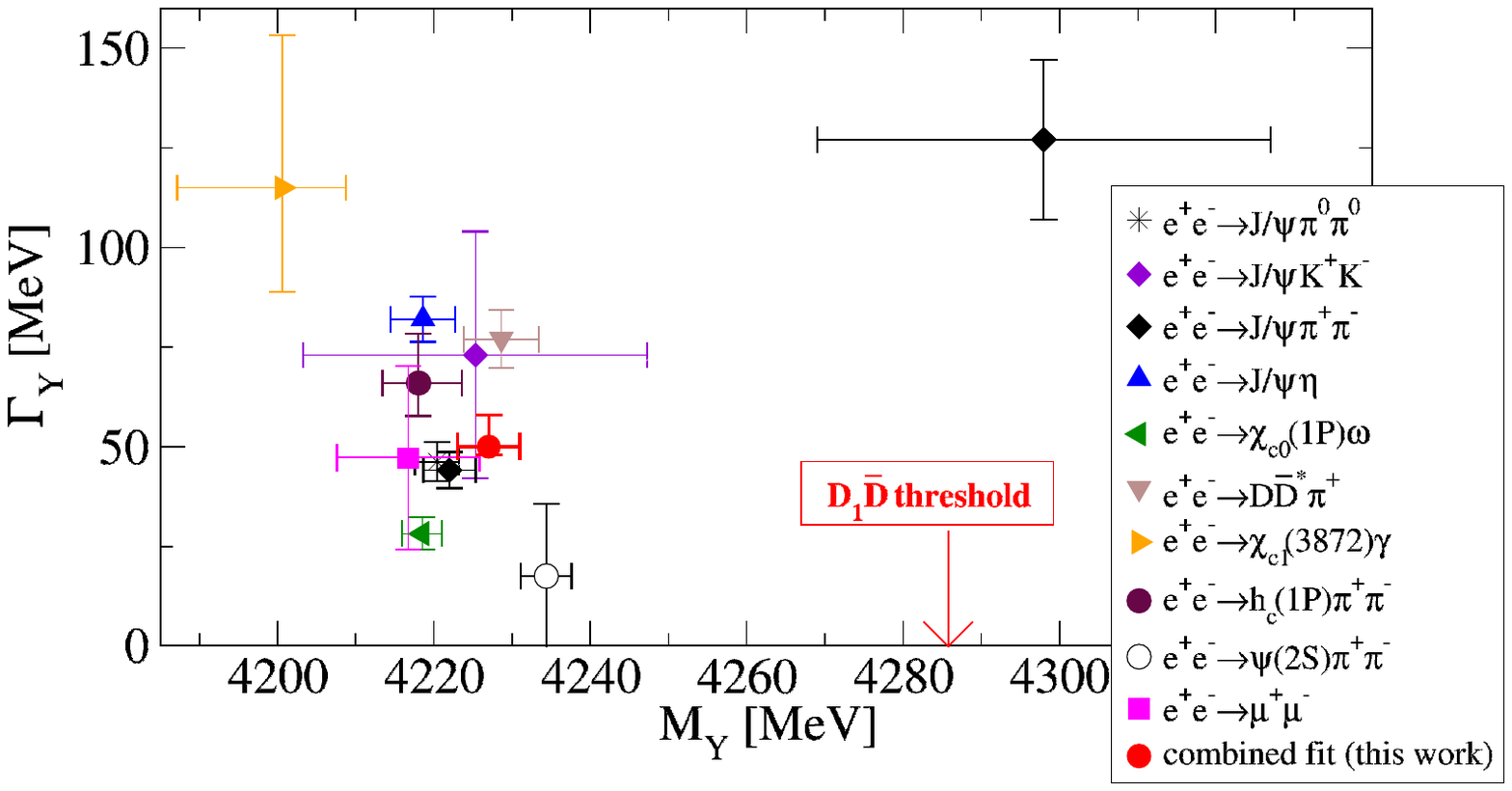}
	\caption{Mass and width of the two $Y$-states discussed
 in the introduction as extracted from
	the experimental analyses of the individual channels shown
	by the labels. All data below a mass value below $4240$ MeV
 is interpreted a $Y(4230)$ and the one data point above
 refers to the $Y(4320)$. The experimental values are taken from \cite{BESIII:2022qal,BESIII:2020oph,BESIII:2020bgb,BESIII:2019gjc,BESIII:2018iea,BESIII:2019qvy,BESIII:2016adj,BESIII:2020peo,BESIII:2021njb,BESIII:2022joj}. 
 The red dot denotes the pole location for the $Y(4230)$ as extracted in this work.\label{Fig:comparison}}
\end{figure*}

Since the discovery of the first exotic state, i.e. $\chi_{c1}(3872)$ also known as $X(3872)$, in the $\bar cc$-sector in 2003, 
a large number of states was discovered in the charmonium and bottomonium mass range that show properties incompatible with
expectations from  quark models that describe mesons as quark-antiquark states. For recent reviews see, e.g.,
Refs.~\cite{Lebed:2016hpi,Esposito:2016noz,Olsen:2017bmm,Guo:2017jvc,Brambilla:2019esw,Chen:2022asf}.
The amount of available data is especially rich in the $J^{PC}=1^{--}$ channel,
since here states containing $\bar cc$ can be generated directly in $e^+e^-$-collisions
and can therefore straightforwardly be studied at experiments like BaBar, Belle and BESIII.
In this work, we focus on vector states in the mass range from $4.2$ to $4.35$ GeV. 
This energy range hosts most prominently the $\psi(4230)$ also known as $Y(4230)$
and potentially one additional state located at $4.32$ GeV. The latter was introduced
in the analyses of the BESIII collaboration for the reaction $e^+e^- \to J/\psi \pi^+\pi^-$ 
to account for the highly asymmetric line shape seen in the experiments reported in Refs.~\cite{BESIII:2016bnd,BESIII:2022qal}.
In particular, the most recent analysis~\cite{BESIII:2022qal} revealed for  the Breit-Wigner mass and width of the $Y(4230)$ in this channel
\begin{equation}\label{eq:expymass}
\begin{aligned}
M_{Y(4230)}&= 4221.4\pm 1.5 \pm 2.0\ \mbox{MeV}\\ \Gamma_{Y(4230)}&=41.8\pm 2.9 \pm 2.7\ \mbox{MeV}  \ ,
\end{aligned}
\end{equation}
and for the $Y(4320)$
\begin{equation}
\label{Eq:Y4230}
\begin{aligned}
M_{Y(4320)}&=  4298\pm 12 \pm 26 \ \mbox{MeV} \\ \Gamma_{Y(4320)}&=127\pm17\pm 10 \ \mbox{MeV} \ ,
\end{aligned}
\end{equation}
 where the first and second uncertainty is statistical and systematic, respectively. 
The  $Y(4320)$ is also needed for analyzing  the $J/\psi\pi^0\pi^0$ channel~\cite{BESIII:2020oph}, and the parameters above
are consistent with the data in this channel.
On the other hand, the $Y(4230)$ is seen in the eight additional channels shown in  Fig.~\ref{Fig:comparison}, admittedly with largely
inconsistent parameters, while the state dubbed  $Y(4320)$ 
 shows up in none of them, at least within the mass range consistent with
Eq.~\eqref{Eq:Y4230},
not even in $e^+e^-\to J/\psi K\bar K$ which is connected to
$e^+e^-\to J/\psi\pi\pi$ by the approximate SU(3) flavor symmetry of QCD.
In experiments by BaBar and Belle a state named $Y(4360)$, with a mass of about $4345$ MeV,  was discovered in the  
$\psi(2S)\pi^+\pi^-$~\cite{BaBar:2012hpr,Belle:2014wyt} final state. However, the recent BESIII
measurement of the same channel revealed that 
the $Y(4360)$ emerges due to a subtle
interference of the $Y(4230)$ and a state at $4390$ MeV with a width of $140$ MeV~\cite{BESIII:2021njb}, which is
thus in a mass range close to the $\psi(4415)$, however,
twice as wide.
A signal at $4390$ MeV with the consistent parameters was also observed by BESIII in the $h_c \pi\pi$ \cite{BESIII:2016adj} and $J/\Psi \eta$ \cite{BESIII:2020bgb} final states.
 Since this state is outside the mass range in focus here, we do not discuss it any further. \\[.3cm]

In the mass
range from $4.2$ to $4.35$ MeV
the findings just described raise the following
{questions}: 
\begin{enumerate}
\item 
Why does the observed width of the $Y(4230)$ deduced from the $J/\psi \pi\pi$ channel, 
differ so significantly from that deduced from the $D^*\bar D\pi$ channel, where the measured width is twice as large~\cite{BESIII:2018iea}?

\item What can we learn from the cross section differences for $Y(4230)$ in 
its various decay channels? Note that the cross section in the $D\bar{D}^*\pi$ channel is about one order-of-magnitude larger than those of hidden charm decays.
\item  
Why is the $Y(4230)$   observed in final states with both $\bar cc$ spin 1 (i.e. $J/\psi\pi\pi$ and $\psi^\prime\pi\pi$ channels) and $\bar cc$ spin 0 (i.e. $h_c\pi\pi$ channel) at a similar rate, 
 despite being produced via a photon, which leads to $\bar cc$ in spin 1 only?
 Can we understand this seemingly large violation of heavy quark spin
symmetry? 
\item Why is  the $Y(4320)$  seen only in
a single channel?
\item Can the apparent asymmetry of the $ J/\psi \pi^+\pi^-$ line shape be generated by the opening
of the $D_1(2420)\bar D$ channel just below the nominal mass of the $Y(4320)$?
Here the $D_1(2420)$ 
is the narrow axial vector with a width of about $30$ MeV, which decays to the $\pi D^*$ channel
predominantly in $D$-wave; 
the nearby $D_1(2430)$ has a width of
about $300$ MeV and  decays to the $\pi D^*$ channel
predominantly in $S$-wave. 
Thus, the broad $D_1$  is not capable of producing structures as narrow as those discussed here, 
although its mixing with the narrow $D_1$, emerging from spin symmetry violation, is relevant for the detailed description of the data.
\end{enumerate}

In this work we address the mentioned issues starting from the 
assumption that the $Y(4230)$ is a $D_1(2420)\bar D$ hadronic molecule, proposed
originally in Ref.~\cite{Ding:2008gr},
and refined in 
Ref.~\cite{Wang:2013cya} by taking into account the the $D_1\bar D$ cut properly and 
in particular the triangle singularity mechanism, which is crucial for the production of the $Z_c(3900)$, accounts for the lineshape of the $J/\psi\pi\pi$
and in this way 
leads to a pole position around $4.23~\mathrm{GeV}$.
Before we discuss in detail the observable consequences of this assumption we present the other
structure assumptions put forward   for this state in the literature, 
namely studies that do not need the pole of the $Y(4230)$ at all
as well as the three types
that do call for a state 
in this mass range, namely the
hybrid, the hadrocharmonium and the compact tetraquark interpretation --- details are given in the following paragraphs.
This allows us to demonstrate that the implications of
a molecular structure for the $Y(4230)$ are very specific and significant.

In Refs.~\cite{Coito:2019cts,Chen:2015bft} no pole for the $Y(4230)$
needs to be introduced. In the 
former reference the
structure near $\sqrt{s}=4230$ MeV
is generated from the $\psi(4160)$
coupling to the $D_s\bar D_s$ channel. While this provides a reasonable description of the $J/\psi\pi\pi$ final state, it is unlikely that the same scenario also allows a description of all the other final states, especially the $D\bar D^*\pi$ channel. 
The same comment applies to 
Ref.~\cite{Chen:2015bft}, where the 
$Y(4230)$ is generated from an interference of the neighboring charmonium
states. We therefore do not consider these mechanisms any further.

In the hadrocharmonium picture an exotic hidden charm state appears
as a compact $\bar cc$ core surrounded by some typically excited light quark cloud~\cite{Dubynskiy:2008mq}~\footnote{The hadrocharmonium picture is contrasted to the molecular one for the $Y(4230)$ in Ref.~\cite{Wang:2013kra}.}.
While this explains naturally that the $Y(4230)$ decays into $J/\psi\pi\pi$ and not $D^{(*)}\bar D^{(*)}$ as would be expected for a $\bar cc$  quark-model state, it appears
at odds with the fact that the $Y(4230)$ is observed also in the $h_c\pi\pi$ final state, since 
heavy quark symmetry calls for a conservation of heavy
quark spin. To overcome this problem it was proposed in Ref.~\cite{Li:2013ssa} that the $Y(4230)$ and the next higher state are in fact emerging from a mixing of two states, one with a spin 0 $\bar cc$ core and one with
a spin 1 core. Thus, this scenario calls for a second
nearby vector state --- a currently good candidate
being the above mentioned $Y(4320)$. Moreover, this mixing scenario implies the existence of four spin symmetry partners~\cite{Cleven:2015era}. For example, there should be two exotic $\eta_c$ states, one in between the two vector states, one significantly lighter
than the $Y(4230)$.

Very early after its discovery, the $Y(4230)$ was proposed to be a hybrid state based either on phenomenological calculations~\cite{Close:2005iz,Kou:2005gt,Kalashnikova:2008qr}
or heavy quark effective field theory~\cite{Berwein:2015vca}.
In the hybrid picture, both quarks and gluons contribute as valence degrees of freedom.
A study of the decays employing heavy quark effective
field theory disfavors a pure hybrid interpretation of the $Y(4230)$~\cite{Brambilla:2022hhi}.
In any case, also the hybrid picture calls for a mixing of two close-by vector states with different spin
of the $\bar cc$ component of the wave functions and thus for the existence of both $Y(4230)$ and $Y(4320)$ to accommodate the decays into final states with both spin 0 and spin 1 for the outgoing charmonium.  In addition, for a hybrid vector the decays into $J/\psi \pi\pi$ and $J/\psi K\bar{K}$ are connected by SU(3) flavor symmetry.  The 
rate in the latter channel deduced in this way is however larger than what one finds experimentally.

In the compact tetraquark picture the states
are typically made of heavy-light diquarks
and anti-diquarks. This approach calls for
four non-strange vector states with masses in the range $4220$ MeV and $4660$ MeV~\cite{Ali:2017wsf,Bhavsar:2020pog}, since the 
diquarks can have either spin one or spin zero
allowing for the following
spin couplings with positive $C$ parity, $[0,0]_0$,
 $[1,0]_1+[0,1]_1$,  $[1,1]_0$,  $[1,1]_2$, with the spins of diquark and antidiquark in the
 brackets and their total spin as subindex outside ---  note that
 a state that contains two spin 1 substructures coupled to total spin 1 has negative $C$ parity. To get the negative
 parity needed for a vector state, an angular momentum of 1 needs to be 
 introduced between the  diquark and antidiquark that in addition
 flips the $C$ parity to the needed -1. For example, the currently preferred
fit of Ref.~\cite{Ali:2017wsf} includes both $Y(4220)$ as well as $Y(4320)$. 
An alternative approach to compact tetraquarks, similar in spirit, but different in the realization, is outlined in Ref.~\cite{Lebed:2017min}. 
Thus, we see that three of the non-molecular scenarios prefer the presence of both $Y(4230)$ and $Y(4320)$
while the remaining
are
challenged by the decay properties of the $Y(4230)$.

\begin{figure}[t!]
	\centering
	\includegraphics[width=1.05\linewidth,trim={30pt 55pt 0pt 93pt}]{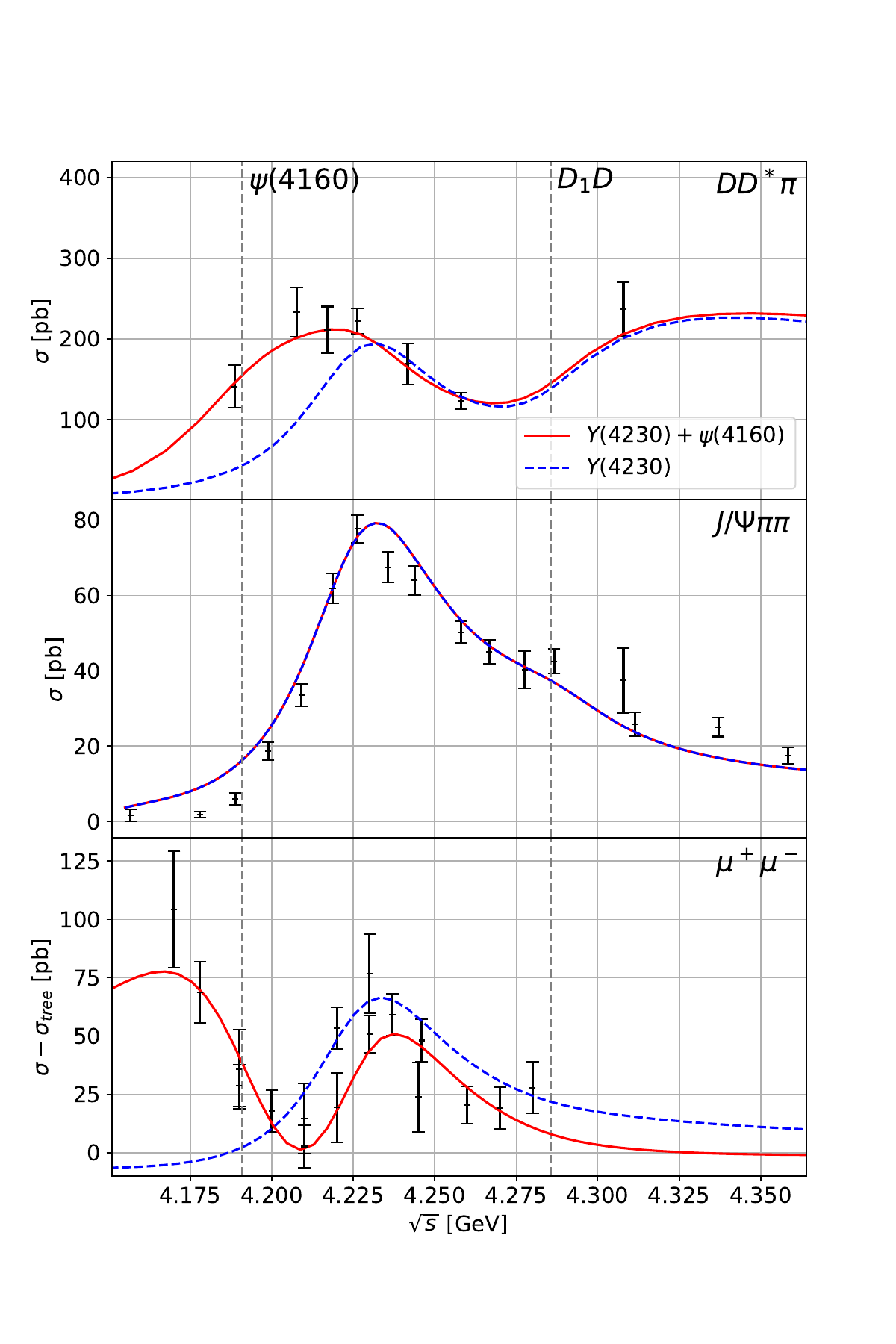}
    \caption{ Fit result for $D^0 D^{* -} \pi^+$, $\jpsi \pi^+ \pi^-$  and $\mu^+ \mu^-$ channels, including the $\psi(4160)$ (solid red line) and omitting it (dashed blue line).
    The data for the $D^0 D^{* -} \pi^+$ channel are from Ref.~\cite{BESIII:2023cmv}, those
    for the $\jpsi \pi^+ \pi^-$ channel from Ref.~\cite{BESIII:2022qal} and for $\mu^+ \mu^-$ from Ref.~\cite{BESIII:2020peo}. The vertical dashed lines indicate the positions of the nominal mass of the $\psi(4160)$ and $D_1 \bar D$ threshold, respectively.
    \label{fig:effectofpsi}}
\end{figure}

In this study, we investigate the feasibility of a combined analysis involving  eight different final states excited in $e^+e^-$ annihilation, namely 
$D^0 D^{* -} \pi$, 
$\jpsi \pi^+ \pi^-$, 
$\jpsi K^+ K^-$, $h_c\pi^+\pi^-$,
$\mu^+ \mu^-$,
$\chi_{c0}(1P) \omega$, $\jpsi \eta$ and $X(3872) \gamma$, in the mass range from 4.2 to 4.35 GeV,
 under the assumption that the $Y(4230)$ is a $D_1\bar D$ molecule.
The main message of this work is
that
the data available in this mass range is consistent with
the presence of a single exotic state predominantly of molecular 
nature, since
such a state necessarily 
has a large
coupling to the $D_1\bar D$ channel.
The  molecular scenario for the $Y(4230)$ was already advocated in  Refs.~\cite{Cleven:2013mka,Qin:2016spb}  based on an analysis of older data  in the $J/\psi \pi\pi$ and $h_c \pi\pi$ channels.
It is crucial to emphasize that, while certain properties of the data  emerge naturally in the current analysis, there are cases where fine-tuned parameters are necessary.
It turns out that in order to obtain a coherent picture, it is unavoidable
to include the interference with an additional vector state whose properties
we fix to those of the well known charmonium state $\psi(4160)$.
This is illustrated in Fig.~\ref{fig:effectofpsi}, where we show the fit results with and without the $\psi(4160)$
for three selected channels: At first glance, the narrow structure in the $J/\psi\pi\pi$ channel 
appears incompatible with the
much broader structure observed in $D\bar D^*\pi$, as well as some other channels
discussed below. However, as shown in the figure this discrepancy can be 
overcome by a simultaneous presence of both
$\psi(4160)$ and $Y(4230)$.
 Based on an analysis of the channels $D\bar D$,
 $D\bar D^*$, $D^*\bar D^*$ and $D\bar D\pi$,
Ref.~\cite{Zhou:2023yjv} puts forward the hypothesis that these two vector states could actually be the same state. However, the data shown in Fig.~\ref{fig:effectofpsi} indicate that this conjecture is not compatible with the data studied in this work. Especially the $\mu^+ \mu^-$ channel shown in the bottom panel of Fig.~\ref{fig:effectofpsi}  clearly
shows two individual peak structures, which
can only be understood by the presence of two
resonance poles.
Moreover, if the higher peak in the $\mu^+ \mu^-$
channel were driven by the $D_1\bar D$ cusp, it
would appear at the $D_1\bar D$ threshold and not
about 60 MeV below.

We regard this work as some exploratory study --- accordingly some
disclaimers need to be given, which will be overcome in subsequent publications:
\begin{itemize}
    \item We treat the effect of the interference of the $\psi(4160)$ with
    the $Y(4230)$ perturbatively. While this simplifies the fitting, it
    violates unitarity
    since only terms linear in the vector propagators are included in the evaluation of the hadronic cross sections.
\item {Also, to accelerate the fitting, we approximate the imaginary parts in the denominators of the resonance propagators for $\psi(4160)$, $Y(4230)$, and $Z_c(3900)$.
Specifically, we keep dynamically the most significant imaginary parts that exhibit strong energy dependence within the considered mass range. Meanwhile, contributions from more distant channels that show minimal changes are replaced by constants.
Accordingly, the complete width of the
$\psi(4160)$ is treated as a constant, and
for the other two states only the $D_1\bar D$ and the $D^*\bar D$
channels, respectively, are kept dynamically.}
 \item This is a phenomenological study. 
    In particular, we cannot estimate uncertainties from a truncation error in some systematic expansion.
   This is appropriate, however, since we
     only aim at demonstrating what is possible with
    a single exotic particle in the mass range of interest. 
  Accordingly, 
    uncertainties
    of e.g. the pole parameters
    of the $Y(4230)$
    were only 
    roughly estimated at this stage.
    \item We focus on the effect of the $D_1\bar D$ intermediate state in the decays of the $Y(4230)$, basically ignoring
    that heavy quark spin symmetry also calls for the coupled channels $D_1\bar D^*$ and the
    $D_2\bar D^*$  --- this is the main limiting factor
    when considering the energy range.
     \item  The channels with two pions or two kaons in the final states necessitate the proper inclusion of  $\pi\pi/K\bar K$ final-state interactions, as discussed in previous works~\cite{Chen:2015jgl,Chen:2016mjn,Baru:2020ywb,Molnar:2019uos,Danilkin:2020kce}.
       In this study we simplify the treatment of these effects.
       While our approximation shows qualitatively very reasonable results, the
       data for $e^+e^- \to \psi(2S)\pi\pi$, exhibit a very unusual energy dependence in the 
       subsystem
       invariant mass distributions at $\sqrt{s}= 4230$ and $4260$ MeV, which seem to require a more refined treatment. 
       Consequently, data from $e^+e^- \to \psi(2S)\pi\pi$ are not included in the current 
       fits. 
 \item 
 The data currently available  do not show apparent peak structures of $Y(4230)$  in $D^{(*)}\bar D^{(*)}$ channels, which must appear in odd partial waves to reach $J^{PC} = 1^{--}$.  
This suggests that the  couplings of $Y(4230)$ to the two-body open charm channels are much smaller than those of the vector charmonium states.  
 In Ref.~\cite{Xue:2017xpu} it was demonstrated that
 the dips seen in the data of $e^+e^-\to D^*\bar{D}^*$ and $D_s^*\bar{D}_s^*$ are consistent with an interference from the $D_1\bar{D}$ molecular nature of the $Y(4230)$.
\end{itemize}
Note that with respect to exploiting 
the implications of the heavy quark spin symmetry there are more advanced studies 
than this one already published~\cite{Peng:2022nrj,Ji:2022blw}.
However, both those works focus solely on the pole locations that emerge from solving the scattering equations for the members of the spin multiplet $\{D_1,D_2\}$ scattering
off those of $\{\bar{D},\bar{D}^*\}$.
No attempt is made
to investigate the resulting line shapes in the various decay channels.
Contrary to those works, we here study the energy
dependence of the cross sections in the various decay channels. 
This allows us to demonstrate that  the inclusion of
the $\psi(4160)$ together with the strong coupling of the $Y(4230)$ to
the $D_1\bar D$ channel
that is a consequence of
its assumed molecular nature, is sufficient to describe all data sets studied
here without the need for an additional exotic state in the mass
range of interest. This conclusion is in line with the preliminary results of this study announced in Ref.~\cite{vonDetten:2023uja}.

While preparing this manuscript we got aware of Ref.~\cite{Nakamura:2023obk}, where, besides various
other ingredients, the $\psi(4160)$, the $Y(4230)$
and the $D_1\bar D$-channel are included.

 The central finding of that work relevant for us is that in total three poles are found in the energy range studied here. While one of them might well represent the $\psi(4160)$, only with somewhat shifted parameters, and another one the $Y(4230)$, there is still a resonance needed close to 4320 MeV, albeit a very broad one (with a width more than 300 MeV) absent in our analysis. From the information provided in Ref.~\cite{Nakamura:2023obk} 
 it is not clear what dynamics drives the appearance of such a pole. 
 
A more detailed comparison with that work will only be possible, once more details are published. 

An alternative analysis to ours that
includes both $\psi(4160)$ as well as $Y(4230)$  in the energy range of interest here but does not call 
for a state located at 4320 MeV is Ref.~\cite{Chen:2017uof}. In this work the asymmetric shape observed in the total cross sections of $\pi\pi J/\psi$  and $D^*\bar D\pi$ can be reproduced as an interference effect between $\psi(4160)$ and the higher energy state $\psi(4415)$---the latter state is beyond the energy range considered in our analysis---combined with a non-resonant background. Then the inclusion of $Y(4230)$ is needed for fine-tuning the agreement with data near 4.2 GeV.
Another striking difference between that work and ours is their omission of any threshold effects. 
As we argue below, the significance of the $D_1\bar D$ threshold in the $Y$ line shapes 
is a direct hint towards its molecular nature --- accordingly Ref.~\cite{Chen:2017uof} argues that their analysis is consistent with a
$\bar cc$ structure of the $Y(4230)$. 
Thus studying observable differences between
the results of that work and ours is important to pin down the nature of the $Y(4230)$. We come back to this 
when discussing the results.
It should also be mentioned that the relatively large cross section seen in $e^+e^-\to h_c \pi\pi$ ---
where the final state contains a $\bar cc$ pair in spin zero, contrary to the production of a $\bar cc$ pair with spin one — suggests a considerable amount of heavy quark spin symmetry (HQSS) violation. This phenomenon, highly unnatural for a $\bar cc$ structure,
is explained naturally by prominent $D_1\bar D$
loops, since the two-meson intermediate state decorrelates
the heavy quark spins.  Therefore, the similarity in
size between the $J/\psi \pi\pi$ and the $h_c\pi\pi$
cross sections suggests a  molecular
structure of the $Y(4230)$.

The paper is structured as follows:
We start with some general considerations about the diagrams to be included in the molecular approach. Then,
in Sec.~\ref{sec:formalism}, we describe in some detail the formalism
employed. Sec.~\ref{sec:results} contains the fitting results
as well as their discussion.
We close with a summary and outlook
in Sec.~\ref{sec:summary}. Additional technical details of the calculations are delegated to Appendices.

\section{General Considerations}
\label{Sec:diagrams}
In the molecular scenario, the coupling of some physical $Y$ states to the nearby continuum channel, $h_1 h_2$, that forms the molecule is maximal~\cite{Weinberg:1965zz,Landau},  see also \cite{Guo:2017jvc} for a review.
In fact, this large coupling is what encodes the molecular nature of a given state.
 Accordingly,  the transition of $Y$ to the channel $h_1 h_2$ dominates over the others and the diagrams containing this coupling appear always
at leading order. 
In effective field theories  (EFT), the relative importance of the diagrams is controlled by power counting rules, as presented for similar systems,
e.g., in Refs.~\cite{Guo:2009wr,Guo:2010ak}. However, our case involves several additional complexities such as 
the presence of the unstable particle $D_1$ in the transition, exploration of a relatively wide energy  range, and the analysis of three-body final states.
 To illustrate the second point, we note that for the energies near the $Y(4230)$ peak, triangle diagram $(c)$ of Fig.~\ref{Fig:DstDpi} is potentially
 important,
as long as we look at $D^*\bar D$ invariant masses close to the mass of the $Z_c(3900)$. Indeed,  not only the $D^*\bar D$ intermediate state is in this case nearly on shell, but also 
the nearby triangle singularity significantly enhances the contribution of this diagram~\cite{Wang:2013hga}.  
However, this diagram is suppressed over a large fraction of the Dalitz plot apart from this range.
Therefore, in what follows, we employ a more pragmatic strategy to
consider the most natural and phenomenologically motivated production mechanisms and investigate their relative importance in different energy regimes. 
Due to this, we postpone an estimation of the theoretical uncertainties   to a later  publication.

\subsection{$e^+e^-\to D^0 D^{* -} \pi^+$}
\begin{figure*}[t!]
\begin{picture}(500,140)
\put(0,80){\includegraphics[width=.44\linewidth]{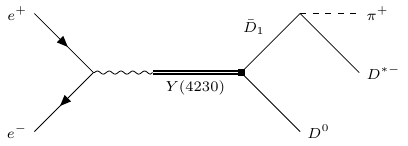}}
\put(45,140){a)}
\put(260,80){\includegraphics[width=.45\linewidth]{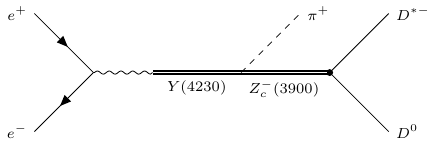}}
\put(305,140){b)}
 \put(0,0){\includegraphics[width=.56\linewidth]{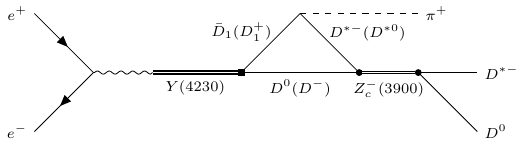}}
 \put(45,60){c)}
\put(260,0){\includegraphics[width=.45\linewidth]{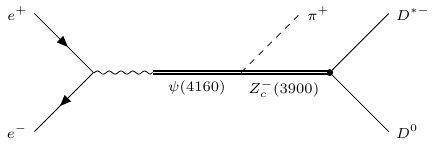}}
 \put(305,60){d)}
\end{picture}
	\caption{Diagram contributing to $e^+e^-\to \bar DD^*\pi$. a) tree level, b) $Y(4230)$ contact term, c) Triangle, d) $\psi(4160)$ contact term, where for the last three the final state interactions in the doubly heavy subsystem are included.}
	\label{Fig:DstDpi}
 \end{figure*}
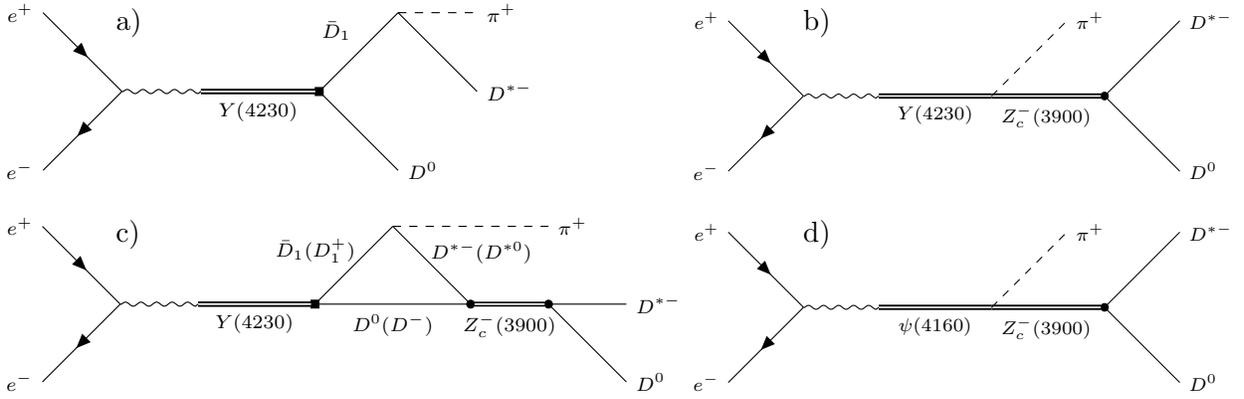

The most direct access to a molecular state is provided by its imprint on the near threshold cross section of the channel that forms the molecular state,
since as outlined above the coupling of the molecule to its constituents is large.
The same reason  is also the origin of the unnaturally large nucleon-nucleon scattering lengths~\cite{Weinberg:1965zz,Guo:2017jvc,Matuschek:2020gqe}. This phenomenon arises from the existence of nearby molecular states in both the spin-1 channel (with the deuteron as a true bound state) and the spin-0 channel (with a closely located virtual state).
The $D_1(2420)$ is unstable with a width of about $30$ MeV.
It decays
predominantly into $D^*\pi$ in $D$-wave, thus the final state with closest connection to a possible molecular
nature of the $Y(4230)$ is the $D^*\bar D\pi$ channel. The corresponding diagrams are shown in Fig.~\ref{Fig:DstDpi}.

Both diagram $(a)$ and diagram $(c)$ scale directly with the large $YD_1D$ coupling. Moreover, they are both enhanced by the
near on-shell $D_1$
propagator --- after all we are near the pole of
a narrow state. 
In our study we also treat the $Z_c(3900)$
as a hadronic molecule,  in line with
Ref.~\cite{Wang:2013cya}. Accordingly the coupling of the $Z_c$ to the $D^*\bar D$ channel is also large.
Moreover, the triangle diagram, which is part of diagram $(c)$, is enhanced by a very close by
triangle singularity~\cite{Wang:2013hga}.
Thus we expect diagrams $(a)$
and $(c)$ to contribute significantly
to the observables. 

The pole of the $Y(4230)$ is located about $60$ MeV
below the nominal $D_1\bar D$ threshold, which is about twice the width of the $D_1(2420)$. This width still allows for a resonance signal 
even at $\sqrt{s}=4230$ MeV~\cite{Braaten:2007dw,Hanhart:2010wh},
however, with a significant kinematic suppression --- the large coupling of the $Y(4230)$ to the $D_1\bar D$
channel with the $D_1\to D^*\pi$ decay in $D$-wave  develops its effect mostly above
the $D_1\bar D$ threshold. However, although violating HQSS,
the narrow $D_1(2420)$ is expected to mix with the much broader $D_1(2430)$. 
In this way the narrow $D_1(2420)$ also gets an $S$-wave decay~\cite{Guo:2020oqk},
which does not so strongly suffer from the above mentioned kinematic suppression allowing it to contribute
significantly to the $Y(4230)$ peak in the $\pi D^*\bar D$ channel. In particular, the decays of the $D_1(2420)$ to  $D^* \pi$ both in $S$- and $D$-waves are therefore included in diagrams a) and c). 
In this paper, whenever referring to $D_1$ without mass number, we talk about the narrow $D_1(2420)$, to simplify notation.

 In addition, because of the large width of the
$D_1(2430)$, its residual effect acts effectively like a very
short ranged contribution. We thus do not calculate loop contributions involving
this broad state explicitly but parameterize it by a point coupling of the
$Y(4230)$ to $\pi D^*\bar D$ with $S$-waves in
all subsystems.
Since the $D^*\bar D$
in $S$-wave also undergoes final state interactions, the just mentioned point
coupling cannot occur in isolated form,
but needs to get dressed
by the $Z_c(3900)$ propagator that
parameterizes the $D^*\bar D$ $S$-wave interaction. This results in an
expression that is represented by 
diagram~\ref{Fig:DstDpi}$(b)$ --- details for the expressions employed are given in Sec.~\ref{sec:formalism}
as well as in the Appendix.
 This construction is automatically consistent with the Watson 
theorem~\cite{Watson:1952ji}.

Finally, in the experimental data
for the $\pi D^*\bar D$ 
channel the width of the structure around $4230$ MeV is notably broader than that observed,
e.g., in the $J/\psi \pi\pi$ channel --- see Fig.~\ref{fig:effectofpsi}. Because of
this, the parameters extracted for the $Y(4230)$ in the two channels by the BESIII collaboration are
inconsistent with each other --- c.f. Fig.~\ref{Fig:comparison}. A possible mechanism that allows for a combined fit of the various channels is that the $\psi(4160)$ also has some
small coupling to the $D^*\bar D\pi$. The
experimental signal observed could then be interpreted as the result of an interference  of the signatures from the two resonances.
Also for the $\psi(4160)$ we assume that the coupling is in $S$-wave with respect to all subsystems and, as before, also
here the direct transition $\psi(4160)\to D^*\bar D\pi$ gets dressed by the
$D^*\bar D$ final state interaction parameterized by the $Z_c(3900)$ propagator. The corresponding diagram is shown in Fig.~\ref{Fig:DstDpi}$(d)$. 
We are aware that, if the $\psi(4160)$ were (predominantly) a $D$-wave charmonium, there should also be angular momenta in the final state as a consequence of HQSS. However, the data do not call for an additional coupling 
structure and we thus omit it from our study.

\subsection{$e^+e^-\to J/\psi(\pi\pi/\bar KK)$}
\label{sec:general_Jpsipipi}

\begin{figure*}
\begin{center}
\begin{picture}(500,360)
\put(0,0){\includegraphics[width=.48\linewidth]{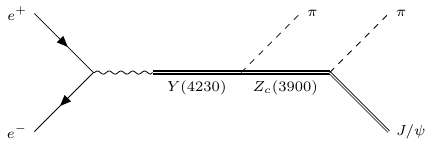}}
\put(45,60){e)}
\put(250,0){\includegraphics[width=.48\linewidth]{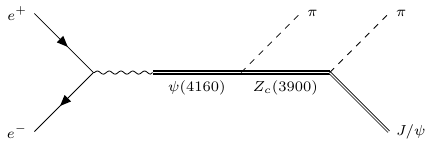}}
\put(295,60){f)}
\put(0,80){\includegraphics[width=.62\linewidth]{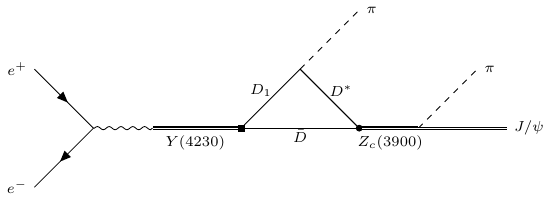}}
\put(45,140){d)}
\put(0,170){\includegraphics[width=.72\linewidth]{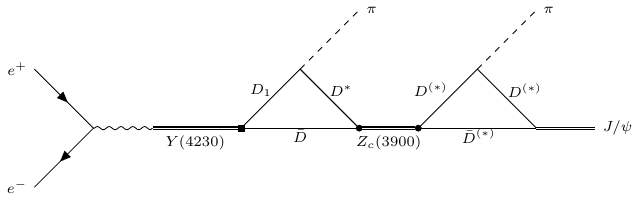}}
\put(45,230){c)}
\put(0,290){\includegraphics[width=.5\linewidth]{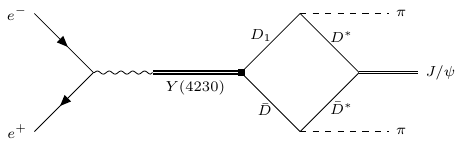}}
\put(45,350){a)}
\put(250,290){\includegraphics[width=.48\linewidth]{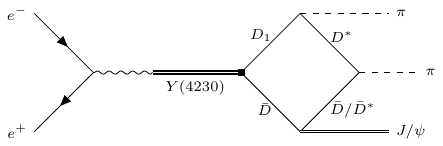}}
\put(295,350){b)}
\end{picture}
	\caption{Diagram contributing to $e^+e^-\to J/\psi\pi\pi$.
	The thin lines in the box and triangle denote $D^*$ or $D$ mesons. a) and b) boxes, c) triangle, d) triangle counter term, e) $Y(4230)$ contact term, f) $\psi(4160)$ contact term, where for the last two the $\jpsi \pi$ final state interactions are included.
    }
	\label{Fig:Jpsipipi}
	\end{center}
\end{figure*}

\begin{figure*}
	\includegraphics[width=.48\linewidth]{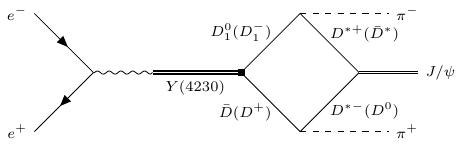}
 \includegraphics[width=.48\linewidth]{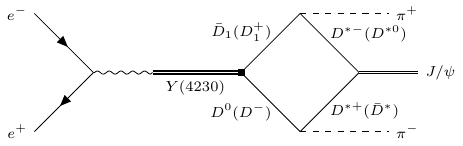}
 \includegraphics[width=.48\linewidth]{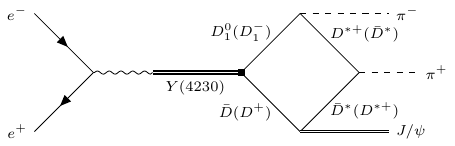}
 \includegraphics[width=.48\linewidth]{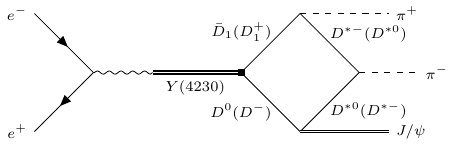}
 \includegraphics[width=.48\linewidth]{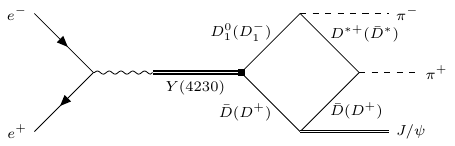}
 \includegraphics[width=.48\linewidth]{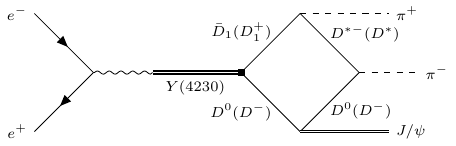}
 \caption{Decomposition for the box topology of $e^+ e^-\rightarrow \jpsi \pi^+ \pi^-$.}
 \label{fig:R2_box_decomp}
\end{figure*}

Next we turn to the discovery channel of the $Y(4230)$, $e^+e^-\to J/\psi\pi\pi$, where the highly asymmetric lineshape lead to the
claim for the existence of an additional state
called $Y(4320)$~\cite{BESIII:2016bnd,BESIII:2022qal}.
Again, driven by the assumed molecular nature of the $Y(4230)$,
contributions that run through the $D_1(2420)\bar D$ intermediate state are sizable and need
to be considered. Then, to reach the 
$J/\psi\pi\pi$ final state, possible 
topologies are either box diagrams
(see Fig.~\ref{Fig:Jpsipipi}, $(a)$
and $(b)$ as well as  Fig.~\ref{fig:R2_box_decomp} for a complete set of box diagrams) or a triangle followed by 
a $Z_c(3900)$ propagator (Fig.~\ref{Fig:Jpsipipi}$(c)$ and $(d)$). As before we need to allow
for additional processes and also here
include a diagram for the contact
transition of the $Y(4230)$ to the
$J/\psi\pi\pi$ final state (Fig.~\ref{Fig:Jpsipipi}$(e)$), as before dressed by the final state interaction that leads to the occurrence of the $Z_c(3900)$ --- see Sec.~\ref{sec:form_factor} for a detailed discussion.
Furthermore also in this channel we allow for a contribution of the $\psi(4160)$, shown in
Fig.~\ref{Fig:Jpsipipi}$(f)$.

To come to the full amplitudes, the $\pi\pi$
final state interaction needs to be taken into account as well. Since the initial photon generates a $\bar cc$-pair, which is isoscalar, and the final $\bar cc$ pair is
isoscalar as well, the pion pair must be isoscalar with even angular momentum (the latter also follows from parity conservation). In the vicinity of the $Y(4230)$ pole, the $\pi\pi$ system is probed in the energy range from its threshold up to about $1.1$ GeV. Since the scalar-isoscalar $\pi\pi$ interaction has a strong coupling to the $\bar KK$ system, the final state interaction is included by employing a formalism that explicitly 
treats the coupled channels.
Since the full treatment of the system is technically very demanding~\cite{Danilkin:2020kce} (see Refs.~\cite{Chen:2015jgl,Chen:2016mjn,Baru:2020ywb} for related studies) because of
the intricate singularity structure of the pertinent integrals, in this exploratory 
study we
employ an approximate treatment that still allows for a sensible description also of the $\pi\pi$ spectra --- 
details are given in the next
section and in Appendix~\ref{sec:pipi_fsi}. 

\begin{figure*}[t]
    \centering
    \includegraphics[width=.48\linewidth]{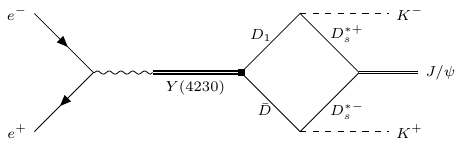}
    \caption{Strange source for $e^+ e^- \rightarrow J/\psi K^+ K^-$.}
    \label{fig:jpsiKK_strange_box}
\end{figure*}

The coupled channel treatment of the $\pi\pi/\bar KK$ final state interaction provides us at the same time access to  $J/\psi\bar KK$ final state.
To make the latter calculation complete,
we also need to take into account strangeness in the source, as shown in Fig.~\ref{fig:jpsiKK_strange_box}. This does not introduce any additional parameters, since 
we demand that the vertices are consistent
with the $SU(3)$-flavor symmetry. Naturally, the
strangeness sources are also included in the calculation of the $J/\psi \pi\pi$ final state.

\subsection{$e^+e^-\to h_c\pi\pi$}
\label{sec:general_hcpipi}

\begin{figure*}[t]
	\includegraphics[width=.48\linewidth]{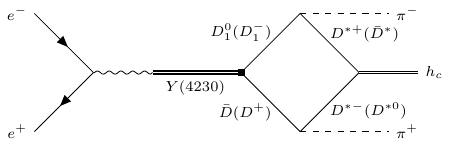}
 \includegraphics[width=.48\linewidth]{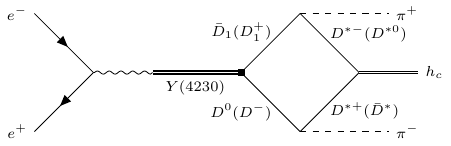}
 \includegraphics[width=.48\linewidth]{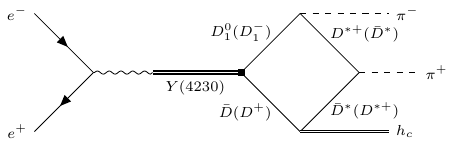}
 \includegraphics[width=.48\linewidth]{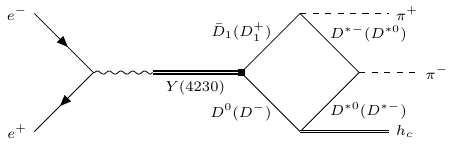}
 \caption{Diagrams contributing to $e^+ e^-\rightarrow h_c \pi^+ \pi^-$.}
 \label{fig:R4_box_decomp}
\end{figure*}

The diagrams contributing here are in principle analogous to those for the $J/\psi \pi\pi$ channel,  shown in Fig.~\ref{Fig:Jpsipipi}.
However, in contrast to that channel, we exclude diagrams containing a $Z_c(3900)$. This is based on the observation that $Z_c(3900)$ does not show a significant contribution to the $h_c\pi$ invariant mass distribution. Additionally, we point out that the $Z_c(4020)$ is not included in this work, since this would require a complete treatment of the $\{D_1 \bar D^{(*)}, D_2 \bar D^{(*)}\}$ coupled channels, and of the $\{D \bar D^*, D^* \bar D^*\}$ sub-systems, which is postponed to future work.
Moreover,  the contact terms
that drive the contributions shown
in diagrams $(e)$ and $(f)$ of Fig.~\ref{Fig:Jpsipipi} in the $J/\psi \pi\pi$ channel
are omitted here as they violate spin symmetry. 
This symmetry violation is overcome by the loop diagrams as a result of the spin symmetry violation that enters through the mass differences of $D$ and $D^*$ as well as $D_1$ and $D_2$ --- the former one being included explicitly in the calculation, the latter one by choosing an energy range where the $D_2$ contribution should be negligible. 
For a detailed discussion on how the spin symmetry gets restored in the heavy quark limit even in the presence of hadronic molecules, see Ref.~\cite{Baru:2022xne}.
In summary, for the $h_c\pi\pi$ channel we only include
the box topologies shown in Fig.~\ref{fig:R4_box_decomp}, expecting some
deviations from experiment as a result of the omission
of the $Z_c(4020)$.  On the other hand, it is not expected that the $Z_c(4020)$ will generate significant structures in the total cross section of $h_c \pi\pi$, which is the focus of the current work, since in this case, the narrow peak from $Z_c(4020)$ in the $\pi h_c$ subsystem is smeared. The same effect is demonstrated explicitly in this work, where the narrow structures of the $Z_c(3900)$ seen in the $J/\psi \pi$ subsystem do not visibly modify the energy dependence of the cross section {for $J/\psi\pi\pi$}.

\subsection{$e^+e^-\to X(3872)\gamma$}

\begin{figure*}[t]
	\begin{center}
	\includegraphics[width=1.\linewidth]{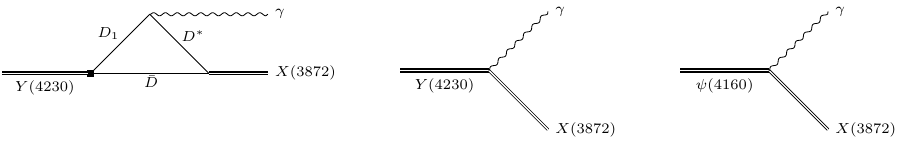}
	\caption{Diagram contributing to $X(3872) \gamma$.
	\label{Fig:X3872gamma}}
	\end{center}
\end{figure*}

If $Y(4230)$ is a $D_1\bar D$ hadronic molecule and both $Z_c(3900)$ and $X(3872)$ are
$D^*\bar D$ hadronic molecules with
$I(J^{PC})=1(1^{+-})$ and $I(J^{PC})=0(1^{++})$, respectively, the production mechanism of the latter pair in $Y(4230)$ decays must be analogous~\cite{Guo:2013zbw}. Only that the
particle radiated off in the course of the
$Y(4230)$ decay must have positive $C$ parity for the transition to the $Z_c$ and negative $C$
parity for the transition to the $X(3872)$. Thus, all that needs to be done to get from the diagram that generates the $Z_c$ in $Y(4230)\to \pi Z_c$ to the one that generates the $X(3872)$, is to replace
the pion in the final state by a photon. The resulting diagrams are shown in Fig.~\ref{Fig:X3872gamma}.

\subsection{$e^+e^- \rightarrow \mu^+ \mu^-$}
\label{sec:intro_ee_mumu}
\begin{figure*}[t!]
	\begin{center}
	\includegraphics[width=1.\linewidth]{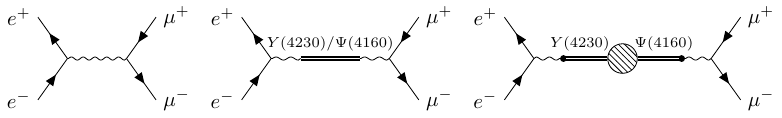}
	\caption{Diagrams contributing to $e^+e^-\to \mu^+ \mu^-$. The hatched 
 circle in the rightmost diagram indicates the mixing of the two vector states  driven by their common decays to the channels $DD^*\pi,\jpsi \pi \pi,\chi_{c0} \omega, \jpsi \eta$ and $X(3872) \gamma$ considered in this analysis  --- for details see text.
	\label{Fig:mumu}}
	\end{center}
\end{figure*}

For each reaction discussed so far the electromagnetic production 
mechanism and the strong decay were entangled in a special way.
What makes the $e^+e^- \rightarrow \mu^+ \mu^-$ especially interesting is,
that here we may isolate production from decay, 
since the total cross section is by far dominated by the real valued tree-level diagram (first diagram in Fig.~\ref{Fig:mumu})
and the hadronic cross sections only contribute significantly through their interference with the mentioned dominating one.
Moreover, the decays of  $Y(4230)$ and $\psi(4160)$ 
 into the same hadronic 
channels induce some 
mixing of these in
the $\gamma^*\to\gamma^*$
transition amplitudes.
The diagrams contributing to the process are shown in Fig.~\ref{Fig:mumu}.
The  mentioned 
mixing of the two vector resonances is depicted here as the hatched blob. The 
imaginary part of this mixing amplitude is given by the respective
interference terms that contribute also
to the various exclusive hadronic channels discussed above. It is
 dominated by
the transitions 
$Y(4230)\to D\bar D^*\pi\to \psi(4160)$,
since the $D\bar D^*\pi$ cross section is by far the largest hadronic cross section.  The details of the calculations can be found in Sec.~\ref{sec:formal_ee_mum}.
Therefore, 
the simultaneous study of the hadronic channels and the $e^+e^- \rightarrow \mu^+ \mu^-$ channel provides a sanity check for
the size of the induced  mixing of the vector states, which turn out to be significant.

\subsection{Further channels}
\begin{figure*}[t!]
	\begin{center}
	\includegraphics[width=1.\linewidth]{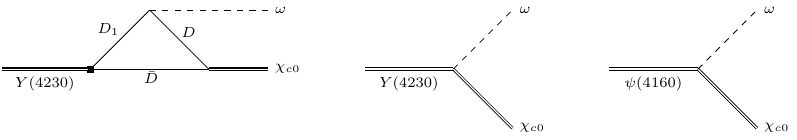}
	\caption{Diagrams contributing to $\chi_{c0} \omega$.
	\label{Fig:chi_c0omega}}
	\end{center}
\end{figure*}
\begin{figure*}[t!]
	\begin{center}
	\includegraphics[width=0.95\linewidth]{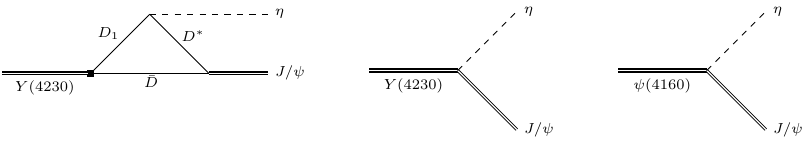}
	\caption{Diagrams contributing to $\jpsi \eta$.
	\label{Fig:jpsieta}}
	\end{center}
\end{figure*}

As shown in Fig.~\ref{Fig:comparison}, in addition to the channels discussed in detail above, the $Y(4230)$ is seen also in the final states  $\omega \chi_{c0}$, $\eta J/\psi$ and $\psi(2S)\pi\pi$. In this work we do not study this last decay channel as the $\psi(2S)\pi$ invariant mass distributions vary so dramatically when the total energy is changed mildly form $4.226$ to $4.258$ MeV~\cite{BESIII:2017tqk} that
there must be some highly non-trivial interplay of different mechanisms at work that to our understanding are not yet understood microscopically (while in Ref.~\cite{Molnar:2019uos} a description of the invariant mass distributions is provided, no attempt is made to understand the energy dependence of the total cross section).

For the first two channels, both triangle diagrams as well as direct transitions contribute as shown in Figs.~\ref{Fig:chi_c0omega} and \ref{Fig:jpsieta}, respectively. Below we discuss the results for these channels as well.

\section{Formalism}
\label{sec:formalism}

In this section the formalism underlying the calculations is
presented in some detail with additional material provided in the
appendix. Those readers most interested in the results and their physical interpretation
may want to jump to Sec.~\ref{sec:results}
immediately.

\subsection{The Y(4230) as a $D_1 \bar D$ state}
\label{sec:Y4230_molecule}

We can write the $D_1 D$ wavefunction as a negative $\mathcal{C}$-eigenstate

$$
|D_1 \bar{D} (\mathcal{C}=-1) \rangle =\frac{1}{\sqrt{2}} \left( |D_1 \bar{D} \rangle- | D \bar{D}_1 \rangle \right)\, .
$$

As the $Y(4230)$ is predominantly produced by $c \bar c$ it must be an iso-singlet. Following the convention 

\begin{equation}
\begin{aligned}
|\tfrac{1}{2},+\tfrac{1}{2} \rangle=\bar{D}^0=\bar{c}u \quad & \quad |\tfrac{1}{2},+\tfrac{1}{2} \rangle = -D^+ =-c \bar{d}\\
|\tfrac{1}{2},-\tfrac{1}{2} \rangle=D^-=\bar{c} d\quad & \quad |\tfrac{1}{2},-\tfrac{1}{2} \rangle=D^0=c \bar{u}\,
\end{aligned}
\end{equation}

the iso-singlet wavefunction is given by ${|I=0\rangle = 1/\sqrt{2} \left( | \uparrow \downarrow \rangle - | \downarrow \uparrow \rangle \right)}$, resulting in

\begin{equation}
\begin{aligned}
&|D_1 \bar{D} (\mathcal{C}=-1,I=0) \rangle=  \\
&\frac{1}{2}\left(  |D_1^+ D^{-} \rangle  {+} |D_1^0 \bar{D}^{0} \rangle{+}  |D^{+} D_1^- \rangle  {+} |D^{0} \bar{D}_1^0 \rangle  \right) \ .
\label{eq:Y_wavefunc}
\end{aligned}
\end{equation}

The effective Lagrangian for the coupling of  $D_1 \bar D$ to $Y(4230)$ and $D_1 \bar D$ self-interactions
reads~\cite{Qin:2016spb}
  \be\label{Eq:LagY}
  \begin{aligned}
\mathcal{L}_Y=&\frac{g_{Y0}}{\sqrt{2}} (\bar{D}^\dagger Y^i D_1^{i \dagger} - \bar{D}_1^{i \dagger} Y^i D^\dagger)\\
+&g_1 \left[ (D_{1}^i \bar{D})^\dagger (D_{1}^i \bar{D}) + (D \bar{D}^i_{1} )^\dagger (D \bar{D}^i_{1} ) \right]\, ,
\end{aligned}
\ee
where the couplings $g_{Y0}$ and $g_1$ 
include the heavy quark mass normalization of the fields.
Typically a proper field redefinition allows one to absorb the effect of non-perturbative hadron-hadron scattering
into a pole term. This is not possible only if there is more than one pole on the physical sheet in the
mass range of interest~\cite{Baru:2010ww}. Since this is not the case here we can safely 
set the parameter $g_1$ to zero\footnote{We checked that the inclusion of this
parameter does not allow us to improve the fit, however, leads to large
correlations between $g_{Y0}$ and $g_1$.}. 
Thus we get for the $D_1\bar D$ scattering potential 

\begin{eqnarray}
    V(E) = -\frac{g_{Y0}^2}{2}G_0(E) \, ,
\end{eqnarray}
    
where the bare $Y$ propagator reads

  \begin{equation}  
   G_0(E) = \frac{1}{2\omega_Y(E-m_0)} \ ,
\end{equation}

with $\omega_Y$ for the on-shell energy of the $Y(4230)$ from the field normalization and 
 $E=\sqrt{s}$. Here we dropped the spin indices although $G_0$ and various other
propagators below refer to the
propagation of a spin one particle. The reason is that in our non-relativistic treatment the spin structure simply refers
to a $\delta^{ij}$ --- the spin simply runs through unchanged.
The relation of the bare propagator $G_0(E)$ to the full propagator
$G_Y(E)$ is given by the Dyson equation

\begin{equation}
    G_Y=G_0 + G_0 g_{Y0} (2\omega_Y\Sigma_{D_1 D}) g_{Y0} G_Y\, .
    \label{dysoneq}
\end{equation}

From this one finds
for the $D_1\bar D$ scattering amplitude 

\begin{equation}
    {\mathcal{M}_{D_1\bar D\to D_1\bar D}}=-\frac{g_{Y0}^2}{2}G_Y(E) \ ,
\end{equation}

with
\begin{eqnarray}\nonumber
G_Y(E)&=&
\frac{1}{2 \omega_Y}\\
& & \hspace{-1.5cm} \times 
\left(E{-}m_0
{-}g_{Y0}^2 \Sigma_{D_1 D}(E) {+}i{ }\Gamma_\text{in}/2\right)^{-1} .
\label{Eq:G}
\end{eqnarray}
 Note that the last term in the denominator was added to account for the contribution to the width of the $Y(4230)$ 
from the various inelastic channels. 
The  self-energy $\Sigma$ for a resonance
$R$ can be derived from the 
standard, scalar one-loop diagram,
which reads
in dimensional regularization  for the intermediate
two-body state $a$,
up to terms irrelevant in what follows

\begin{widetext}
    
\begin{equation}
\begin{aligned}
\begin{split}
 2\omega_R\tilde\Sigma_{a}(s)=&\frac{1}{(4 \pi)^2} \left[ \frac{m_{a2}^2-m_{a1}^2+s}{2s} \log\left(\frac{m_{a1}^2}{m_{a2}^2}\right) \right.\\
&\left. \hspace{2cm}+\frac{\lambda^{1/2}(s,m_{a2}^2,m_{a1}^2)}{2s}\log\left(\frac{m_{a2}^2+m_{a1}^2-s+\lambda^{1/2}(s,m_{a2}^2,m_{a1}^2)}{m_{a2}^2+m_{a2}^2-s-\lambda^{1/2}(s,m_{a2}^2,m_{a1}^2)}\right)\right]\, .
\end{split}
\end{aligned}
\end{equation}

\end{widetext}

Here $s=E^2$.
The masses in the expression
refer to the masses of the two particles
propagating in channel $a$.
To come from this to
the expression for the self-energies
employed in the propagators,
we use

\begin{equation}
    \Sigma_a(s)
    =\tilde \Sigma_a(s)
    -{\mbox Re}(\tilde \Sigma_a(m_0^2)) \ .
\end{equation}

With this subtraction, the real part of the inverse $Y$ propagator vanishes at $E= m_0$ and it
reduces significantly the correlations between couplings and bare masses~\cite{Baru:2021ldu}.

\begin{figure*}[htbp]
	\begin{tikzpicture}
	\begin{feynman}[small]
	\vertex [] (g_i) {};
	\vertex [right=of g_i] (g_f);
	\vertex [right=of g_f] (Y_f);
	\vertex [below right=of Y_f] (D_f);
	\vertex [above right=of Y_f] (D1_f);
	\vertex [right=of Y_f] (e1) {=};
 
	\vertex [right=of e1] (g1_i);
	\vertex [crossed dot, right=of g1_i] (g1_f) {};
	\vertex [above right=of g1_f] (D1_2f);
	\vertex [below right=of g1_f] (D_2f);
	\vertex [right=of g1_f] (p1) {+};

	\vertex [right=of p1] (g3_i); 
	\vertex [crossed dot, right=of g3_i] (l_i) {};
	\vertex [dot, right=of l_i] (l_f) {};
	\vertex [square dot, right=of l_f] (y3_f) {};
	\vertex [below right=of y3_f] (D_4f);
	\vertex [above right=of y3_f] (D1_4f);

	\diagram*{
		(g_i) -- [photon] (g_f),
		(g_f) -- [double, edge label'=$G$] (Y_f),
		(Y_f) -- [] (D_f), 
		(Y_f) -- [] (D1_f), 
		
		(g1_i) -- [photon] (g1_f),
		(g1_f) -- [] (D_2f), 
		(g1_f) -- [] (D1_2f), 
		
		(g3_i) -- [photon] (l_i),
		(y3_f) -- [] (D_4f), 
		(y3_f) -- [] (D1_4f),
		
		(l_i) -- [quarter left] (l_f),
		(l_i) -- [quarter right, edge label'=$\Sigma$] (l_f),
		(l_f) -- [double, edge label'=$G$] (y3_f)
		
	};
	\end{feynman}
	\end{tikzpicture}
    \begin{tikzpicture}
    \begin{feynman}[small]
    \vertex [] (g1_i1) {};
    \vertex [right=of g1_i1] (g1_i) {};
	\vertex [crossed dot, right=of g1_i] (g1_f) {};
	\vertex [above right=of g1_f] (D1_2f);
	\vertex [below right=of g1_f] (D_2f);
	\vertex [right=of g1_f] (p1) {=};

    \vertex [right=of p1] (g1_i2);
	\vertex [empty dot, right=of g1_i2] (g1_f2) {};
	\vertex [above right=of g1_f2] (D1_2f2);
	\vertex [below right=of g1_f2] (D_2f2);
	\vertex [right=of g1_f2] (p12) {+};
 
    \vertex [right=of p12] (g2_i);
	\vertex [empty dot, right=of g2_i] (g2_f) {};
	\vertex [dot, right=of g2_f] (y2_f) {};
	\vertex [below right=of y2_f] (D_3f) ;
	\vertex [above right=of y2_f] (D1_3f);

 	\diagram*{
		
		(g1_i) -- [photon] (g1_f),
		(g1_f) -- [] (D_2f), 
		(g1_f) -- [] (D1_2f), 

		(g1_i2) -- [photon] (g1_f2),
		(g1_f2) -- [] (D_2f2), 
		(g1_f2) -- [] (D1_2f2), 
		
		(g2_i) -- [photon] (g2_f),
		(g2_f) -- [edge label'=$G_0$] (y2_f),
		(y2_f) -- [] (D_3f), 
		(y2_f) -- [] (D1_3f),

	};
    \end{feynman}
 \end{tikzpicture}
 \caption{\label{fig:MprodY} The $Y(4230)$ induced production of the $D_1 \bar D$ pairs from a pointlike source. The solid lines denote $D_1$ and $D$ mesons as well as the bare propagator $G_0$, double line stands for the dressed propagator of the Y(4230), and the wiggly line corresponds to the initial photon.}
\end{figure*}
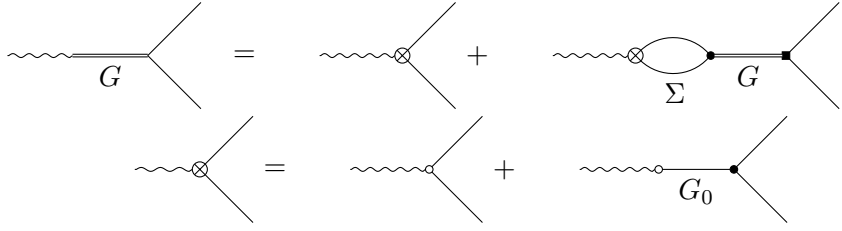

 Using  the $D_1 \bar D$ scattering amplitude and the $Y(4230)$ propagator $G_Y$, one is in the position to derive the pointlike  production operator ${\cal M}_Y$ via the 
 $Y(4230)$ to $D_1 \bar D$ (see Fig.~\ref{fig:MprodY}  for the graphical illustration)
 
\begin{equation}
\begin{aligned}
{\cal M}_Y&= (c- \alpha G_0 g_{Y0})\\
& \qquad \times(1+ 2\omega_Y\Sigma_{D_1 D} G_Y g_{Y0}^2)\ ,
\label{eq:3pt_func}     
\end{aligned}
\end{equation}

where $c$ is the direct coupling of the photon to $D_1 \bar D$ in the quantum numbers $J^{PC}=1^{--}$,
which vanishes in the HQSS limit, and $\alpha$ is the source term coupling of the photon to the bare $Y$ state. 
Eq.~(\ref{eq:3pt_func}) gives the impression as if it had a pole at the bare mass $m_0$, however,
from Eq.~(\ref{dysoneq}) one gets that 
\begin{equation}
\begin{aligned}
  &(1+ 2\omega_Y\Sigma(E^2) G_Y(E) g_{Y0}^2) \\
  & \qquad\qquad\qquad\qquad = G_0(E)^{-1}G_Y(E) \ , 
\end{aligned}
\end{equation}
which allows us to rewrite Eq.~(\ref{eq:3pt_func}) as
\be
{\cal M}_Y= (a+E b) G_Y(E) g_{Y0} \ .
\label{eq:gammaVfinal}
\ee
Here, in the purely one channel $D_1\bar D$ problem, unitarity requires the parameters  to be real.

However, allowing for additional complex phases
at the photon-resonance couplings enables us to effectively include other effects, such as   interference between $\psi(4040)$  and $\psi(4160)$,
as will be discussed below.

\subsection{Production in the
presence of coupled channel final state interactions}
\label{sec:form_factor}
Through hadronic final state interactions, unitarity links contact terms to resonance propagators --- a
special example of this was already demonstrated above: Eq.~(\ref{eq:3pt_func}) contains both a 
contact term to the final state as well as the resonance contributions collecting the interactions
in that final state. As demonstrated there, employing unitarity makes the tree level production
term vanish and the final amplitude, Eq.~(\ref{eq:gammaVfinal}), is proportional to the dressed
resonance propagator. 

Analogously one cannot discuss other tree-level operators or contact terms involved in the decay transitions without the inclusion
of the non-perturbative final state interactions in the relevant subsystem parameterized via the pertinent resonance propagators. 
In analogy to the $Y$ propagator provided in Eq.~(\ref{Eq:G}) we find for the propagator
of the $Z_c(3900)$ from solving the related Dyson equation
\begin{equation}
G_Z=\frac{1}{2\omega_Z}\frac1{E-m_0-\sum_i g_i \Sigma_i g_i} \, ,
\end{equation}
where the sum in the denominator runs over all relevant channels, which for the $Z_c(3900)$
 are $D^*\bar D$ and $J/\psi \pi$ ~\cite{ParticleDataGroup:2022pth} (denoted as channels 1 and 2 respectively) and $E$ is the energy in these subsystems.
Furthermore, $g_i$ stands for the couplings of the $Z_c(3900)$ with the channel $i$, and $\Sigma_i$ refers to the self energy in the corresponding channel.  
As before the trivial spin structure of the
propagator is not shown. 
As the energy range studied in this work is far above the threshold of $\jpsi \pi$, the contribution
of this channel to the self energy is well approximated by a constant whose real part
can  be absorbed into the bare mass $m_0$.

\begin{figure*}
\begin{tikzpicture}
\begin{feynman}[small]
	\vertex [] (gi) at (-1.5, 0);
	\vertex [dot] (gf) at (0,0) {};
	\vertex [] (p1) at (1,0);
    \vertex [] (p2) at (1,-1);
    \vertex [] (p3) at (1,1);
    \vertex [] (lb2) at (1.4,0) {$2$};
	\diagram*{
		(gi)  -- [photon](gf),
		(gf) -- [] (p1),
        (gf) -- [] (p2),
        (gf) -- [scalar] (p3),
	};
\end{feynman}
\end{tikzpicture}
$\qquad$
\begin{tikzpicture}
\begin{feynman}[every blob={/tikz/fill=gray!30,/tikz/inner sep=2pt},small]
	\vertex [] (gi) at (-1.5, 0);
	\vertex [dot] (gf) at (0,0) {};
	\vertex [blob] (T) at (1.5,0) { $T_{22}$};
    \vertex [] (p1) at (2.5,1);
    \vertex [] (lb1) at (0.75,0) {$2$};
    \vertex [] (p2) at (2.5,-1);
    \vertex [] (lb2) at (2.4,0) {$2$};
    \vertex [] (p3) at (1.5,1.);
	\diagram*{
		(gi)  -- [photon](gf),
		(gf) -- [quarter left] (T),
        (gf) -- [quarter right] (T),
        (T) -- [] (p1),
        (T) -- [] (p2),
        (gf) -- [scalar,quarter left] (p3),
	};
\end{feynman}
\end{tikzpicture}
$\qquad$
\begin{tikzpicture}
\begin{feynman}[every blob={/tikz/fill=gray!30,/tikz/inner sep=2pt},small]
	\vertex [] (gi) at (-1.5, 0);
	\vertex [dot] (gf) at (0,0) {};
	\vertex [blob] (T) at (1.5,0) { $T_{12}$};
    \vertex [] (p1) at (2.5,1);
    \vertex [] (lb1) at (0.75,0) {$1$};
    \vertex [] (p2) at (2.5,-1);
    \vertex [] (lb2) at (2.4,0) {$2$};
    \vertex [] (p3) at (1.5,1.);
	\diagram*{
		(gi)  -- [photon](gf),
		(gf) -- [quarter left] (T),
        (gf) -- [quarter right] (T),
        (T) -- [] (p1),
        (T) -- [] (p2),
        (gf) -- [scalar,quarter left] (p3),
	};
\end{feynman}
\end{tikzpicture}
\caption{Feynman diagram for production of channel 2. The scattering amplitudes $T_{ij}$ in the channels $ij\, (i,j=1,2)$ are related to the $Z_c$ propagator as $T_{ij}= g_i G_Z g_j$.}
\label{fig:Feynman_formfactor_2}
\end{figure*}
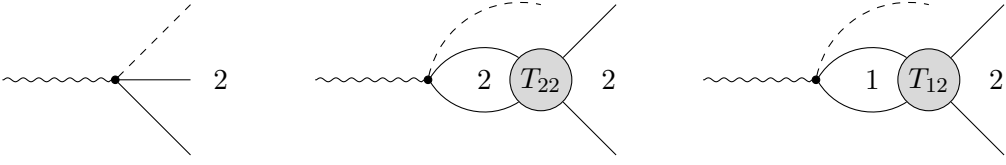

\begin{figure*}[t]
	\begin{tikzpicture}
	\begin{feynman}[small]
	\vertex [](z_start) at (-5.5,0);
	\vertex [](j_start) at (-4,0);
	\vertex [label=right: {\scriptsize \( \jpsi \)}](j_end) at (-3,0);
	\vertex [label=right: {\scriptsize$\pi$}](pi_end) at (-3,1);
    \vertex [label=right: {\scriptsize$\pi$}](pi_end2) at (-3,-1);
	\vertex [](sign) at (-2,0) {$+$};
	\vertex [](a) at (0,0) ;
	\vertex [] (yi) at (-1.5, 0);
    \vertex [label=right: {\scriptsize$\pi$}] (pi_end3) at (1., -1);
	\vertex [] (d) at (2,0);
	\vertex [] (d1) at (1,1) ;
	\vertex [label=right: {\scriptsize$\pi$}] (pi) at (3,1);
	\vertex [label=right: {\scriptsize \( \jpsi \)}] (zf) at (3,0);
	
	\diagram*{
		(a)  -- [double, edge label'= {\scriptsize\(Y\)}, line width=.7pt](yi) ,
		(d) -- [edge label= {\scriptsize\( \bar{C} \)}] (a),
		(a) -- [edge label= {\scriptsize\(A\)}] (d1),
		(d1) -- [edge label= {\scriptsize\(B \)}] (d),
		  (d1) -- [scalar] (pi),
		(d) -- [double, line width=.6pt] (zf),
		(z_start) -- [double, edge label= {\scriptsize \( Y \)}, line width=.6pt] (j_start),
		(j_start) -- [double, line width=.6pt] (j_end),
		(j_start) -- [scalar] (pi_end),
        (j_start) -- [scalar] (pi_end2),
        (a) -- [scalar] (pi_end3),
		
	};
	\end{feynman}
	\end{tikzpicture}
	\centering

 \includegraphics[width=.58\linewidth]{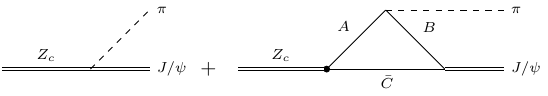}
\begin{equation*}
(\text{A,B,C})=\{(D^*,D^*,D), (D^*,D,D), (D, D^*, D^*)\}
\end{equation*}
\caption{Upper line: Feynman diagrams for production operator for $Y(4230) \to \jpsi \pi \pi$. In the full amplitude the $J/\psi$ and one of the pions undergo final state interactions driven by the $Z_c$.
Lower line: The corresponding 
transition $Z_c\to J/\psi \pi$.
\label{fig:Y_to_jpsipipi}}
\end{figure*}

The production amplitude $F_2$ for channel 2, shown in Fig.~\ref{fig:Feynman_formfactor_2}, can now be expressed as
\begin{equation}
\begin{aligned}\label{eq:prodF2}
    F_2&=M_2 (1+\Sigma_2 g_2 G_Z g_2) + M_1 \Sigma_1 g_1 G_Z g_2\\
    &=G_Z\left(M_2 (E{-}m_0{-}g_1^2 \Sigma_1) {+} M_1 \Sigma_1 g_1 g_2\right) \, ,
\end{aligned}
\end{equation}
with $M_i$ denoting the production operator for channel $i$.  The expression for $F_1$ can be easily obtained from Eq.~\eqref{eq:prodF2} by interchanging 1 and 2. 
Defining $M_1=g_1 \hat{M}_1$ and $M_2=g_2 \hat{M}_2$, 
the form factor $\Vec{F}$ can be therefore expressed by

\begin{equation}
\begin{aligned}
\Vec{F}{=}G_Z \!\!
\left(\!\!
\begin{array}{c}
g_1 \left[ (E-m_0) \hat{M}_1 {+}g_2^2 \Sigma_2 (\hat{M}_1{-}\hat{M}_2) \right]\\[.3cm]
g_2 \left[ (E-m_0) \hat{M}_2 {+}g_1^2 \Sigma_1 (\hat{M}_2{-}\hat{M}_1) \right]\\
\end{array}\!\!
\right) ,
\label{eq:ZFF}
\end{aligned}
\end{equation}

 which takes the form already expressed in the Feynman diagrams in Sec.~\ref{Sec:diagrams}, namely that it is given by some vertex structure for the source term, times the $Z_c$ propagator times the 
respective channel coupling.
As indicated
 in the lower line of
Fig.~\ref{fig:Y_to_jpsipipi}, the effective coupling of the $Z_c$ to the $J/\psi \pi$ channel,
here abbreviated as $g_2$, contains besides a contact term also a triangle 
topology. The same is true  for 
$M_2$, as shown by the upper line  of
Fig.~\ref{fig:Y_to_jpsipipi}. 
The triangles for $M_2$ and $g_2$ in this figure are essentially identical, except for the couplings of $Y$ and $Z_c$ to $D$-mesons, which are evidently different. In particular, they  incorporate the $D^{(*)}\bar D^{(*)} J/\psi$ vertex in P-wave, causing the principal value part of these triangles to depend on a regulator that must be renormalized by a contact term, consistent in both cases.  In the picture advocated here, where the decay of $Y(4230)$ is predominantly governed by diagrams involving the $D_1\bar D$ intermediate state rather than those depicted in Fig. 14, it is reasonable to assume that the overall coefficient $\hat{M}_2$ connecting $M_2$ and $g_2$ is real-valued.
While formally present in the transition amplitude, we observed 
that the fits to the experimental data do not need the term proportional to $(\hat M_2-\hat M_1)$, since it was consistently found to be zero. We thus omit the corresponding terms from the start and employ for the production amplitude

\begin{equation}
\begin{aligned}
\Vec{F}=G_Z 
\left(
\begin{array}{c}
g_1 \alpha_1^{(1)}(\alpha_2^{(1)} + E )\\
g_2 \alpha_1^{(2)}(\alpha_2^{(2)} + E) \\
\end{array}
\right) \, ,
\end{aligned}
\label{eq:ffpara}
\end{equation}

with $\alpha_i^{(j)}$ being free parameters to be determined in the fit. These form factors appear in both the $Y$ and the $\psi$ decays.
The corresponding strength parameters of the latter resonance
are denoted as $\beta_i^{(j)}$.

\subsection{Observables}

\subsubsection{$e^+ e^- \to D^0 D^{* -} \pi^+$} 

With $D^0 D^{* -} \pi^+$ being the channel with the most direct access to the molecular nature of the $Y(4230)$, one expects the tree-level decay, shown in Fig.~\ref{Fig:DstDpi}a, to provide the most significant contribution followed by the triangle loop and contact interactions. As argued in Appendix \ref{Sec:AppA}, see the discussion below Eq.~(\ref{hpdef}), 
the $D_1(2420)$ can decay into $D^* \pi$ in both $S$- and $D$-wave, such that the spin structure of the $Y\rightarrow D^0 D^{* -} \pi^+$ amplitude can be written as

\begin{widetext}

\begin{equation}    
\begin{aligned}
\mathcal{M}_{Y \rightarrow D D^* \pi}^{i}&=G_\text{Y}\left\{
\left(\mathcal{M}_{Y \ \text{CT}}^{D D^* \pi} \right)^{ij}-
\frac{g_{Y0}}{\sqrt{2}}
\left(h^\pi_{1d}\left( 3 p_\pi^i p_\pi^j- p_\pi^2 \delta^{ij} \right)-h^\pi_{1s} \omega_\pi \delta^{ij}
\right)
\right.\\
&\left.\phantom{\left(\mathcal{M}_{Y \ \text{CT}}^{D D^* \pi} \right)^{ij}\frac{g_{Y0}}{\sqrt{2}}}
\qquad \times 
\left[
G_{D_1}(E_{D^*\pi})-2 g_{Z0}^2 \mathcal{T}_{D_1 D D^*} G_Z(E_{DD^*}) \right]\right\}
\epsilon_{\Ds}^{* \ j} \, ,
\label{eq:R1_spin_structure}
\end{aligned}
\end{equation}

\end{widetext}
where we introduced as short hand notation $h^\pi_{1s}{=}h_s' \sqrt{m_{D_1} m_{D^*}}/(\sqrt{3} f_\pi)$
and  $h_{1d}^\pi{=}\sqrt{2/3} h^\prime \sqrt{m_{D_1} m_{D^*}}/f_\pi$. The $D_1{\to}D^*\pi$ couplings $h_s'$
and $h^\prime$ are fixed from
the $D_1$ decay properties --- details are given in App.~\ref{Sec:AppA}.
To respect the Goldstone theorem, stating that the pion amplitude has to vanish in the chiral limit for $p_\pi\rightarrow0$, 
the $S$-wave vertex and other amplitudes below scale with the on-shell pion energy $\omega_\pi=\sqrt{m_\pi^2+p_\pi^2}$.
The indices $i,j$ are the spin indices and a summation over $j$ is assumed.
The tree-level diagram is shown in Fig.~\ref{Fig:DstDpi}a. Additionally, the produced $D \bar{D}^*$ pair can re-scatter into the $Z_c(3900)$ shown in diagram \ref{Fig:DstDpi}c via a triangle loop. 
The Lagrangian further allows for a direct point-like transition of the $Y(4230) {\to} D^0 D^{* -} \pi^+$ in an $S$-wave, corresponding to diagram in Fig.~\ref{Fig:DstDpi}b. The phase of this
diagram is fixed by the re-scattering of the $D \bar{D}^*$ pair into the $Z_c(3900)$, where the formalism described in section \ref{sec:form_factor} is used,

\begin{eqnarray}
\nonumber
{\left(\mathcal{M}_{Y \ \text{CT}}^{D D^* \pi} \right)^{kj}}\hspace{-0.2cm}&{=}& G_Z(E_{DD^*}) g_{Z0} \omega_\pi \\
 &\times&\! \left[ \alpha_1^{(1)} (\alpha_2^{(1)}{+}E_{DD^*}) \delta^{kj} 
\right] \, .
\end{eqnarray}

 This expression agrees to Eq.~(\ref{eq:ffpara}), only that the generic name $g_1$ used there was now adapted to the notation employed in this section.

The relative factor 2 between the tree-level and $Z_c$ contributions comes from the isospin coefficient of the $Y(4230)$ wave function shown in Eq.~\eqref{eq:Y_wavefunc}, as the $Z_c^-(3900) \pi^+$ pair is produced via $Y(4230) \to D_1^+ D^-$ and $Y(4230) \to D \bar{D}_1$. The coefficients of the iso-triplet production of the $Z_c(3900)$ from $D^* \bar{D}$ and the $\mathcal{C}=-1$ eigenstate are absorbed into the coupling $g_{Z0}$ 
of the $Z_c(3900)$ with  $D\bar D^*$. 

Formally, the $Y(4230)$ should be treated as emerging from a $D\bar D^*\pi$ three-body system,
which can be most conveniently handled using time-ordered-perturbation-theory (TOPT)~\cite{Zhang:2021hcl}.
In preparation for this more complete treatment, that we will attack in a subsequent publication,
we evaluate also the  loop integrals in this work using the same formalism. 
Thus, the scalar triangle with pion emission is given by
\begin{eqnarray}\nonumber
\mathcal{T}_{D_1 D D^*}&=&\int \frac{\text{d}^3 l}{(2 \pi)^3} \frac{1}{8 \omega_{D_1} \omega_D \omega_{D^*}} \\ & &\hspace{-1.5cm}\times
\frac{1}{E{-}\omega_{D_1}{-}\omega_D} \frac{1}{E{-}\omega_\pi{-}\omega_{D^*}{-}\omega_D}\, ,
\end{eqnarray}
where the $D_1$ energy is given by ${\omega_{D_1}=\sqrt{(m_{D_1}-i\Gamma_{D_1}/2)^2+l^2}}$, with $m_{D_1}$ and $\Gamma_{D_1}$ for
the mass and width of the $D_1$, respectively. The other particle energies are defined analogously,
however, with their widths neglected. The width of the $D_1$ can here be treated as constant,
since the $D_1$ pole is sufficiently high above the $\pi D^*$ threshold~\cite{Hanhart:2010wh}.
We checked that the energy dependences of the various loop diagrams included in this study agree to the analogous loops evaluated covariantly.
As argued above, a simultaneous treatment of $Y\to J/\psi \pi\pi$ and $Y\to D\bar D^*\pi$ is
possible only if also the interference with the $\psi(4160)$ is included.
The contribution of $\psi(4160) \to D^0 D^{* -} \pi^+$ is parameterized as  
\begin{equation}
\begin{aligned}
\mathcal{M}_{\psi \rightarrow D D^* \pi}^{i}&= G_\psi g_{Z0} G_Z \omega_\pi \\
&\times \left[ \beta_1^{(1)} (\beta_2^{(1)} {+} E_{D D^*}) \delta^{ij} \right]\epsilon^{* \ j}_{D^*}\, ,
\end{aligned}
\end{equation}
again in line with Eq.~(\ref{eq:ffpara}).
The free parameters that appear in the equations above
were fixed in a fit to data --- the
resulting values 
are listed in Tab.~\ref{tab:fit_params}.

Here, a comment is in order.
As our focus lies in examining the interference effect between $Y(4230)$ and $\psi(4160)$ on the line shapes in various channels—specifically, for $e^+ e^-\to D^0 D^{* -} \pi$, $\jpsi \pi^+ \pi^-$, $\jpsi K^+ K^-$, $h_c \pi^+ \pi^-$, $\chi_{c0}(1P) \omega$, $\jpsi \eta$, and $X(3872) \gamma$—we derive the corresponding observables by multiplying the amplitudes ${\cal M}_Y$ and ${\cal M}_{\psi}$, discussed in this and subsequent sections, by the same  complex couplings $g_{\gamma R}=\exp(i \delta_{R \gamma}) e m_R^2/f_R$ of the photon with the resonance $R$ as defined in Eq.~\eqref{eq:photon_VMD}, where $R$ represents both $Y(4230)$ and $\psi(4160)$.
 Clearly, 
 for all channels listed above, only the relative phase of the two resonances plays a role. This is, however, not the case for 
 $e^+ e^-\to \mu^+ \mu^-$, where the two phases enter individually, see 
 Sec.~\ref{Sec:Form_ee_mumu} for details. 

\subsubsection{$e^+ e^- \rightarrow J/\psi \pi^+ \pi^-$}
The Feynman diagrams for $e^+ e^- \rightarrow J/\psi \pi^+ \pi^-$ are shown in figure \ref{Fig:Jpsipipi}. The dominant contributions corresponding to the molecular nature of the $Y(4230)$ are the box, below denoted as $\mathcal{M}^\square$, and triangle, $\mathcal{M}^\triangle$, topologies, since those contain the $D_1\bar D$ intermediate state.
As the second triangle in figure \ref{Fig:Jpsipipi} c) is divergent, due to the internal $P$-wave vertex
that is connected to a $J/\psi$ coupling to a pair of $D^{(*)}$ mesons, 
a counter term $\mathcal{M}^\triangle_\text{CT}$ is also introduced
\begin{widetext}
    
\begin{equation}
\begin{aligned}
	\mathcal{M}_{Y \rightarrow \jpsi \pi \pi}^{i}=& G_Y  \left[ \left(\mathcal{M}_{Y \ \text{CT}}^{\jpsi \pi \pi} \right)^{il}- \frac{g_{Y0}}{\sqrt{2}}\left( h^\pi_{1 d} (3 p_{\pi 1}^i p_{\pi 1}^j-\delta^{ij} p_{\pi 1}^2) - h^\pi_{1 s} \omega_{\pi 1} \delta^{ij} \right)\right.\\
 & \hspace{4cm} \times \left. 2 \left( (\mathcal{M}^\square)^{jl} + (\mathcal{M}^\triangle)^{jl} + (\mathcal{M}^\triangle_\text{CT})^{jl} \right) {\color{white} \frac{1}{1}}\right] \epsilon_{\jpsi}^{* \ l} \\
 & \hspace{4cm} + \left( p_{\pi_1} \leftrightarrow p_{\pi_2}\right) \\
     \mathcal{M}_{\psi \rightarrow \jpsi \pi \pi}^{i}=&G_\psi  \left[ \beta_1^{(2)} (\beta_2^{(2)} + E_{\jpsi \pi_1} ) \right] g^{Z \, il}_{\jpsi \pi} \omega_{\pi_1}  G_Z(E_{\jpsi \pi_1}) \epsilon_{\jpsi}^{* \ l} + \left( p_{\pi_1} \leftrightarrow p_{\pi_2}\right) \, ,
    \label{eq:contactterm}
\end{aligned}
\end{equation}

\end{widetext}
 where $g^{Z_c}_{\jpsi \pi}$ is the coupling of $Z_c \to \jpsi \pi$, given by the triangle transition shown in Fig.~\ref{fig:Y_to_jpsipipi}
 
\begin{equation}
        g^{Z_c \ ik}_{\jpsi \pi}=g_{Z0} (\mathcal{M}^\triangle_2)^{ik} + \omega_{\pi_2} c_\text{CT}^\triangle \delta^{ik}  \, .
\end{equation}
To reduce the run-time of the numerical evaluation of the loop integrals only two out of four contributions of the $Y(4230)$ wave function shown in Eq.~\eqref{eq:Y_wavefunc} with $p_{\pi_1} \leftrightarrow p_{\pi_2}$ are considered, as the differences due to isospin breaking are negligible small. For example, for the box topologies shown in figure~\ref{fig:R2_box_decomp} only the particle content spelled out first at each line in the boxes is evaluated explicitly, while those in brackets are included via a multiplication with a factor 2.
The $Y(4230)$ contact term $\mathcal{M}_{Y \ \text{CT}}^{\jpsi \pi \pi}$ has two contributions, one in the $\pi \pi$ invariant mass from the subtraction polynomial of the $\pi \pi$ final state interaction and the other in $\jpsi \pi$ from the chiral contact term and intermediate $Z_c(3900)$

\begin{equation}
\begin{aligned}
    \left(\mathcal{M}_{Y \ \text{CT}}^{\jpsi \pi \pi} \right)^{ik}&= G_{Z} g^{Z \, ik}_{\jpsi \pi} \omega_{\pi_1} \\
    &\times \left[ \alpha_1^{(2)} (\alpha_2^{(2)} + E_{\jpsi \pi}) \right]\\
    & + \Omega_{11} M_{0}^{\pi \pi} + \frac{2}{\sqrt{3}} \Omega_{12}  M_{0}^{K K}  \, ,
\end{aligned}
\end{equation}
 with $c_\text{CT}^\triangle$ denoting the free parameter of the triangle counterterm. The amplitudes of the loop diagrams are given below, where the notation and numerical implementation are discussed in the appendix \ref{sec:loop_calc}
 
 \begin{widetext}
     
\begin{equation}
\begin{aligned}
    (\mathcal{M}^\square)^{jl}&=\mathcal{B}^\text{I}(g^\pi_1 g_{\jpsi D^* D^*},q_\text{I}^l p_{\pi_2}^j -p_{\pi_2}^l q_\text{I}^j- \delta^{lj} (p_{\pi_2} \cdot q_\text{I}))+\mathcal{B}^\text{II}(g^\pi_2 g_{\jpsi D D^*},p_{\pi_2}^l q_\text{II}^j - \delta^{jl} (p_{\pi_2} \cdot q_\text{II}))\\
    &\quad +\mathcal{B}^\text{III}(g^\pi_1 g_{\jpsi D D},p_{\pi_2}^j q_\text{III}^l)\\[.3cm]
    (\mathcal{M}^\triangle)^{jl}&=\mathcal{T}_{D_1 D D^*} g_{Z0}^2  G_Z(E_{\jpsi \pi}) (\mathcal{M}^\triangle_2)^{jl}\\
    (\mathcal{M}^\triangle_2)^{jl}&= \mathcal{T}_2^1 ( g_2^\pi g_{\jpsi D D^*},p_{\pi_2}^l q_\text{I}^{\prime \ j} - \delta^{lj} (p_{\pi_2} \cdot q_\text{I}^\prime) )+\mathcal{T}_2^2(g_1^\pi g_{\jpsi D D},p_{\pi_2}^j q_\text{II}^{\prime \ l})\\
    &\quad +\mathcal{T}_2^3 ( g_2^\pi g_{\jpsi D^* D^*}, q_\text{III}^{\prime \ l} p_{\pi_2}^j-p_{\pi_2}^l q_\text{III}^{\prime \ j} - \delta^{lj} (q_\text{III}^\prime \cdot p_{\pi_2}) ) \\[.3cm]
(\mathcal{M}^\triangle_\text{CT})^{jl}&=\mathcal{T}_{D_1 D D^*} g_{Z0}  G_Z(s_{\jpsi \pi}) c_\text{CT}^\triangle \delta^{jl} \omega_{\pi_2} \, ,
\end{aligned}
\end{equation}

 \end{widetext}
 where the $q_\text{I},q_\text{II},q_\text{III},q_\text{I}^\prime,q_\text{II}^\prime,q_\text{III}^\prime$ denote the relative momenta at the $J/\psi D^{(*)} D^{(*)}$ vertex for the different box and triangle topologies. Additional free parameters come from the production polynomials of the $Y(4230)$ and $\Psi(4160)$ contact terms, namely $\alpha_1^{(2)},\alpha_2^{(2)}$ and $\beta_1^{(2)},\beta_2^{(2)}$ respectively, as well as the triangle counterterm $c_\text{CT}^\triangle$. The inclusion of the $\pi \pi- \bar{K} K$ final state interaction is discussed in Appendix~\ref{sec:pipi_fsi}.

\subsubsection{$e^+ e^- \rightarrow h_c \pi^+ \pi^-$}

In general, one expects the diagrams for $h_c \pi^+ \pi^-$ to be analogous to $\jpsi \pi \pi$, apart from the fact that the $Y(4230)$ contact term is omitted as it violates HQSS. 
In addition, since the $h_c \pi$ subsystem does not show any prominent signal of the $Z_c(3900)$, no triangle operators are included in this study. 
Meanwhile, the $h_c \pi$  subsystem shown in Ref.~\cite{BESIII:2013ouc} shows a strong peak from the $Z_c(4020)$, which would, however, require to include the coupling of $Y(4230) \to D_1 D^*$, as the $Z_c(4020)$ couples strongly to $D^* \bar{D}^*$. 
On the other hand,  the $Z_c(4020)$ is not anticipated to generate significant structures in the total cross section of $h_c \pi\pi$, which is part of the current analysis. 
The  inclusion of this state will be postponed for the upcoming full coupled channel analysis, such that for now we only consider the box topologies, where the free parameters are fixed by $D^0 D^{* -} \pi^+$, $\jpsi \pi^+ \pi^-$ and the two-body final states. The amplitude therefore reads

\begin{equation}
\begin{aligned}
	\mathcal{M}_{Y \to \hc \pi \pi}^{i}=& \frac{G_Y g_{y0}}{\sqrt{2}} \\
 & \hspace{-1cm} \times \left[h_{1s}^\pi \omega_{\pi_1} \delta^{ij} {-} h_{1 d}^\pi(3 p_{\pi 1}^i p_{\pi 1}^j {-} \delta^{ij} p_{\pi 1}^2) \right]\\
 & \hspace{-1cm}\times \mathcal{M}^{h_c \pi \pi}_\square m_{h_c} \epsilon_{ljm} p_{\pi_2}^m  \epsilon_{\hc}^j \, ,
\end{aligned}
\end{equation}
with $\mathcal{M}^{h_c \pi \pi}_\square$ given by
\begin{equation}
\begin{aligned}
    \mathcal{M}^{h_c \pi \pi}_\square&=&\frac{4 g m_{D^*}^{3/2}\sqrt{m_D m_{h_c} m_{\chi_{c0}}}}{\sqrt{3} f_\pi f_{\chi_{c0}}}\\
 &\times&  \left( \mathcal{B}^{h_c \pi \pi}_\text{I}{+}\mathcal{B}^{h_c \pi \pi}_\text{II} \right) \, .
\end{aligned}
\end{equation}
\subsubsection{$e^+ e^- \rightarrow \jpsi K^+ K^-$}

With $\jpsi \pi \pi$ included in the study, we can also easily access $\jpsi K \bar K$, as the main contribution is expected to go via the $\pi \pi \to K \bar K$ final state interaction in the $S$-wave, where no new parameters need to be introduced. Here the contributions of the triangle topologies are negligible, as the partial wave projection on the $\pi \pi$ system
contains a tiny $S$-wave piece due to the presence of the near on-shell $Z_c(3900)$ in the $\jpsi \pi$ subsystem.  The amplitude is given by

\begin{widetext}
    
\begin{equation}
\begin{aligned}
    \mathcal{M}_{Y \rightarrow \jpsi K K}^i=& G_Y \left( \left(\mathcal{M}_{Y \ \text{CT}}^{\jpsi K K} \right)^{il}  - \left( h^\pi_{1 d} (3 p_1^i p_1^j-\delta^{ij} p_1^2) + h^\pi_{1 s} \omega_{\pi_1} \delta^{ij} \right) \right. \\ 
    & \hspace{4cm} \times \left. \left[ \mathcal{M}_{Y \rightarrow \jpsi K K}^{jl} +  \left[\left(\mathcal{M}_{\jpsi \pi \pi}^{\text{loop}}\right)^{jl} \right]^\text{FSI}_{\pi \pi \to K K}  \right] \right) \epsilon_{\jpsi}^l \, ,
\end{aligned}
\end{equation}

\end{widetext}
where we collected the loop diagrams in the amplitude
\begin{equation}
\left(\mathcal{M}_{\jpsi \pi \pi}^{\text{loop}}\right)^{jl}=
    (\mathcal{M}^\square)^{jl} + (\mathcal{M}^\triangle)^{jl}
    \end{equation}
    and
\begin{equation}
\left(\mathcal{M}_{Y \ \text{CT}}^{\jpsi K K} \right)^{kl}{=}
     \left( \Omega_{21} M_{0}^{\pi \pi} + \frac{2}{\sqrt{3}} \Omega_{22}  M_{0}^{K K} \right) \delta^{kl} \, .
\end{equation}
Furthermore, $\mathcal{M}_{Y \rightarrow \jpsi K K}^{jl}$ is a strange source shown in Fig.~\ref{fig:jpsiKK_strange_box}. 
We postpone the inclusion of strange triangles, including the $Z_{cs}(4000)$, to a later, more complete analysis. In this sense we regard this channel in this analysis as a consistency check. 
On the other hand, the $Z_{cs}(4000)$ can only appear in conjunction with an additional kaon within the triangular mechanism. Consequently, this state is expected to contribute significantly    
only in the energy range around 4470 MeV, well exceeding the energy range of interest in this study, even when accounting for the $Z_{cs}$ width.

\subsubsection{$e^+ e^- \rightarrow\chi_{c0} \omega$}

The Feynman diagrams are shown in figure \ref{Fig:chi_c0omega}. The main contribution is expected from the triangle, which scales like the scalar triangle as both the $D_1 \rightarrow D \omega$ and $D D \rightarrow \chi_{c0}$ are $S$-wave at leading order. Additionally, there are two $S$-wave contact terms for the $Y(4230)$ and $\psi(4160)$ respectively
\begin{equation}
\begin{aligned}
 	\mathcal{M}_{Y \rightarrow \chi_{c0} \omega}^{i}{=}& G_Y \left(c_{\chi_{c0} \omega}^\triangle m_\jpsi m_D \mathcal{T}_{\chi_{c0} \omega} {+} c^Y_{\chi_{c0} \omega} \right) \epsilon_\omega^i\\
    \mathcal{M}_{\psi \rightarrow \chi_{c0} \omega}^{i}{=}&G_\psi c^\psi_{\chi_{c0} \omega} \epsilon_\omega^i \, ,
\end{aligned}
\end{equation}
where $c_{\chi_{c0} \omega}^\triangle,c^Y_{\chi_{c0} \omega}$ and $c^\psi_{\chi_{c0} \omega}$ are free parameters. The width of the $\omega$ is 
included by convolving the cross section for a fixed $\omega$ mass with
the $\omega$ spectral function --- see, e.g., Ref.~\cite{Hanhart:2001ft}.

\subsubsection{$e^+ e^- \rightarrow \jpsi \eta$}

For $\jpsi \eta$, the couplings of the triangle shown in figure \ref{Fig:jpsieta} are fixed. The vector-vector-axial vector vertex of the contact terms must couple via $\epsilon^{\mu \nu \rho \sigma}$
which reduces to a three dimensional $\epsilon^{mjl}$ in the rest frame of the incoming particles:

\begin{widetext}

\begin{equation}
\begin{aligned}
\mathcal{M}_{Y \rightarrow \jpsi \eta}^{i}&{=} G_Y \left( - \frac{1}{\sqrt{6}}\left[ h^\pi_{1d}(3 p_\eta^i p_\eta^j - \delta^{ij} p_\eta^2) - h^\pi_{1 s} \omega_\eta \delta^{ij} \right] \mathcal{T}_{\jpsi \eta}\left(g^{DD^*}_\jpsi,q_l\right) + c^Y_{\jpsi \eta}  p_\eta^l \right) \epsilon^{mjl}\epsilon_\jpsi^m \\
\mathcal{M}_{\psi \rightarrow \jpsi \eta}^{i}&{=} G_\psi c^\psi_{\jpsi \eta} \epsilon^{ijl} p_\eta^j \epsilon_{\jpsi}^l \, ,
\end{aligned}
\end{equation}

\end{widetext}
where $q$ denotes the relative momentum at the $\jpsi$ vertex and $c^Y_{\jpsi \eta}$ and $ c^\psi_{\jpsi \eta}$ are free parameters. We do not consider the mixing of the singlet $\eta_1$ and octet $\eta_8$ to the physical $\eta$ and $\eta^\prime$ states, but just match $\eta_8=\eta$, as the mixing effects are small.

\subsubsection{$e^+ e^- \rightarrow X(3872) \gamma$}

\label{Sec:Form_ee_mumu}

The diagrams for $Y(4230)\to X(3872) \gamma$ are shown in Fig.~\ref{Fig:X3872gamma}, and are analogous to $Y(4230) \to Z_c\pi$ as well as $\jpsi \eta$. However, the quality of data for $X(3872) \gamma$ does not allow one to distinguish between the triangle and contact transition of $Y(4230) \to X(3872) \gamma$, such that we omit the latter from the start\footnote{We can get equally good fits to the data by replacing the triangle by the contact term, since the quality of the data does not allow one to see the different energy dependences of the two amplitudes.}. The vector-vector-axial-vector coupling of $D_1 \to D^* \gamma$ scales with $\epsilon^{kjl}$, such that the amplitude is given by
\begin{equation}
\begin{aligned}
\mathcal{M}_{Y \rightarrow X \gamma}^{i}&=  G_Y c^Y_{X \gamma} \mathcal{T}_{X\gamma} \epsilon^{ijl} \epsilon_\gamma^j \epsilon_{X}^l\\
\mathcal{M}_{\psi \rightarrow X\gamma}^{i}&= G_\psi c^\psi_{X\gamma} \epsilon^{ijl} \epsilon_\gamma^j \epsilon_{X}^l \, ,
\end{aligned}
\end{equation}
with $c^Y_{X \gamma}$ and $c^\psi_{X \gamma}$ being free parameters to be determined in a fit.

\subsubsection{$e^+ e^- \rightarrow\mu^+ \mu^-$}
\label{sec:formal_ee_mum}

 As already explained in section \ref{sec:intro_ee_mumu} we consider three main contributions for $e^+ e^- \to \mu^+ \mu^-$, namely 
\begin{eqnarray}\nonumber
    \sigma_{e^+ e^- \to \mu^+ \mu^-}&=&\sigma_{e^+ e^- \to \mu^+ \mu^-}^\text{tree} \\
    & & \quad \times\left|  1{+}
    \mathcal{A}_R
    {+}\mathcal{A}_{\rm mix}\right|^2
    \label{mumucs}
   \end{eqnarray}
   with
   \begin{equation}
 \sigma_{e^+ e^- \to \mu^+ \mu^-}^\text{tree} =   \frac{4 \pi \alpha^2}{3 s}   
   \end{equation}
   for the tree level amplitude   and we introduced
   \begin{equation}
     \mathcal{A}_R=  \sum_{R=Y,\psi} g_{\gamma R} G_R g_{\gamma R}
   \end{equation}
   and
   \begin{equation}
   \mathcal{A}_{\rm mix}=    \sum_{R \neq R^\prime} g_{\gamma R} G_R \mathcal{M}_\text{mix}^{R R^\prime} G_{R^\prime} g_{\gamma R^\prime} \ ,
   \label{defAmix}
   \end{equation}
where $g_{\gamma R}= \exp(i \delta_{R \gamma}) e m_R^2/f_R$ defined in Eq.~\eqref{eq:photon_VMD} with $\delta_{R \gamma}$ denoting a phase factor discussed in Sec.~\ref{sec:results}. 
The individual terms in Eq.~\eqref{mumucs} represent the different diagrams shown in Fig.~\ref{Fig:mumu}.
The imaginary part of $\mathcal{M}_\text{mix}^{R R^\prime}$ is fixed by unitarity and can be reconstructed from the optical theorem
\begin{equation}
\begin{aligned}
    &\text{Im} \, \mathcal{M}^{R R^\prime} _\text{mix} =\\
    &\quad \frac{1}{2} \sum_f \int \text{d} \Pi_f \,\,\, \mathcal{M}^*(R^\prime \to f) \mathcal{M}(R \to f) \, ,    
\end{aligned}
\end{equation}
where $f=DD^*\pi,\chi_{c0} \omega, \jpsi \eta$ and $X(3872) \gamma$ are all final states with significant contributions from both $\psi(4160)$ and $Y(4230)$ studied in
this work. Note that the sum runs over all allowed final states with the given particle content --- accordingly $f=[DD^*\pi]$ should be 
understood as
\begin{equation}
    \begin{aligned}
        [DD^*\pi]=&\{D^- D^{* 0} \pi^+,D^- D^{* +} \pi^0,D^+ \bar{D}^{* 0} \pi^-,\\
        &D^+ D^{* -} \pi^0, \bar D^0 D^{* +} \pi^-, \bar D^0 D^{* 0}\pi^0,\\
        &  D^0 D^{* -} \pi^+, D^0 \bar{D}^{* 0} \pi^0\}\, .
    \end{aligned}
    \label{DDstpichannels}
\end{equation}
Since all those channels are connected via isospin symmetry,
they can be included via a proper multiplicity factor --- clearly for
that we need to neglect, e.g., the mass differences between the different channels. For example,
for $D D^* \pi$ we denote decay amplitudes for the transition of $Y(4230)$ and $\psi(4160)$ to the experimentally measured channel $D^0 D^{* -} \pi^+$ as $A$ and $B$, respectively,
\begin{equation}
    \begin{aligned}
        \mathcal{M}(Y \to D^0 D^{* -} \pi^+) =&A \\
        \mathcal{M}(\psi \to D^0 D^{* -} \pi^+) =&B \,,
    \end{aligned}
\end{equation}
where in accordance to Eq.~\eqref{defAmix} $A$ and $B$ do not contain the resonance propagators, but only the decay vertices. Summing over all  channels 
one therefore obtains
\begin{equation}
\begin{aligned}
   &\text{Im} \, \mathcal{M}^{Y(4230) \psi(4160)}_\text{mix} \\
   &= \frac{1}{2} \sum_{f} \int \text{d} \Pi_{f} \, \mathcal{M}^*( \psi \to {f}) \mathcal{M}(Y \to {f}) \\
   &=\frac{1}{2} \int \text{d} \Pi \quad  4 \left(B^*A + \frac{1}{2} B^*A  \right)\\
   &=\frac{1}{2} \int \text{d} \Pi \quad 6 B^*A  \, ,
\end{aligned}
\label{eq:mumu_factors}
\end{equation}
where ${f}\in [DD^*\pi]$
 was defined in Eq.~\eqref{DDstpichannels}.
The factor $4$ in Eq.~\eqref{eq:mumu_factors} arises from the four different decay modes of the $Y(4230)$ wave function given in Eq.~\eqref{eq:Y_wavefunc}. For each mode the subsequent $D_1$ decay can produce a charged or a neutral pion, e.g. $D_1^0$ can decay into $D^{* \, 0} \pi^0$ and $D^{* \, +} \pi^-$, where the amplitudes scale as $1$ and $1/\sqrt{2}$, respectively, due to the isospin factors.
The additional factors arising in the other channels are $3/2$ for $\jpsi \pi \pi$ and $1$ for $\chi_{c0} \omega, \jpsi \eta$ and $X(3872) \gamma$.
The real part of $\mathcal{M}_\text{mix}^{R R^\prime}$ can in principle also be constructed dispersively, however, there is still freedom in the subtraction constant. So for now we just approximate it via a real constant

\begin{equation}
\begin{aligned}
     &\mathcal{M}_\text{mix}^{R R^\prime}=\frac{c_\text{mix}}{2}\\
     & \quad + \frac{i}{2} \sum_f \int \text{d} \Pi_f \,\,\, \mathcal{M}^*(R^\prime \to f) \mathcal{M}(R \to f)\, .
\end{aligned}
\end{equation}

\section{Fit Strategy, Results and Discussion}
\label{sec:results}

In 2022 and 2023 BESIII published new XYZ data sets for $\jpsi \pi^+ \pi^-$ \cite{BESIII:2022qal} and $D^0 D^{* -} \pi$ \cite{BESIII:2023cmv}
with very impressive statistics.
Those data clearly highlight the asymmetric lineshapes of the total cross sections in these two channels.
It turns out that from those channels most of the parameters specific for the $Y(4230)$ are fixed.
 The $Z_c(3900)$ shows up prominently 
  only in the $D \bar{D}^*$ \cite{BESIII:2015pqw} and $\jpsi \pi^\pm$ \cite{BESIII:2017bua} subsystems of those channels.
 To get a better constraint on the light quark SU(3) singlet and octet components (for details see Appendix \ref{sec:pipi_fsi}) we also include $\jpsi K^+ K^-$ in the first fit. This may overestimate the contributions of the contact term in $\jpsi K^+ K^-$ to some extend as it needs to compensate for  a possible contribution from the missing $Z_{cs}(4000)$ triangle, but allows us to reduce the correlation of the parameters.
 We do not include the data for the $\jpsi \pi^0 \pi^0$ channel in the fit, due to their reduced statistics in comparison to $\jpsi \pi^+ \pi^-$. Since $\mu^+ \mu^-$ is the only channel showing a clear separation of the $Y(4230)$ and $\psi(4160)$ signals and their interference, 
 it is also included in the first fit. This further allows us to properly 
 separate photon and
 strong couplings, since
 in the hadronic channels they only appear as a product. 
 With this in mind, our fit strategy is the following:
 
\begin{enumerate}
    \item The resonance parameters of the $Y(4230)$ and $Z_c(3900)$, as well as the channel dependent parameters of $D^0 D^{* -} \pi$, $\jpsi \pi^+ \pi^-$, $\jpsi K^+ K^-$ and $\mu^+ \mu^-$ are fitted simultaneously to the $D^0 D^{* -} \pi$, $\jpsi \pi^+ \pi^-$, $\jpsi K^+ K^- $ and $\mu^+ \mu^-$ total cross sections, the $D {\bar D^{*}}$, $\jpsi \pi^\pm$ and $\pi^+ \pi^-$ invariant mass distributions, and the pion Jackson angle  extracted from $D^0 D^{* -} \pi^+$.
    \item With the resonance and channel dependent parameters of $D^0 D^{* -} \pi^+$, $\jpsi \pi^+ \pi^-$ and $\mu^+ \mu^-$ being fixed, the remaining parameters in the channels $\chi_{c0} \omega, \jpsi \eta$ and $X(3872) \gamma$
    are fitted to the corresponding cross sections data.
    \item At last, the parameters obtained in the previous steps are used as initial parameters for a global fit to all observables. 
\end{enumerate}

If we were working with a complete formalism, with all relevant channels dynamical and unitarity imposed, all parameters would necessarily be real. However, here some ingredients are approximated. e.g. as shown
in Ref.~\cite{Li:2013yka} the direct transition of a photon to the $D_1(2420)\bar D$ intermediate state that predominantly couples
to the $Y(4230)$, if it is a hadronic molecule, is suppressed by heavy quark spin symmetry, since this narrow $D_1$ state has a light quark cloud with
$j=3/2$. On the other hand, there is no such suppression for the transition of the photon to $D_1(2430)\bar D$, where the broad $D_1(2430)$ has its light quark cloud
with $j=1/2$. The $D_1(2430)\bar D$ intermediate state may thus act as a doorway state to feed the production of the molecule. This effect can be included effectively via a complex coupling of the $Y(4230)$ to the photon. Moreover, the $\psi(4160)$ production from a photon sits in the tail of the $\psi(4040)$~\cite{BESIII:2020peo} 
--- an effect which may also be included by allowing for a complex coupling.
 It is worth noting that, while all hadronic cross sections are sensitive  to the difference of those two phases only, the leptonic cross section $e^+e^-\to \mu^+\mu^-$ probes the phases individually, as shall be discussed below.
It turns out that all other parameters of the model 
can be chosen real valued. 

\begin{figure*}[t]
    \centering
    \includegraphics[width=\linewidth,trim={0pt 25pt 0pt 0pt}]{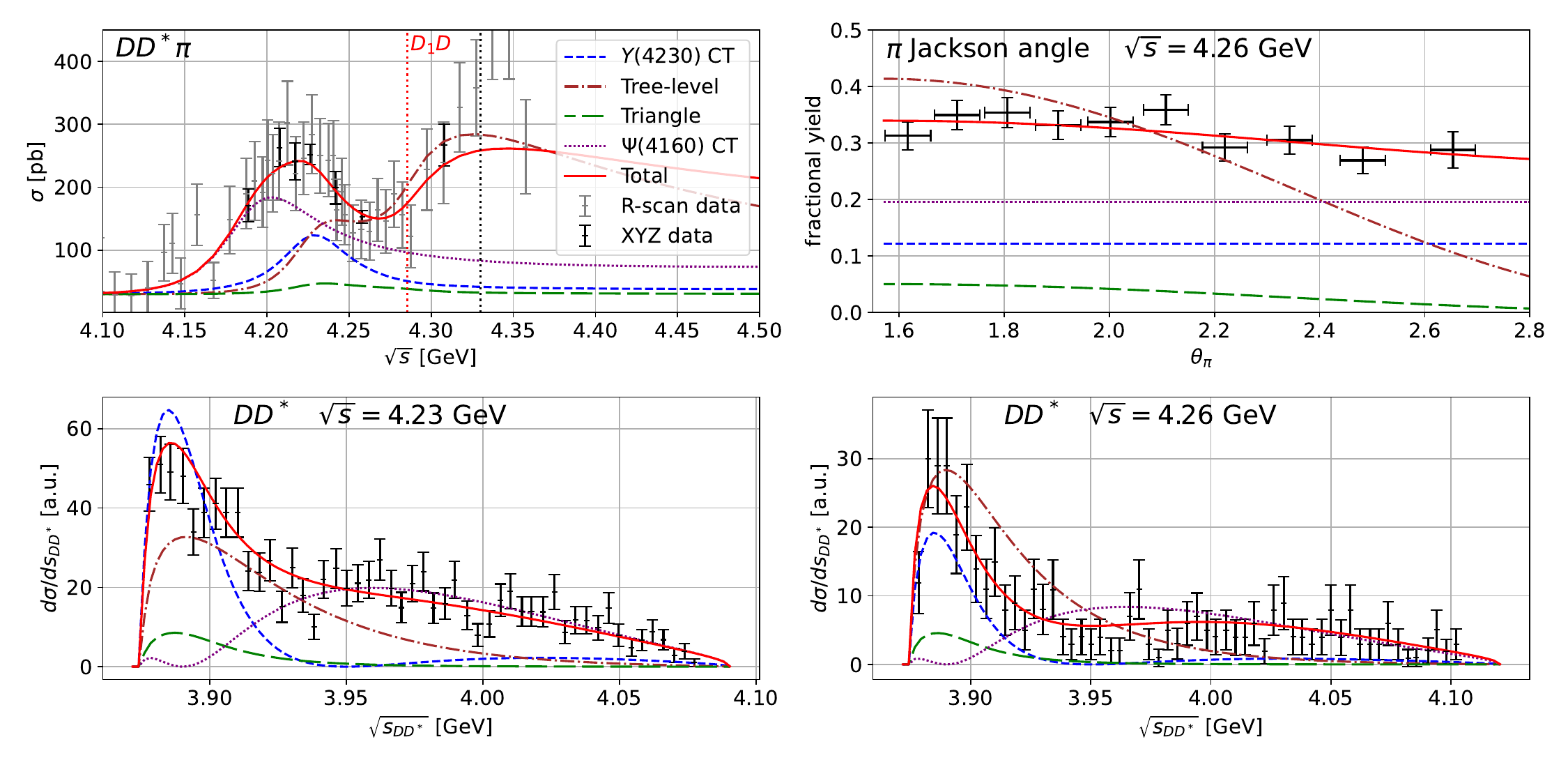}
    \caption{Fit results for the $D^0 D^{* -} \pi$ cross section, the $D^0 D^{* -}$ invariant mass distribution and the pion Jackson angle. $D^0 D^{* -} \pi^+$ R-scan and XYZ data are from Ref.~\cite{BESIII:2023cmv}, $D^0 D^{* -}$ invariant mass distribution is from Ref.~\cite{BESIII:2015pqw}.}
    \label{fig:Fit_DDspi}
\end{figure*}

\begin{figure*}[t]
    \centering
    \includegraphics[width=0.49\linewidth,trim={0pt 25pt 0pt 0pt}]{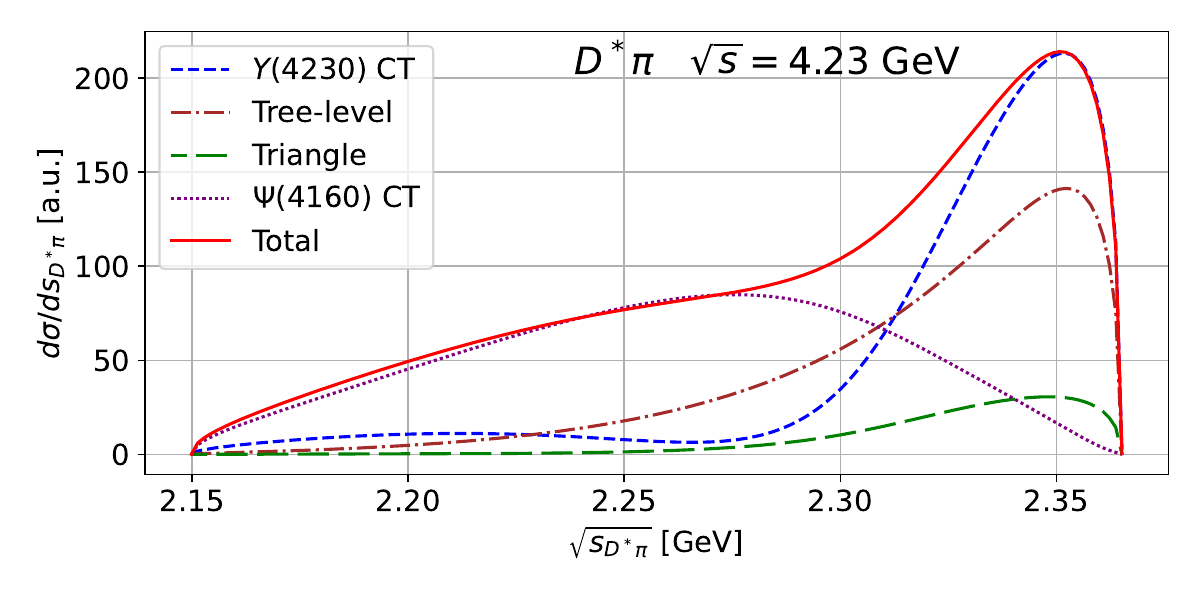}
    \includegraphics[width=0.49\linewidth,trim={0pt 25pt 0pt 0pt}]{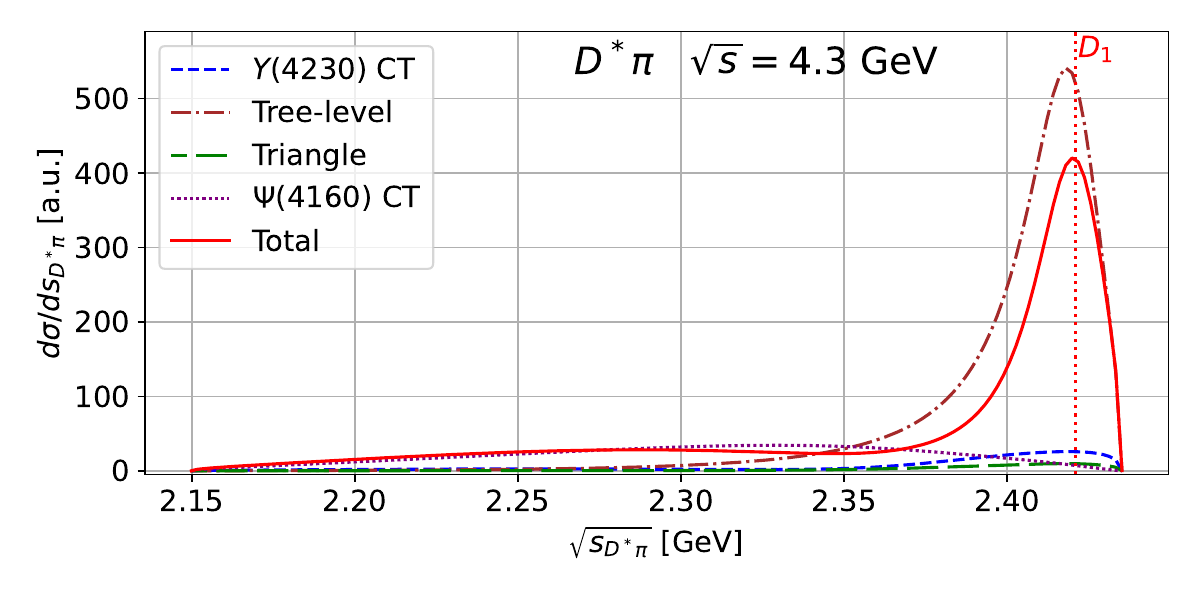}
        \caption{Predictions for the $D^*\pi$ invariant mass distributions to be measured in $e^+e^-\to \bar DD^*\pi$.
        The left (right) panel shows our prediction at 4230 (4300) MeV.}
    \label{fig:Pred_Dstpi}
\end{figure*}

\begin{figure*}[t]
\includegraphics[width=\linewidth,trim={30pt 15pt 5pt 5pt}]{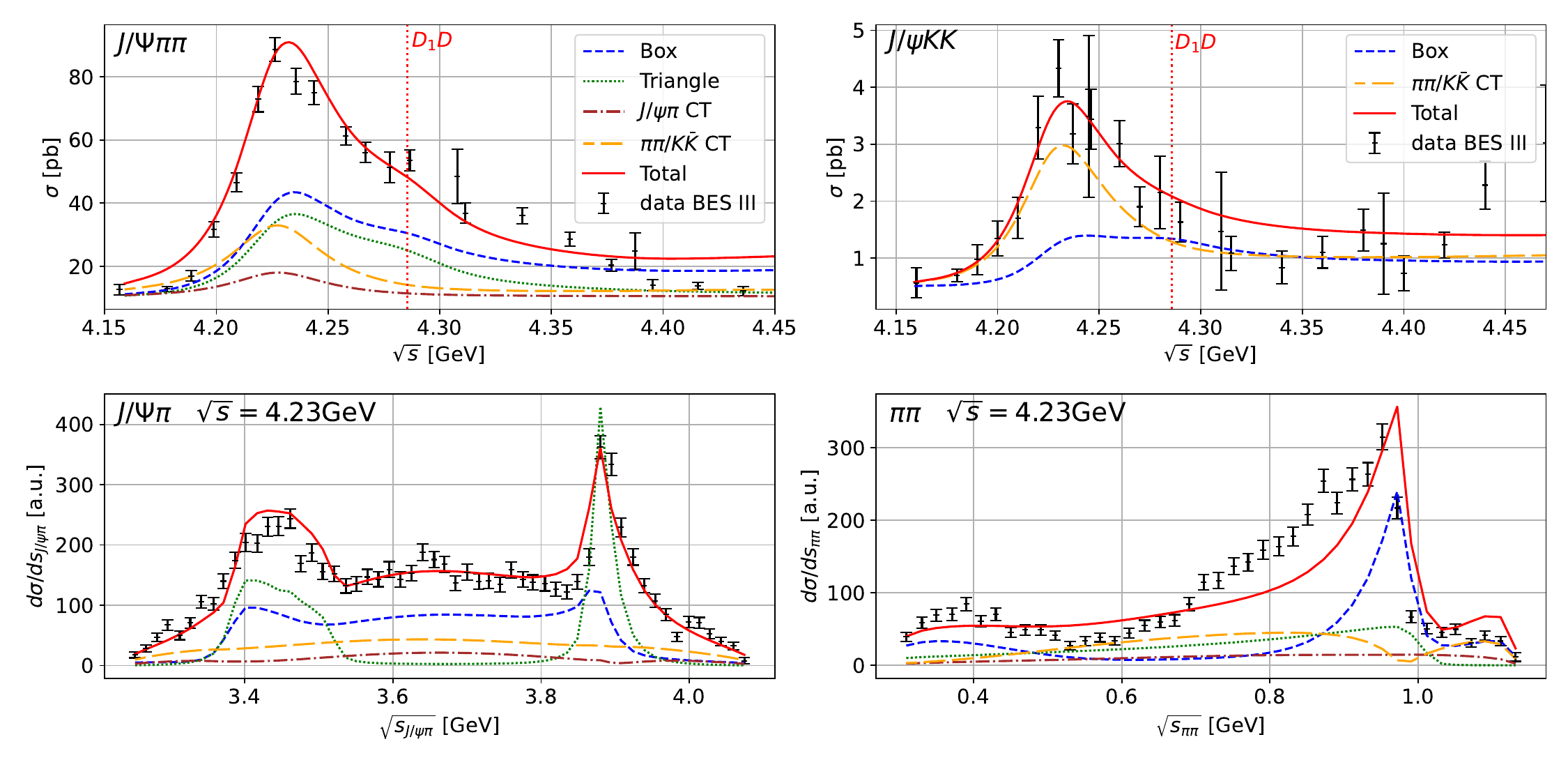}
\caption{Fit results for the $\jpsi \pi^+ \pi^-$ cross section and the $\jpsi \pi^\pm$ and $\pi^+ \pi^-$ invariant mass distributions. $\jpsi \pi^+ \pi^-$ XYZ data from Ref.~\cite{BESIII:2022qal}, $\jpsi \pi^\pm$ and $\pi^+ \pi^-$ invariant mass distribution from Ref.~\cite{BESIII:2017bua}.
The data for the $J/\psi K\bar K$ channel are taken from Ref.~\cite{BESIII:2022joj}.}
\label{fig:Fit_jpsipipi}
\end{figure*}

\begin{figure}[t]
    \centering
    \includegraphics[width=1.1\linewidth,trim={15pt 15pt 0pt 40pt}]{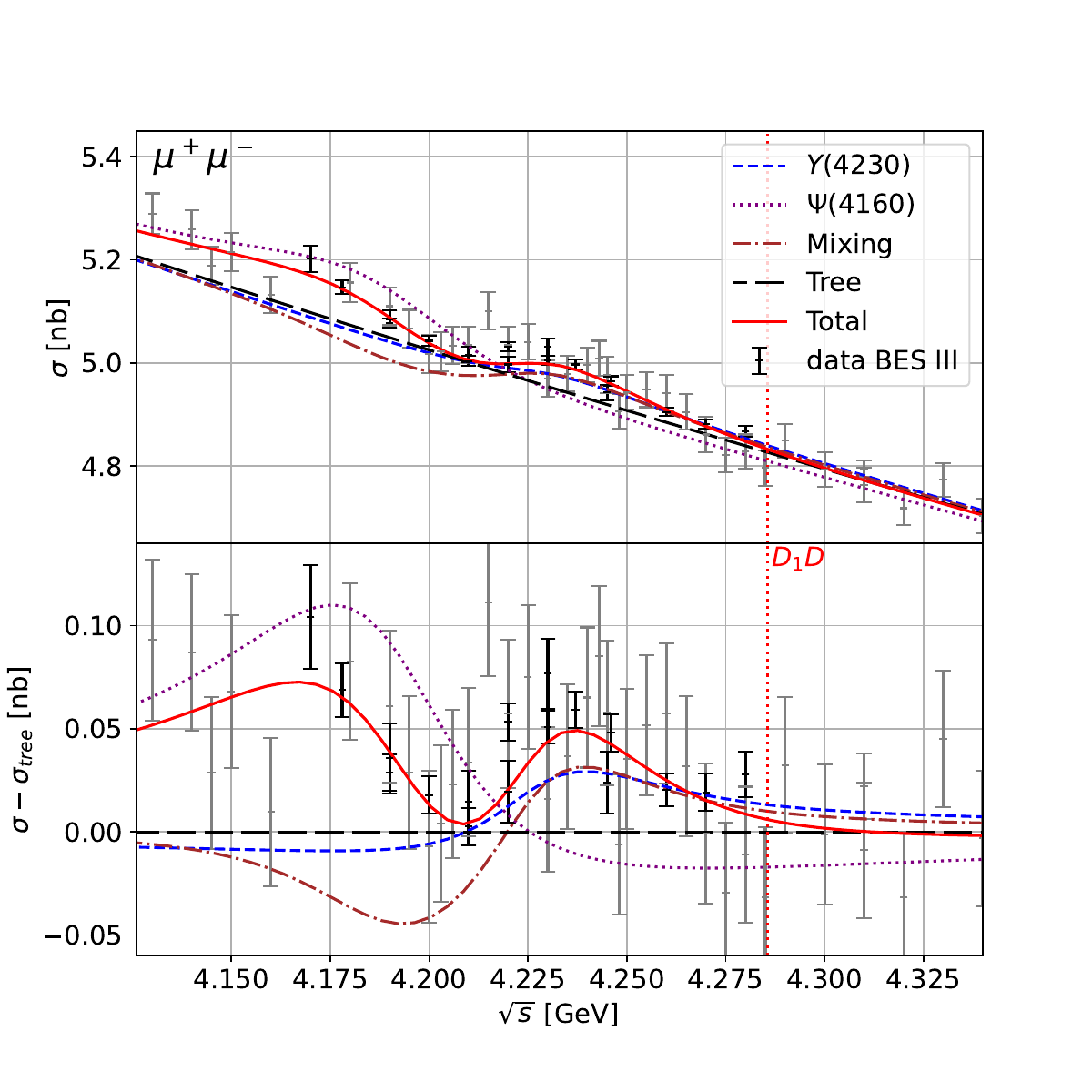}
    \caption{Fit results for the $\mu^+ \mu^-$ cross section. Upper panel: The measured born cross section; lower panel: The same, however, with the cross section from the tree level amplitude subtracted. The data are taken from Ref.~\cite{BESIII:2020peo}, where the data points with an uncertainty smaller than $32$ pb are shown in black to better highlight the structure in the data.}
    \label{fig:Fit_mumu}
\end{figure}

\begin{figure}[t]
    \centering
    \includegraphics[width=\linewidth]{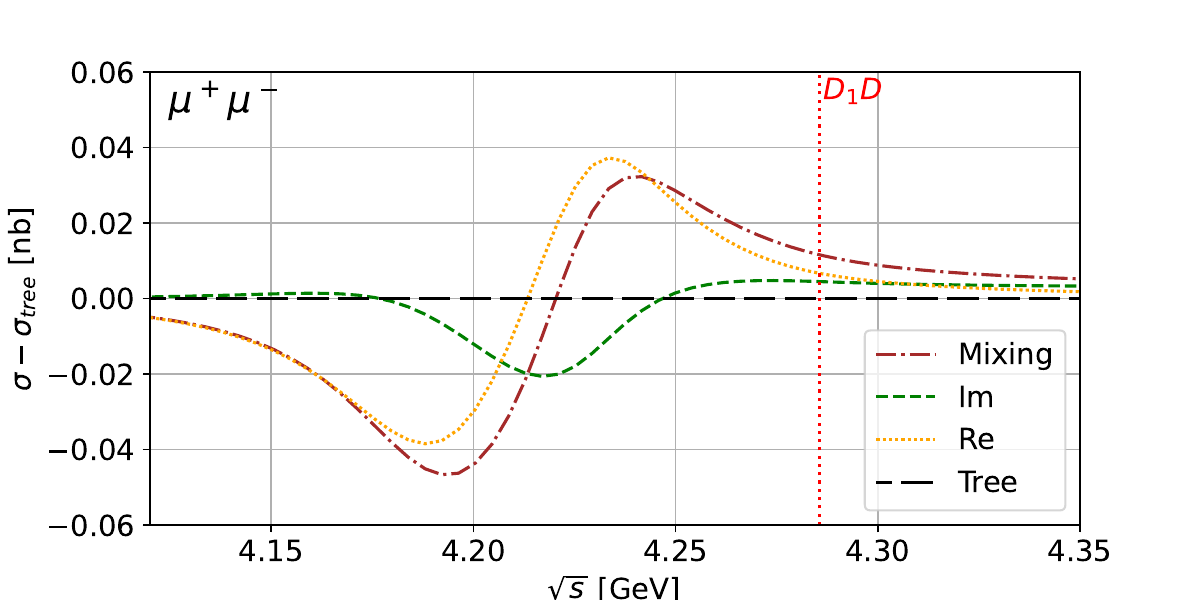}
    \caption{Contributions to the cross section difference
    from the real and imaginary parts of the mixing of $Y(4230)$ and $\Psi(4160)$ in $e^+e^- \to \mu^+ \mu^-$, denoted by 
     $\cal A_{\rm mix}$ in Eq.~\eqref{defAmix}.  The brown dash-dotted curves here and in Fig.~\ref{fig:Fit_mumu} are identical.}
    \label{fig:mumu_interference}
\end{figure}

\begin{figure}[t!]
    \centering
    \begin{minipage}[t]{0.496\textwidth}
        \includegraphics[width=\linewidth,trim={30pt 15pt 5pt 5pt}]{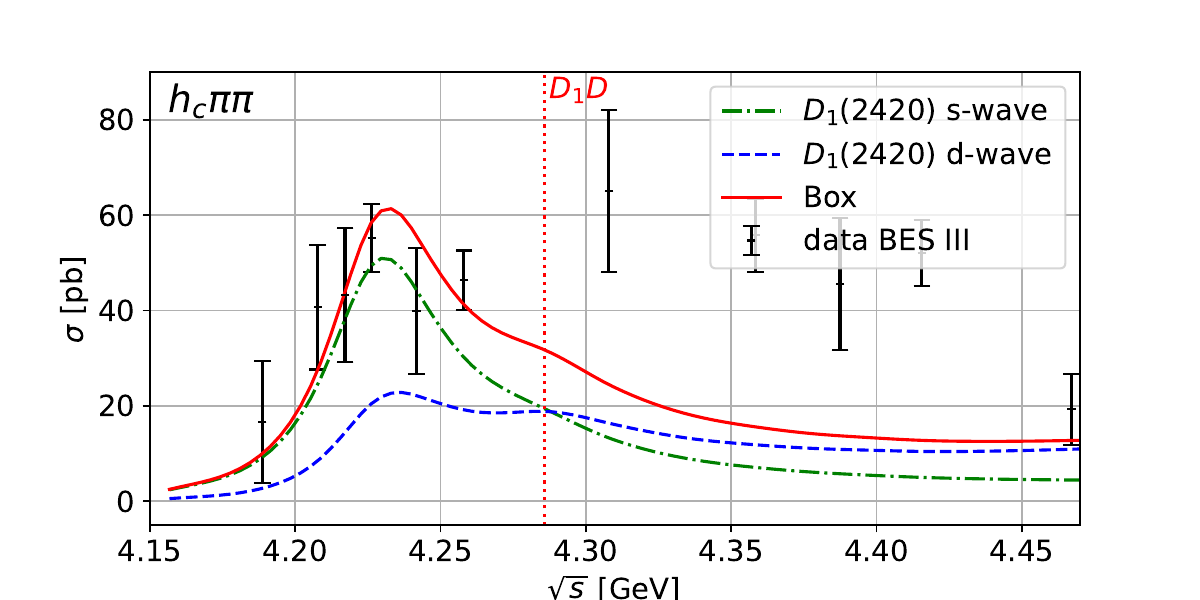}
    \end{minipage}
    \begin{minipage}[t]{0.496\textwidth}
      
    \end{minipage}
    \caption{Prediction for the $h_c \pi^+ \pi^-$ cross section. The data are taken from Ref.~\cite{BESIII:2016adj}.}
    \label{fig:Fit_hc}
\end{figure}

The results of the fits are shown in Fig.~\ref{fig:Fit_DDspi} to Fig.~\ref{fig:Fit_hc}, the parameters and statistical uncertainties that emerge from the fit are listed in Table \ref{tab:fit_params}. The numerical inputs for the particles masses and widths are given in Table \ref{tab:input_params}. We see that the interplay of the $\psi(4160)$ and $Y(4230)$ is important
and shows a non-trivial impact in almost all final states. This naturally explains the large scatter of the resonance parameters of the $Y(4230)$ in the single channel analyses of BESIII --- c.f. Fig.~\ref{Fig:comparison}.
\begin{center}
\begin{table}[ht]
\begin{tabular}{ l l r l }
\hline
 & Name \hspace{.cm} &  & Value \\ 
\hline
$Y$ \hspace{.cm}  & $m_{Y}$ && $\left(4227\pm 0.4 \right) \mev$ \\
& $g_{Y0}$ & $-$&$\left( 10.4\pm 0.2 \right) \gev$ \\
& $\Gamma_\text{in}^Y$ && $\left(54\pm1\right) \mev$ \\
& $1/f_Y$ & $-$&$\left( 0.012 \pm 0.001 \right)$\\
& $\delta_{Y \gamma}$ && $\left( 17.1\pm 0.1 \right)^o$  \\
\hline
$\psi$&  $1/f_\psi$ & $-$&$\left( 0.023 \pm 0.003\right)$\\
& $\delta_{\psi \gamma}$ && $\left(67 \pm2 \right)^o$ \\
\hline
$Z_c$ & $m_{Z}$ &  & $\left( 3884 \pm 1\right) \mev$\\
& $g_{Z0}$ &  & $\left( 4.15 \pm 0.06 \right) \gev$\\
& $\Gamma_\text{in}^Z$ & & $\left( 48 \pm 1 \right) \mev $  \\
\hline
$D \bar D^{*} \pi $ & $\alpha_1^{(1)}$ & $-$ &$\left( 128 \pm 12\right) $\\
& $\alpha_2^{(1)}$ & $-$ & $\left(3.95\pm 0.01 \right) \gev$ \\
& $\beta_1^{(1)}$ & $-$& $\left(202 \pm 18 \right)  $\\
& $\beta_2^{(1)}$ & $-$ & $\left(3.89\pm 0.1 \right) \gev$\\
\hline
$\jpsi \pi^+ \pi^-$ & $\alpha_1^{(2)}$ & $-$ & $\left( 133.9\pm 4 \right)$	\\
& $g_1$ &  $-$ & $\left( 14.9\pm 0.9\right) 10^{-3}$\\
& $g_8$ &  & $\left( 24\pm 1\right) 10^{-3}$\\
& $h_1$ & $-$ & $\left( 16.8\pm 2.4\right) 10^{-3}$\\
& $h_8$ &  & $\left( 15 \pm 0.7\right) 10^{-3}$\\
& $\beta_1^{(2)}$ && $\left( 0 \pm 0.1 \right)$\\
& $c^\triangle_\text{CT}$ &$-$& $\left( 0.4\pm 0.1\right) \gev^2 $ \\
& $f_\jpsi$ && $456 \mev$  \\
\hline
$\chi_{c0} \omega$ & $c_{\chi_{c0} \omega }^\triangle$ && $\left( 1.469 \pm 0.015\right) \gev^2$\\
& $c^Y_{\chi_{c0} \omega }$ && $\left(0.36 \pm 0.07  \right) 10^{-3}$\\
& $c^\psi_{\chi_{c0} \omega }$ & $-$ & $\left(16 \pm 0.5  \right) 10^{-3}$\\
\hline
$\jpsi \eta$ & $c^Y_{\jpsi \eta }$ && $\left( 67.3 \pm 3.4\right) 10^{-3} \gev^{-1}$\\
& $c^\psi_{\jpsi \eta}$ && $\left(298\pm11 \right) 10^{-3} \gev^{-1}$\\
\hline
$X \gamma$ & $c^Y_{X \gamma}$ & & $\left(0.71\pm0.15 \right) \gev^2$\\
& $c^\psi_{X \gamma}$ & & $\left( 0.017\pm0.003\right) \gev$\\
\hline
$\mu^+ \mu^-$ & $c_\text{mix}$ & & $\left( 0.6\pm0.01 \right)$\\
\end{tabular}
\caption{Parameters of the model as determined in the fit. 
We find the value of $f_{\jpsi}$ to be strongly dependent on the fit range in $D^{0} D^{* -} \pi^+$, such
that we did not assign an uncertainty to this quantity.}
\label{tab:fit_params}
\end{table}
\end{center}

\begin{center}
\begin{table}
\begin{tabular}{ l l }
\hline 
Parameter \,\,\,& Value [MeV] \\ 
\hline
$m_\pi^0$ & 135\\
$m_\pi^\pm$ & 139.6\\
$m_K^\pm$ & 493.7\\
$m_\eta$ & 547.9\\
$m_\omega$ & 782.7\\
$m_D^0$ & 1864.8 \\
$m_D^\pm$ & 1869.7 \\
$m_D^{* 0}$ & 2006.9\\
$m_D^{* \pm}$ & 2010.3\\
$m_{D_1}$ & 2420.8 \\
$m_\jpsi$ & 3096.9\\
$m_{\chi_{c0}}$ & 3414.7\\
$m_{h_c}$ & 3525.9\\
$m_{X(3872)}$ & 3871.7\\
$m_{\psi(4160)}$ & 4191\\
\hline
$\Gamma_{D^*}$ & 83.4$\times 10^{-3}$\\
$\Gamma_{D_1}$ & 31.7\\
$\Gamma_{\psi(4160)}$ & 70\\

\end{tabular}
\caption{Input values for masses and widths used in this work, taken from the central value of the Review of Particle Physics by the Particle Data Group~\cite{ParticleDataGroup:2022pth}.}
\label{tab:input_params}
\end{table}
\end{center}

With the central values of the parameters fixed in the fits,
the pole parameters of the $Y(4230)$ can be extracted from its propagator. We find
\begin{equation}
\sqrt{s_\text{pole}^{Y(4230)}}=\left( 4227{\pm} 4 {-} \frac{i}{2}\left(50^{+8}_{-2}\right) \right) \mev \, ,
\end{equation}
where the uncertainty estimation is described in Appendix~\ref{sec:pole_uncertainty}.

We now discuss the results for the various channels in some detail.
The results for the $D^0 D^{* -} \pi$ channel are shown in Fig.~\ref{fig:Fit_DDspi}.
The apparent peak structure around $4.22 \gev$ emerges
in our study from the interplay of the $\psi(4160)$ and $Y(4230)$.
  Remarkably, this interplay  manifests differently in   the $D^0 D^{* -} \pi$ and  $J/\psi\pi\pi$ channels ---  we refer to Fig.~\ref{fig:effectofpsi} for an illustration.
In addition to this, we find a strong enhancement at the $D_1 \bar{D}$ threshold 
in the cross section, mainly driven by the prominent $D_1$  decay in $D$-wave. 
The deviations of our results
from the data, starting around $4.35 \gev$, 
are expected, as the molecular scenario  predicts an additional bound state in the $D_2 \bar{D}^*$ channel~\cite{Wang:2013kra,Ji:2022blw,Peng:2022nrj}\footnote{Another bound state is expected in the $D_1 \bar{D}^*$ channel, however, this channel does predominantly decay into $D^*\bar D^*\pi$ 
and not into the channel studied here, $D^*\bar D\pi$.}, which will be included in a
subsequent study.
 The peak at low  $D \bar D^{*}$ invariant masses is generated by
 the interplay of the tree-level two-step decay  $Y(4230)\to D_1 \bar D \to D^* \pi \bar D$, 
the contact mechanism and  the triangle operator. The last two mechanisms involve the rescattering of $D \bar{D}^*$ into the $Z_c(3900)$. The resonance parameters of the $Z_c(3900)$ are very poorly constrained. The fit seems to prefer masses slightly above the $D \bar{D}^*$ threshold, however, for
 the whole mass range
  of approximately $m_Z \in \left[ 3.86, 3.9 \right] \gev$, 
  the data are described with similar quality. In the current fit the pole closest to the real axis of the $Z_c(3900)$ appears at the $+-$ sheet with respect to the $\jpsi \pi$ and $D \bar D^*$ channels, respectively (where $+(-)$ denotes the sign of the imaginary part of the three-momentum in each channel), with $\sqrt{s^{Z_c(3900)}_\text{pole}}=\left( 3884- i 44/2 \right) \mev$. In comparison to Ref.~\cite{Chen:2023def} we find a slightly higher mass, however, double the width for the $Z_c(3900)$.  It remains to be seen if this feature is caused by the incomplete $\pi\pi-K \bar{K}$ final state interaction, used in this work. The data for the pion Jackson angle are 
  also reproduced well. 
  Contrary to Ref.~\cite{Cleven:2013mka}, in this study
the $S$-wave is more prominent due to the presence of the $\psi(4160)$ as
well as the $S$-wave decay of the $D_1(2420)$.

Naturally, a prominent contribution from the $D_1\bar D$ intermediate state not only influences strongly the energy dependence of the total cross section but also the $D^*\pi$ invariant mass distributions. Our predictions for those
at total energies near the $Y(4230)$ pole location and near the nominal $D_1\bar D$ threshold
are shown in the left and right panel of Fig.~\ref{fig:Pred_Dstpi}, respectively. 
In both panels the peak from the $D_1$ is clearly visible at the upper end of the spectrum. While the data currently available do not allow us to provide an unambiguous determination of the various parameters leaving some freedom in the 
actual height of the $D_1$ signal, the presence of such a peak is a model independent prediction of the 
molecular scenario.
Any model that does not account for the $D_1\bar D$ as a prominent component of the $Y(4230)$ wave function will not show such a structure --- as such this invariant mass distribution is a crucial observable to either support or disprove the molecular picture.

The results for the $\jpsi \pi \pi$ final state are shown in Fig.~\ref{fig:Fit_jpsipipi}. 
A linear non-interfering background of $9\, \text{pb}$ is added due to the presence of the $\jpsi \pi$ continuum.
The loop contributions, dominant in the molecular scenario, enhance the cross section at the $D_1 \bar{D}$ threshold, 
allowing for a description of the highly asymmetric lineshape with just a single pole --- in the experimental analyses of Refs.~\cite{BESIII:2016bnd,BESIII:2022qal} this asymmetry was
generated by the additional state $Y(4320)$ as described in the introduction. 
It should be noted that also in Ref.~\cite{Chen:2017uof} an analysis is presented, where, in particular,  
the $\jpsi \pi \pi$  data 
is described with essentially the same resonance content 
as presented here, along with the addition of a $\psi(4415)$ state, but without including threshold effects. In this case,  the asymmetry of the peak in the total cross section is driven by interferences, predominantly involving  $\psi(4160)$ and $\psi(4415)$. While we regard our explanation of the data as more natural, since the data indicate some structure right at the $D_1\bar D$ threshold, at some point experiment will 
allow us to choose between the two explanations, not only since the energy dependences in the mentioned energy range are different (but not sufficiently to be distinguished given the current quality of the data), but also since an analysis of the type presented in Ref.~\cite{Chen:2017uof} will provide completely different $D \bar D^*$ and $D^*\pi$ invariant mass distributions compared to the ones shown in Figs.~\ref{fig:Fit_DDspi} and \ref{fig:Pred_Dstpi}, respectively.

The $\jpsi \pi^\pm$ invariant mass distribution shows a prominent peak, generated
by the  $Z_c(3900)$ pole, the $D^* \bar{D}$ threshold 
and the nearby triangle
singularity, and its reflection. In principle, the $\jpsi \pi^\pm$ and $D^0 D^{* -}$ lineshape can also be described by just  
including the triangles and introducing a contact interaction for $D \bar D^* \to D \bar{D}^*$, where the 
cusp is then generated simply by kinematic effects of the $D \bar{D}^*$ rescattering without any resonance 
structure. However, we find that the strength of the  $D \bar D^* \to D \bar{D}^*$
transition potential, necessary for
producing the pronounced structure
in the $D^*\bar D$ invariant mass distribution,
becomes too large to justify a perturbative approach.
We confirm the observation made in 
 Ref.~\cite{Guo:2014iya} that with this strength parameter
 the next order in $D \bar{D}^*$ scattering becomes larger
 than the perturbative rescattering; moreover, resumming
 the scattering series generates a pole in the 
 subsystem. Based on this, we argue that the 
 existing data calls for the presence of a $Z_c(3900)$
 pole.

As pointed out in the introduction,  the $\psi(4160)$ needs to be included
to get a consistent simultaneous description
of the $J/\psi \pi\pi$ and $D^*\bar D\pi$
final states. We allow this well established
vector charmonium state
to contribute to both of these channels (as well as to all other channels included in
the analysis), however, the fit reveals that no significant coupling of the $\psi(4160)$
to the $J/\psi \pi\pi$ is needed.
Indeed, the fit puts the parameter $\beta_1^{(2)}$, introduced in Eq.~(\ref{eq:contactterm}),
for the production of the $\psi(4160)$ to a value consistent with zero.

The results for the $\mu^+ \mu^-$ final state
are shown in the upper panel of Fig.~\ref{fig:Fit_mumu}.
The cross section is completely dominated
by the real tree level amplitude, shown as the
first diagram in Fig.~\ref{Fig:mumu}. 
Accordingly, following Eq.~\eqref{mumucs}, the signal of interest to us
reads to very good approximation 
\begin{eqnarray}\label{eq:XSmumudiff}\nonumber
    \sigma_{e^+e^-\to \mu^+\mu^-}
    - \sigma_{e^+e^-\to \mu^+\mu^-}^{\rm tree}&\approx&\\
    & &\hspace{-4cm}
    2\sigma_{e^+e^-\to \mu^+\mu^-}^{\rm tree}{\rm Re}(\mathcal{A}_R+\mathcal{A}_{\rm mix}) \ .
\end{eqnarray}
This quantity is shown in the lower panel of Fig.~\ref{fig:Fit_mumu}.
As argued
above, we allow for complex
couplings at the resonance-photon vertices.
Contrary to all observables
studied so far, where only the relative phase of the $Y(4230)$ and $\psi(4160)$ contributions played a role, this leptonic cross section is 
 sensitive to the
 individual phases of these resonance contributions.  The phases of those couplings are thus fixed by the
fit to the $\mu^+\mu^-$
cross section.

In Ref.~\cite{BESIII:2020peo} a sum of Breit-Wigner terms  with complex couplings is used to parameterize the data including the 
$\psi(4040), \psi(4160), S(4220)\equiv Y(4230)$ (and the $\psi(4415)$, which is, however
outside our energy range of interest). We find that the complex phase 
$\delta_{\psi \gamma}$ called for
by the fit
in the production vertex of the $\psi$ to the photon agrees within uncertainties to the one of the
experimental paper.  
With this phase included we can reproduce the $\mu^+\mu^-$ lineshape in the energy range studied. We see that the contribution of the $\psi(4160)$ is dominant in comparison to that of the $Y(4230)$  at least in the energy below 4.2 GeV, as expected in the molecular scenario, since
the coupling of a photon to the $D_1\bar D$ channel violates spin symmetry~\cite{Li:2013ssa}.   One should, however,  keep in mind that there should also be some suppression
of the coupling of the $\psi(4160)$ to the photon, if
it indeed is predominantly a $D$-wave charmonium state.
The peak 
in the data near
4230 MeV in our fits emerges from both
the interference of the two resonances
and the $Y(4230)$ itself. The main contribution to the imaginary part of the pertinent mixing matrix element of the $\psi(4160)$ and the $Y(4230)$  is generated from the $D \bar{D}^{*} \pi$ intermediate 
state --- this part is fixed completely by the data for $e^+e^-\to D\bar D^*\pi$. As outlined above, the corresponding real part is here taken as a free parameter that is adjusted in the fit. It is reassuring, however, that the real and the imaginary part of the mixing amplitude contribute with comparable strength, as shown in figure \ref{fig:mumu_interference}.
Although the energy dependence emerging from the real and imaginary part of the mixing amplitude, $\cal{A}_{\rm mix}$, resembles that of a single resonance structure, it emerges from an interplay of the different resonance propagators as well as the mixing amplitude,
$\mathcal{M}_\text{mix}^{R R^\prime}$, as outlined in  Eq.~\eqref{mumucs} and below. One may naively expect that the imaginary part of the mixing matrix element does not contribute to the total cross section significantly, as only interferences of the strong amplitudes with the real tree level amplitude matter quantitatively.
However, the phases of the resonance propagators in Eq.~\eqref{mumucs} non-trivially mix real and imaginary parts of the mixing amplitude, allowing both contributions to interfere with the tree level amplitude.

The fit result for $\jpsi K^+ K^-$ is shown in the top right of Fig.~\ref{fig:Fit_jpsipipi}.
Note that the line shape emerging for this channel is closely linked 
to that of the $\jpsi \pi\pi$
channel --- there are no new independent parameters entering for this hidden strangeness channel.
In our fit the contact term is the dominant contribution. A possible reason is that it needs to absorb the effects of 
the $Z_{cs}(4000)$ and the corresponding triangle diagrams not included in this
analysis, though their main effect is expected at the energies above those considered here. The boxes again show a very strong enhancement in the cross section at the $D_1 \bar D$ threshold explaining the apparent asymmetry in the data.
We find the $Y \to \jpsi \pi^+ \pi^- \to \jpsi K^+ K^-$ contribution generated by the $\pi\pi/K\bar K$ FSI to be by far dominant in the studied energy range,  in comparison to the  box with strange $D$-mesons, as shown in Fig.~\ref{fig:jpsiKK_strange_box}, where only the $D_1 \bar{D}$ cut goes on-shell. At higher energies, above the $D^*_s \bar{D} K$ threshold at about $4.47 \gev$, 
the $D^*_s \bar{D} K$  intermediate state in this box will go on shell.  Consequently, we expect a more pronounced contribution from the strange source
in this mass range. Starting from this energy also the $Z_{cs}(4000)$ generated via the triangle mechanism should contribute considerably. 

It should be stressed that in our analysis the unusual energy dependences of the $\jpsi \pi\pi$ and $\jpsi K\bar K$
cross sections emerge from the same physics, which is natural given the approximate SU(3) flavor symmetry of QCD, while in the experimental analyses
the former is driven by an interference of the $Y(4230)$
with the new resonance $Y(4320)$~\cite{BESIII:2022qal} and the latter predominantly by the shifted threshold
for $\jpsi \bar KK$ vs. $\jpsi\pi\pi$
with some small distortion from
an interference with another new resonance, called $Y(4500)$~\cite{BESIII:2022joj}.

To complete the discussion of the final states with three hadrons, in Fig.~\ref{fig:Fit_hc}
we show the cross sections with an $h_c$
in the final state. These rates are of particular interest, since the $h_c$ has its $\bar cc$ pair in the spin singlet state, which was originally produced in a spin triplet via its coupling to the photon.
Thus, in this transition the heavy quark spin changes, at odds with heavy quark spin symmetry.
However, besides violations of that symmetry due to the relatively small charm quark mass, spin symmetry conservation can also be circumvented by the presence of hadronic molecules: In the molecular picture it is natural to 
expect the $h_c\pi\pi$
and the $J/\psi \pi\pi$
cross sections of similar order of magnitude as is confirmed by the data, since only in the
presence of a molecule the two-meson loops that
decorrelate the heavy quark spins appear at leading
order for both channels as explained above.
Moreover, by using  values for both the $J/\psi D^{(*)}\bar D^{(*)}$
and the $h_c D^{(*)}\bar D^{(*)}$
couplings available in the literature (details on how the various couplings were determined are given in App.~\ref{Sec:AppA}), we can describe the cross sections
in both channels, providing additional support for the molecular
picture. 
In the $h_c \pi\pi$ channel we observe a discrepancy
between the data and our prediction starting 
already at around 4.3 GeV. We think  this 
reflects the omission of the $D_1\bar D^*$ channel
in our study: Only once this channel is incorporated we can include
the $Z_c(4020)$ which might be responsible at least for some part of this discrepancy.

\begin{figure}[t!]
    \centering

        \includegraphics[width=\linewidth,trim={30pt 15pt 5pt 5pt}]{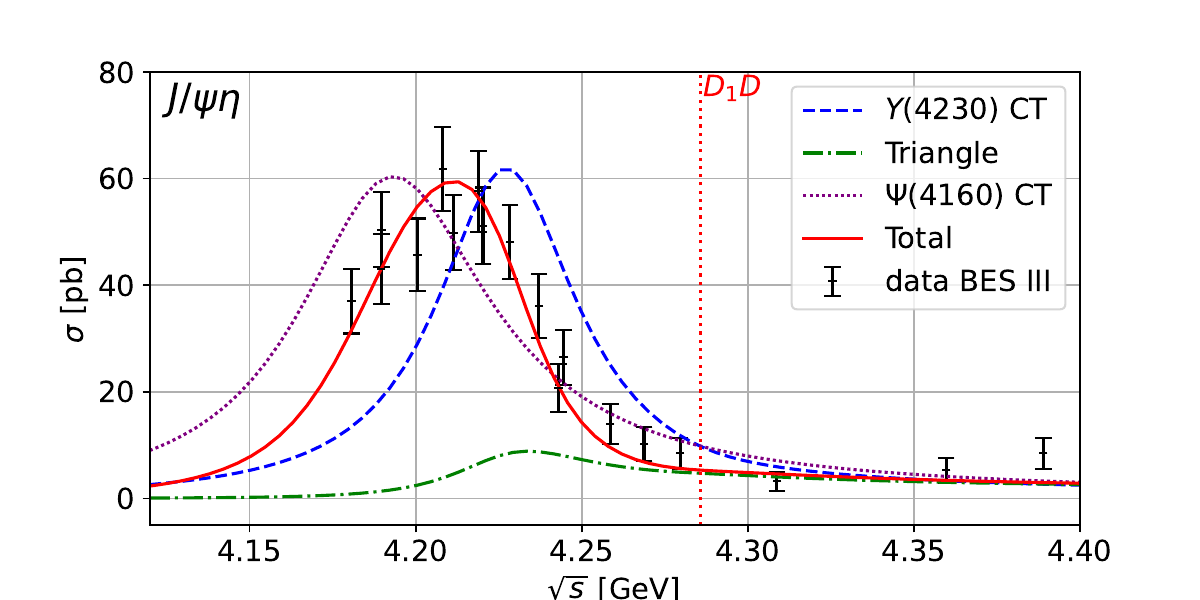}
        \includegraphics[width=\linewidth,trim={30pt 15pt 5pt 5pt}]{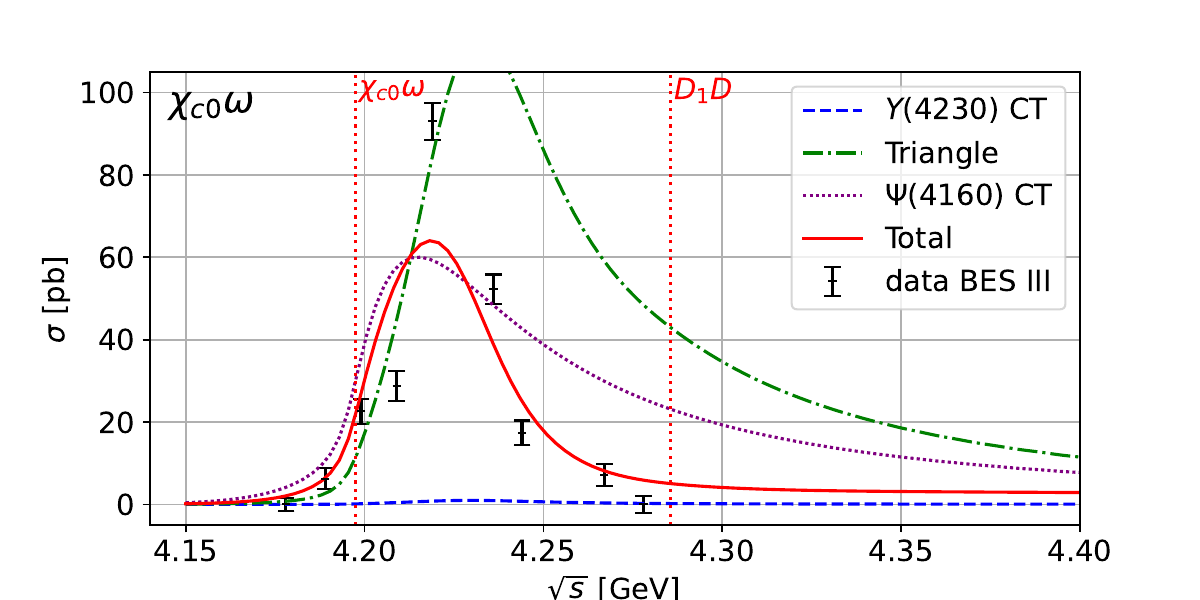}
        \includegraphics[width=\linewidth,trim={30pt 15pt 5pt 5pt}]{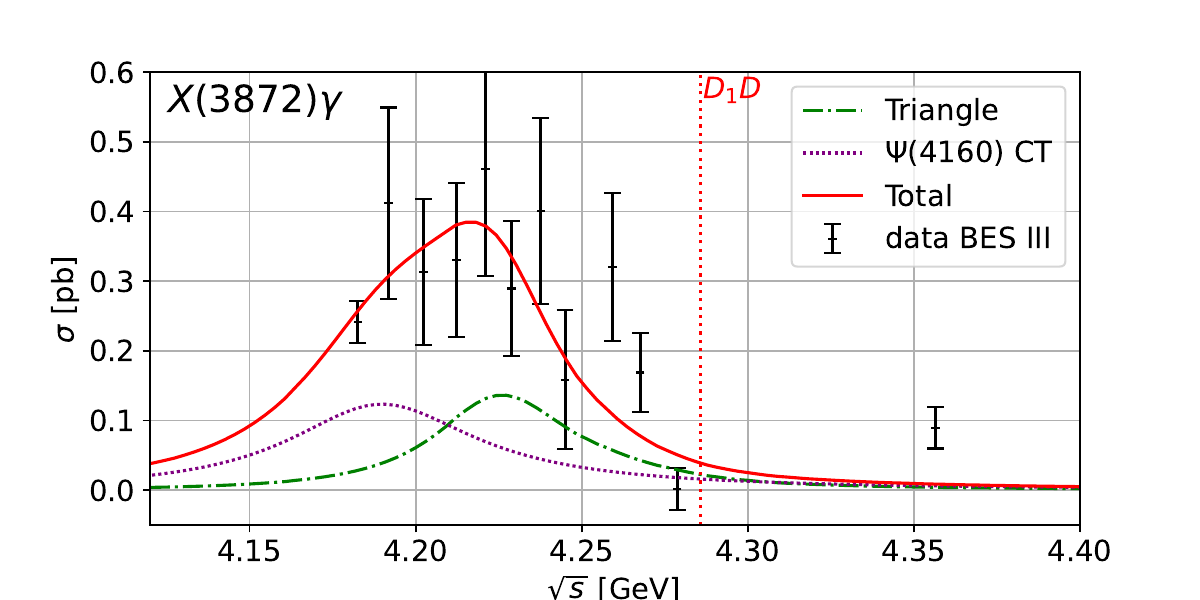}

    \caption{Fit results for the $\jpsi \eta$, $\chi_{c0} \omega$ and $X(3872) \gamma$ cross section. The data sets are from
    Ref.~\cite{BESIII:2020bgb}, Ref.~\cite{BESIII:2019gjc}, and Ref.~\cite{BESIII:2019qvy},
    respectively.}
    \label{fig:Fit_2body}
\end{figure}

We now turn to a discussion of remaining hadronic two-particle final states,
also included in 
Fig.~\ref{fig:Fit_2body}.
As one can read from the figure, 
the energy dependences of the $\chi_{c0} \omega$, the $J/\psi \eta$,  and the $X(3872)\gamma$ cross section are
rather different: While the 
first one shows a very narrow structure, the structure in the second is already a lot broader and the one in the last is more than four times as broad as the first --- this is also reflected in the  resonance parameters extracted in the single channel analyses of the \mbox{BESIII} collaboration 
collected in Fig.~\ref{Fig:comparison}.
In contrast to this, our model allows us to describe all three cross sections with consistent resonance parameters as a result of an interplay of 
the two vector resonances
 $\psi(4160)$and $Y(4230)$: the narrow peak in the $\chi_{c0} \omega$ channel emerges from a destructive interference of the triangle diagram shown in Fig.~\ref{Fig:chi_c0omega}a
and the $\psi(4160)$ contact term, shown  in Fig.~\ref{Fig:chi_c0omega}c, since the energy dependences of the two contributions are quite different, as can be clearly seen in Fig.~\ref{fig:Fit_2body} --- we included the width of the $\omega$ by a convolution of the
cross section with the omega spectral function as explained above which is the origin of the not vanishing cross section below the nominal $\chi_{c0}\omega$ cross section. 
The mechanism we propose here is different to that studied in Ref.~\cite{Cleven:2016qbn}, however, the energy dependence found there appears inconsistent with that of the newest data set 
for this channel measured at BESIII~\cite{BESIII:2019gjc}.
Also for the
 $\jpsi \eta$ and 
$X(3872)\gamma$ final states
the interplay of the two resonances is crucial,
but less dramatic.

\section{Summary and Outlook}
\label{sec:summary}

In this work we simultaneously analyzed the lineshapes of the 
cross sections for 
$e^+e^-$ to seven hadronic channels in the mass range from
$4.2$ to $4.35$ GeV as well as
data
for $e^+e^-\to \mu^+\mu^-$.
We show that a description of
all those channels is possible 
with consistent resonance parameters for a single $Y$ state with the pole
\begin{equation}
 \nonumber\sqrt{s_\text{pole}^{Y(4230)}}=\left( 4227{\pm} 4 {-} \frac{i}{2}\left(50^{+8}_{-2}\right) \right) \mev \, .
\end{equation}

This was made possible mainly because
of two features of our model: We included
the interference of the exotic
$Y(4230)$ with the more conventional 
$\psi(4160)$ and considered the effects of
the $D_1\bar D$ intermediate states. The prominence of the latter is natural in a scenario, where the $Y(4230)$ is a hadronic 
molecule in this channel.
The 
interference of the  
$Y(4230)$ and 
$\psi(4160)$ is especially 
important to get a consistent description of both channels $D^0 D^{* -} \pi^+$ and $\jpsi \pi^+ \pi^-$,  making a substantial impact on the former. It is at the same time necessary to describe
the $\mu^+ \mu^-$ cross section,
where the 
mixing of the two vector states deduced from the
strong decay channels is in fact consistent with what is needed for the leptonic final state.
We interpret this as providing additional
support for the  mixing scenario advocated 
here. The
explicit inclusion of the $D_1 \bar{D}$ intermediate states
reflects itself in a significant distortion of line shapes, which are especially prominent in final states with a $J/\psi$ and two light mesons. 
In particular, contrary to the experimental analyses,
in our study the energy dependences of the total cross sections for $e^+e^-\to J/\psi\pi\pi$ and 
$e^+e^-\to J/\psi K\bar K$ 
emerge from the same physics as expected from the approximate SU(3) flavor symmetry of QCD.

For the other final states, within our model especially the energy dependence of the $\chi_{c0}\omega$ cross section is non-trivial,
which  emerges from the distinct energy dependences of the triangle diagram, influenced by the $D_1\bar D$ intermediate state, and the $\psi(4160)$ contact term, which does not.
Moreover, within our analysis we understand that the very different lineshapes of the $\chi_{c0}\omega$ cross section and, e.g., the $J/\psi \eta$ cross section emerge naturally through the interference of the $Y(4230)$ with the $\psi(4160)$.

To summarize, we have demonstrated that the data for electron-positron annihilation 
in to various final states in the mass range from 4.2 to 4.35 GeV are consistent with
the existence of 
just  a single vector charmonium-like state 
in this mass range --- the $\psi(4230)$ also known as $Y(4230)$, with properties consistent
with it being a $D_1(2420)\bar D$ 
hadronic molecule.
 Moreover, we show that a consistent description of all channels with the same resonance content is possible only, if we allow for an interference with
 the conventional $\psi(4160)$ resonance.

 The non-trivial insights of this work were possible only
 because we studied various final states simultaneously --- to get access to reliable resonance parameters this appears to be unavoidable, while single channel analyses
 have the tendency to provide resonance parameters with a wild scatter as shown in Fig.~\ref{Fig:comparison}.

\begin{acknowledgments}
We thank Ryan Mitchell for the idea to allow for the interference with lower lying vector states to achieve
a simultaneous description of the various channels.
This work is supported in part by the
Deutsche Forschungsgemeinschaft (DFG) through the funds provided to the Sino-German Collaborative Research Center TRR110 ``Symmetries and the Emergence of Structure in QCD'' (NSFC Grant No. 12070131001, DFG Project-ID 196253076),
by the NSFC under Grants No.~11835015, No.~12047503, No.~11961141012, No.~12035007, and No.~12375073 and by the Chinese Academy of Sciences under Grants
No. QYZDB-SSW-SYS013, No. XDB34030000, No.~XDPB15 and No.~2020VMA0024.
The work of Q.W. is also supported by Guangdong Major Project of Basic and Applied Basic Research under Grant No.~2020B0301030008, by the
Science and Technology Program of Guangzhou under Grant No.~2019050001, and by
Guangdong Provincial funding under Grant No.~2019QN01X172.
\end{acknowledgments}

\bibliography{refs.bib}

\begin{thebibliography}{85}%
\makeatletter
\providecommand \@ifxundefined [1]{%
 \@ifx{#1\undefined}
}%
\providecommand \@ifnum [1]{%
 \ifnum #1\expandafter \@firstoftwo
 \else \expandafter \@secondoftwo
 \fi
}%
\providecommand \@ifx [1]{%
 \ifx #1\expandafter \@firstoftwo
 \else \expandafter \@secondoftwo
 \fi
}%
\providecommand \natexlab [1]{#1}%
\providecommand \enquote  [1]{``#1''}%
\providecommand \bibnamefont  [1]{#1}%
\providecommand \bibfnamefont [1]{#1}%
\providecommand \citenamefont [1]{#1}%
\providecommand \href@noop [0]{\@secondoftwo}%
\providecommand \href [0]{\begingroup \@sanitize@url \@href}%
\providecommand \@href[1]{\@@startlink{#1}\@@href}%
\providecommand \@@href[1]{\endgroup#1\@@endlink}%
\providecommand \@sanitize@url [0]{\catcode `\\12\catcode `\$12\catcode `\&12\catcode `\#12\catcode `\^12\catcode `\_12\catcode `\%12\relax}%
\providecommand \@@startlink[1]{}%
\providecommand \@@endlink[0]{}%
\providecommand \url  [0]{\begingroup\@sanitize@url \@url }%
\providecommand \@url [1]{\endgroup\@href {#1}{\urlprefix }}%
\providecommand \urlprefix  [0]{URL }%
\providecommand \Eprint [0]{\href }%
\providecommand \doibase [0]{http://dx.doi.org/}%
\providecommand \selectlanguage [0]{\@gobble}%
\providecommand \bibinfo  [0]{\@secondoftwo}%
\providecommand \bibfield  [0]{\@secondoftwo}%
\providecommand \translation [1]{[#1]}%
\providecommand \BibitemOpen [0]{}%
\providecommand \bibitemStop [0]{}%
\providecommand \bibitemNoStop [0]{.\EOS\space}%
\providecommand \EOS [0]{\spacefactor3000\relax}%
\providecommand \BibitemShut  [1]{\csname bibitem#1\endcsname}%
\let\auto@bib@innerbib\@empty
\bibitem [{\citenamefont {Lebed}\ \emph {et~al.}(2017)\citenamefont {Lebed}, \citenamefont {Mitchell},\ and\ \citenamefont {Swanson}}]{Lebed:2016hpi}%
  \BibitemOpen
  \bibfield  {author} {\bibinfo {author} {\bibfnamefont {R.~F.}\ \bibnamefont {Lebed}}, \bibinfo {author} {\bibfnamefont {R.~E.}\ \bibnamefont {Mitchell}}, \ and\ \bibinfo {author} {\bibfnamefont {E.~S.}\ \bibnamefont {Swanson}},\ }\href {\doibase 10.1016/j.ppnp.2016.11.003} {\bibfield  {journal} {\bibinfo  {journal} {Prog. Part. Nucl. Phys.}\ }\textbf {\bibinfo {volume} {93}},\ \bibinfo {pages} {143} (\bibinfo {year} {2017})},\ \Eprint {http://arxiv.org/abs/1610.04528} {arXiv:1610.04528 [hep-ph]} \BibitemShut {NoStop}%
\bibitem [{\citenamefont {Esposito}\ \emph {et~al.}(2017)\citenamefont {Esposito}, \citenamefont {Pilloni},\ and\ \citenamefont {Polosa}}]{Esposito:2016noz}%
  \BibitemOpen
  \bibfield  {author} {\bibinfo {author} {\bibfnamefont {A.}~\bibnamefont {Esposito}}, \bibinfo {author} {\bibfnamefont {A.}~\bibnamefont {Pilloni}}, \ and\ \bibinfo {author} {\bibfnamefont {A.~D.}\ \bibnamefont {Polosa}},\ }\href {\doibase 10.1016/j.physrep.2016.11.002} {\bibfield  {journal} {\bibinfo  {journal} {Phys. Rept.}\ }\textbf {\bibinfo {volume} {668}},\ \bibinfo {pages} {1} (\bibinfo {year} {2017})},\ \Eprint {http://arxiv.org/abs/1611.07920} {arXiv:1611.07920 [hep-ph]} \BibitemShut {NoStop}%
\bibitem [{\citenamefont {Olsen}\ \emph {et~al.}(2018)\citenamefont {Olsen}, \citenamefont {Skwarnicki},\ and\ \citenamefont {Zieminska}}]{Olsen:2017bmm}%
  \BibitemOpen
  \bibfield  {author} {\bibinfo {author} {\bibfnamefont {S.~L.}\ \bibnamefont {Olsen}}, \bibinfo {author} {\bibfnamefont {T.}~\bibnamefont {Skwarnicki}}, \ and\ \bibinfo {author} {\bibfnamefont {D.}~\bibnamefont {Zieminska}},\ }\href {\doibase 10.1103/RevModPhys.90.015003} {\bibfield  {journal} {\bibinfo  {journal} {Rev. Mod. Phys.}\ }\textbf {\bibinfo {volume} {90}},\ \bibinfo {pages} {015003} (\bibinfo {year} {2018})},\ \Eprint {http://arxiv.org/abs/1708.04012} {arXiv:1708.04012 [hep-ph]} \BibitemShut {NoStop}%
\bibitem [{\citenamefont {Guo}\ \emph {et~al.}(2018)\citenamefont {Guo}, \citenamefont {Hanhart}, \citenamefont {Mei\ss{}ner}, \citenamefont {Wang}, \citenamefont {Zhao},\ and\ \citenamefont {Zou}}]{Guo:2017jvc}%
  \BibitemOpen
  \bibfield  {author} {\bibinfo {author} {\bibfnamefont {F.-K.}\ \bibnamefont {Guo}}, \bibinfo {author} {\bibfnamefont {C.}~\bibnamefont {Hanhart}}, \bibinfo {author} {\bibfnamefont {U.-G.}\ \bibnamefont {Mei\ss{}ner}}, \bibinfo {author} {\bibfnamefont {Q.}~\bibnamefont {Wang}}, \bibinfo {author} {\bibfnamefont {Q.}~\bibnamefont {Zhao}}, \ and\ \bibinfo {author} {\bibfnamefont {B.-S.}\ \bibnamefont {Zou}},\ }\href {\doibase 10.1103/RevModPhys.90.015004} {\bibfield  {journal} {\bibinfo  {journal} {Rev. Mod. Phys.}\ }\textbf {\bibinfo {volume} {90}},\ \bibinfo {pages} {015004} (\bibinfo {year} {2018})},\ \Eprint {http://arxiv.org/abs/1705.00141} {arXiv:1705.00141 [hep-ph]} \BibitemShut {NoStop}%
\bibitem [{\citenamefont {Brambilla}\ \emph {et~al.}(2020)\citenamefont {Brambilla}, \citenamefont {Eidelman}, \citenamefont {Hanhart}, \citenamefont {Nefediev}, \citenamefont {Shen}, \citenamefont {Thomas}, \citenamefont {Vairo},\ and\ \citenamefont {Yuan}}]{Brambilla:2019esw}%
  \BibitemOpen
  \bibfield  {author} {\bibinfo {author} {\bibfnamefont {N.}~\bibnamefont {Brambilla}}, \bibinfo {author} {\bibfnamefont {S.}~\bibnamefont {Eidelman}}, \bibinfo {author} {\bibfnamefont {C.}~\bibnamefont {Hanhart}}, \bibinfo {author} {\bibfnamefont {A.}~\bibnamefont {Nefediev}}, \bibinfo {author} {\bibfnamefont {C.-P.}\ \bibnamefont {Shen}}, \bibinfo {author} {\bibfnamefont {C.~E.}\ \bibnamefont {Thomas}}, \bibinfo {author} {\bibfnamefont {A.}~\bibnamefont {Vairo}}, \ and\ \bibinfo {author} {\bibfnamefont {C.-Z.}\ \bibnamefont {Yuan}},\ }\href {\doibase 10.1016/j.physrep.2020.05.001} {\bibfield  {journal} {\bibinfo  {journal} {Phys. Rept.}\ }\textbf {\bibinfo {volume} {873}},\ \bibinfo {pages} {1} (\bibinfo {year} {2020})},\ \Eprint {http://arxiv.org/abs/1907.07583} {arXiv:1907.07583 [hep-ex]} \BibitemShut {NoStop}%
\bibitem [{\citenamefont {Chen}\ \emph {et~al.}(2023{\natexlab{a}})\citenamefont {Chen}, \citenamefont {Chen}, \citenamefont {Liu}, \citenamefont {Liu},\ and\ \citenamefont {Zhu}}]{Chen:2022asf}%
  \BibitemOpen
  \bibfield  {author} {\bibinfo {author} {\bibfnamefont {H.-X.}\ \bibnamefont {Chen}}, \bibinfo {author} {\bibfnamefont {W.}~\bibnamefont {Chen}}, \bibinfo {author} {\bibfnamefont {X.}~\bibnamefont {Liu}}, \bibinfo {author} {\bibfnamefont {Y.-R.}\ \bibnamefont {Liu}}, \ and\ \bibinfo {author} {\bibfnamefont {S.-L.}\ \bibnamefont {Zhu}},\ }\href {\doibase 10.1088/1361-6633/aca3b6} {\bibfield  {journal} {\bibinfo  {journal} {Rept. Prog. Phys.}\ }\textbf {\bibinfo {volume} {86}},\ \bibinfo {pages} {026201} (\bibinfo {year} {2023}{\natexlab{a}})},\ \Eprint {http://arxiv.org/abs/2204.02649} {arXiv:2204.02649 [hep-ph]} \BibitemShut {NoStop}%
\bibitem [{\citenamefont {Ablikim}\ \emph {et~al.}(2017{\natexlab{a}})\citenamefont {Ablikim} \emph {et~al.}}]{BESIII:2016bnd}%
  \BibitemOpen
  \bibfield  {author} {\bibinfo {author} {\bibfnamefont {M.}~\bibnamefont {Ablikim}} \emph {et~al.} (\bibinfo {collaboration} {BESIII}),\ }\href {\doibase 10.1103/PhysRevLett.118.092001} {\bibfield  {journal} {\bibinfo  {journal} {Phys. Rev. Lett.}\ }\textbf {\bibinfo {volume} {118}},\ \bibinfo {pages} {092001} (\bibinfo {year} {2017}{\natexlab{a}})},\ \Eprint {http://arxiv.org/abs/1611.01317} {arXiv:1611.01317 [hep-ex]} \BibitemShut {NoStop}%
\bibitem [{\citenamefont {Ablikim}\ \emph {et~al.}(2022{\natexlab{a}})\citenamefont {Ablikim} \emph {et~al.}}]{BESIII:2022qal}%
  \BibitemOpen
  \bibfield  {author} {\bibinfo {author} {\bibfnamefont {M.}~\bibnamefont {Ablikim}} \emph {et~al.} (\bibinfo {collaboration} {BESIII}),\ }\href {\doibase 10.1103/PhysRevD.106.072001} {\bibfield  {journal} {\bibinfo  {journal} {Phys. Rev. D}\ }\textbf {\bibinfo {volume} {106}},\ \bibinfo {pages} {072001} (\bibinfo {year} {2022}{\natexlab{a}})},\ \Eprint {http://arxiv.org/abs/2206.08554} {arXiv:2206.08554 [hep-ex]} \BibitemShut {NoStop}%
\bibitem [{\citenamefont {Ablikim}\ \emph {et~al.}(2020{\natexlab{a}})\citenamefont {Ablikim} \emph {et~al.}}]{BESIII:2020oph}%
  \BibitemOpen
  \bibfield  {author} {\bibinfo {author} {\bibfnamefont {M.}~\bibnamefont {Ablikim}} \emph {et~al.} (\bibinfo {collaboration} {BESIII}),\ }\href {\doibase 10.1103/PhysRevD.102.012009} {\bibfield  {journal} {\bibinfo  {journal} {Phys. Rev. D}\ }\textbf {\bibinfo {volume} {102}},\ \bibinfo {pages} {012009} (\bibinfo {year} {2020}{\natexlab{a}})},\ \Eprint {http://arxiv.org/abs/2004.13788} {arXiv:2004.13788 [hep-ex]} \BibitemShut {NoStop}%
\bibitem [{\citenamefont {Lees}\ \emph {et~al.}(2014)\citenamefont {Lees} \emph {et~al.}}]{BaBar:2012hpr}%
  \BibitemOpen
  \bibfield  {author} {\bibinfo {author} {\bibfnamefont {J.~P.}\ \bibnamefont {Lees}} \emph {et~al.} (\bibinfo {collaboration} {BaBar}),\ }\href {\doibase 10.1103/PhysRevD.89.111103} {\bibfield  {journal} {\bibinfo  {journal} {Phys. Rev. D}\ }\textbf {\bibinfo {volume} {89}},\ \bibinfo {pages} {111103} (\bibinfo {year} {2014})},\ \Eprint {http://arxiv.org/abs/1211.6271} {arXiv:1211.6271 [hep-ex]} \BibitemShut {NoStop}%
\bibitem [{\citenamefont {Wang}\ \emph {et~al.}(2015)\citenamefont {Wang} \emph {et~al.}}]{Belle:2014wyt}%
  \BibitemOpen
  \bibfield  {author} {\bibinfo {author} {\bibfnamefont {X.~L.}\ \bibnamefont {Wang}} \emph {et~al.} (\bibinfo {collaboration} {Belle}),\ }\href {\doibase 10.1103/PhysRevD.91.112007} {\bibfield  {journal} {\bibinfo  {journal} {Phys. Rev. D}\ }\textbf {\bibinfo {volume} {91}},\ \bibinfo {pages} {112007} (\bibinfo {year} {2015})},\ \Eprint {http://arxiv.org/abs/1410.7641} {arXiv:1410.7641 [hep-ex]} \BibitemShut {NoStop}%
\bibitem [{\citenamefont {Ablikim}\ \emph {et~al.}(2021)\citenamefont {Ablikim} \emph {et~al.}}]{BESIII:2021njb}%
  \BibitemOpen
  \bibfield  {author} {\bibinfo {author} {\bibfnamefont {M.}~\bibnamefont {Ablikim}} \emph {et~al.} (\bibinfo {collaboration} {BESIII}),\ }\href {\doibase 10.1103/PhysRevD.104.052012} {\bibfield  {journal} {\bibinfo  {journal} {Phys. Rev. D}\ }\textbf {\bibinfo {volume} {104}},\ \bibinfo {pages} {052012} (\bibinfo {year} {2021})},\ \Eprint {http://arxiv.org/abs/2107.09210} {arXiv:2107.09210 [hep-ex]} \BibitemShut {NoStop}%
\bibitem [{\citenamefont {Ablikim}\ \emph {et~al.}(2017{\natexlab{b}})\citenamefont {Ablikim} \emph {et~al.}}]{BESIII:2016adj}%
  \BibitemOpen
  \bibfield  {author} {\bibinfo {author} {\bibfnamefont {M.}~\bibnamefont {Ablikim}} \emph {et~al.} (\bibinfo {collaboration} {BESIII}),\ }\href {\doibase 10.1103/PhysRevLett.118.092002} {\bibfield  {journal} {\bibinfo  {journal} {Phys. Rev. Lett.}\ }\textbf {\bibinfo {volume} {118}},\ \bibinfo {pages} {092002} (\bibinfo {year} {2017}{\natexlab{b}})},\ \Eprint {http://arxiv.org/abs/1610.07044} {arXiv:1610.07044 [hep-ex]} \BibitemShut {NoStop}%
\bibitem [{\citenamefont {Ablikim}\ \emph {et~al.}(2020{\natexlab{b}})\citenamefont {Ablikim} \emph {et~al.}}]{BESIII:2020bgb}%
  \BibitemOpen
  \bibfield  {author} {\bibinfo {author} {\bibfnamefont {M.}~\bibnamefont {Ablikim}} \emph {et~al.} (\bibinfo {collaboration} {BESIII}),\ }\href {\doibase 10.1103/PhysRevD.102.031101} {\bibfield  {journal} {\bibinfo  {journal} {Phys. Rev. D}\ }\textbf {\bibinfo {volume} {102}},\ \bibinfo {pages} {031101} (\bibinfo {year} {2020}{\natexlab{b}})},\ \Eprint {http://arxiv.org/abs/2003.03705} {arXiv:2003.03705 [hep-ex]} \BibitemShut {NoStop}%
\bibitem [{\citenamefont {Ablikim}\ \emph {et~al.}(2019{\natexlab{a}})\citenamefont {Ablikim} \emph {et~al.}}]{BESIII:2018iea}%
  \BibitemOpen
  \bibfield  {author} {\bibinfo {author} {\bibfnamefont {M.}~\bibnamefont {Ablikim}} \emph {et~al.} (\bibinfo {collaboration} {BESIII}),\ }\href {\doibase 10.1103/PhysRevLett.122.102002} {\bibfield  {journal} {\bibinfo  {journal} {Phys. Rev. Lett.}\ }\textbf {\bibinfo {volume} {122}},\ \bibinfo {pages} {102002} (\bibinfo {year} {2019}{\natexlab{a}})},\ \Eprint {http://arxiv.org/abs/1808.02847} {arXiv:1808.02847 [hep-ex]} \BibitemShut {NoStop}%
\bibitem [{\citenamefont {Ding}(2009)}]{Ding:2008gr}%
  \BibitemOpen
  \bibfield  {author} {\bibinfo {author} {\bibfnamefont {G.-J.}\ \bibnamefont {Ding}},\ }\href {\doibase 10.1103/PhysRevD.79.014001} {\bibfield  {journal} {\bibinfo  {journal} {Phys. Rev. D}\ }\textbf {\bibinfo {volume} {79}},\ \bibinfo {pages} {014001} (\bibinfo {year} {2009})},\ \Eprint {http://arxiv.org/abs/0809.4818} {arXiv:0809.4818 [hep-ph]} \BibitemShut {NoStop}%
\bibitem [{\citenamefont {Wang}\ \emph {et~al.}(2013{\natexlab{a}})\citenamefont {Wang}, \citenamefont {Hanhart},\ and\ \citenamefont {Zhao}}]{Wang:2013cya}%
  \BibitemOpen
  \bibfield  {author} {\bibinfo {author} {\bibfnamefont {Q.}~\bibnamefont {Wang}}, \bibinfo {author} {\bibfnamefont {C.}~\bibnamefont {Hanhart}}, \ and\ \bibinfo {author} {\bibfnamefont {Q.}~\bibnamefont {Zhao}},\ }\href {\doibase 10.1103/PhysRevLett.111.132003} {\bibfield  {journal} {\bibinfo  {journal} {Phys. Rev. Lett.}\ }\textbf {\bibinfo {volume} {111}},\ \bibinfo {pages} {132003} (\bibinfo {year} {2013}{\natexlab{a}})},\ \Eprint {http://arxiv.org/abs/1303.6355} {arXiv:1303.6355 [hep-ph]} \BibitemShut {NoStop}%
\bibitem [{\citenamefont {Coito}\ and\ \citenamefont {Giacosa}(2020)}]{Coito:2019cts}%
  \BibitemOpen
  \bibfield  {author} {\bibinfo {author} {\bibfnamefont {S.}~\bibnamefont {Coito}}\ and\ \bibinfo {author} {\bibfnamefont {F.}~\bibnamefont {Giacosa}},\ }\href {\doibase 10.5506/APhysPolB.51.1713} {\bibfield  {journal} {\bibinfo  {journal} {Acta Phys. Polon. B}\ }\textbf {\bibinfo {volume} {51}},\ \bibinfo {pages} {1713} (\bibinfo {year} {2020})},\ \Eprint {http://arxiv.org/abs/1902.09268} {arXiv:1902.09268 [hep-ph]} \BibitemShut {NoStop}%
\bibitem [{\citenamefont {Chen}\ \emph {et~al.}(2016{\natexlab{a}})\citenamefont {Chen}, \citenamefont {Liu}, \citenamefont {Li},\ and\ \citenamefont {Ke}}]{Chen:2015bft}%
  \BibitemOpen
  \bibfield  {author} {\bibinfo {author} {\bibfnamefont {D.-Y.}\ \bibnamefont {Chen}}, \bibinfo {author} {\bibfnamefont {X.}~\bibnamefont {Liu}}, \bibinfo {author} {\bibfnamefont {X.-Q.}\ \bibnamefont {Li}}, \ and\ \bibinfo {author} {\bibfnamefont {H.-W.}\ \bibnamefont {Ke}},\ }\href {\doibase 10.1103/PhysRevD.93.014011} {\bibfield  {journal} {\bibinfo  {journal} {Phys. Rev. D}\ }\textbf {\bibinfo {volume} {93}},\ \bibinfo {pages} {014011} (\bibinfo {year} {2016}{\natexlab{a}})},\ \Eprint {http://arxiv.org/abs/1512.04157} {arXiv:1512.04157 [hep-ph]} \BibitemShut {NoStop}%
\bibitem [{\citenamefont {Ablikim}\ \emph {et~al.}(2019{\natexlab{b}})\citenamefont {Ablikim} \emph {et~al.}}]{BESIII:2019gjc}%
  \BibitemOpen
  \bibfield  {author} {\bibinfo {author} {\bibfnamefont {M.}~\bibnamefont {Ablikim}} \emph {et~al.} (\bibinfo {collaboration} {BESIII}),\ }\href {\doibase 10.1103/PhysRevD.99.091103} {\bibfield  {journal} {\bibinfo  {journal} {Phys. Rev. D}\ }\textbf {\bibinfo {volume} {99}},\ \bibinfo {pages} {091103} (\bibinfo {year} {2019}{\natexlab{b}})},\ \Eprint {http://arxiv.org/abs/1903.02359} {arXiv:1903.02359 [hep-ex]} \BibitemShut {NoStop}%
\bibitem [{\citenamefont {Ablikim}\ \emph {et~al.}(2019{\natexlab{c}})\citenamefont {Ablikim} \emph {et~al.}}]{BESIII:2019qvy}%
  \BibitemOpen
  \bibfield  {author} {\bibinfo {author} {\bibfnamefont {M.}~\bibnamefont {Ablikim}} \emph {et~al.} (\bibinfo {collaboration} {BESIII}),\ }\href {\doibase 10.1103/PhysRevLett.122.232002} {\bibfield  {journal} {\bibinfo  {journal} {Phys. Rev. Lett.}\ }\textbf {\bibinfo {volume} {122}},\ \bibinfo {pages} {232002} (\bibinfo {year} {2019}{\natexlab{c}})},\ \Eprint {http://arxiv.org/abs/1903.04695} {arXiv:1903.04695 [hep-ex]} \BibitemShut {NoStop}%
\bibitem [{\citenamefont {Ablikim}\ \emph {et~al.}(2020{\natexlab{c}})\citenamefont {Ablikim} \emph {et~al.}}]{BESIII:2020peo}%
  \BibitemOpen
  \bibfield  {author} {\bibinfo {author} {\bibfnamefont {M.}~\bibnamefont {Ablikim}} \emph {et~al.} (\bibinfo {collaboration} {BESIII}),\ }\href {\doibase 10.1103/PhysRevD.102.112009} {\bibfield  {journal} {\bibinfo  {journal} {Phys. Rev. D}\ }\textbf {\bibinfo {volume} {102}},\ \bibinfo {pages} {112009} (\bibinfo {year} {2020}{\natexlab{c}})},\ \Eprint {http://arxiv.org/abs/2007.12872} {arXiv:2007.12872 [hep-ex]} \BibitemShut {NoStop}%
\bibitem [{\citenamefont {Ablikim}\ \emph {et~al.}(2022{\natexlab{b}})\citenamefont {Ablikim} \emph {et~al.}}]{BESIII:2022joj}%
  \BibitemOpen
  \bibfield  {author} {\bibinfo {author} {\bibfnamefont {M.}~\bibnamefont {Ablikim}} \emph {et~al.} (\bibinfo {collaboration} {(BESIII),, BESIII}),\ }\href {\doibase 10.1088/1674-1137/ac945c} {\bibfield  {journal} {\bibinfo  {journal} {Chin. Phys. C}\ }\textbf {\bibinfo {volume} {46}},\ \bibinfo {pages} {111002} (\bibinfo {year} {2022}{\natexlab{b}})},\ \Eprint {http://arxiv.org/abs/2204.07800} {arXiv:2204.07800 [hep-ex]} \BibitemShut {NoStop}%
\bibitem [{\citenamefont {Dubynskiy}\ and\ \citenamefont {Voloshin}(2008)}]{Dubynskiy:2008mq}%
  \BibitemOpen
  \bibfield  {author} {\bibinfo {author} {\bibfnamefont {S.}~\bibnamefont {Dubynskiy}}\ and\ \bibinfo {author} {\bibfnamefont {M.~B.}\ \bibnamefont {Voloshin}},\ }\href {\doibase 10.1016/j.physletb.2008.07.086} {\bibfield  {journal} {\bibinfo  {journal} {Phys. Lett. B}\ }\textbf {\bibinfo {volume} {666}},\ \bibinfo {pages} {344} (\bibinfo {year} {2008})},\ \Eprint {http://arxiv.org/abs/0803.2224} {arXiv:0803.2224 [hep-ph]} \BibitemShut {NoStop}%
\bibitem [{\citenamefont {Wang}\ \emph {et~al.}(2014)\citenamefont {Wang}, \citenamefont {Cleven}, \citenamefont {Guo}, \citenamefont {Hanhart}, \citenamefont {Mei\ss{}ner}, \citenamefont {Wu},\ and\ \citenamefont {Zhao}}]{Wang:2013kra}%
  \BibitemOpen
  \bibfield  {author} {\bibinfo {author} {\bibfnamefont {Q.}~\bibnamefont {Wang}}, \bibinfo {author} {\bibfnamefont {M.}~\bibnamefont {Cleven}}, \bibinfo {author} {\bibfnamefont {F.-K.}\ \bibnamefont {Guo}}, \bibinfo {author} {\bibfnamefont {C.}~\bibnamefont {Hanhart}}, \bibinfo {author} {\bibfnamefont {U.-G.}\ \bibnamefont {Mei\ss{}ner}}, \bibinfo {author} {\bibfnamefont {X.-G.}\ \bibnamefont {Wu}}, \ and\ \bibinfo {author} {\bibfnamefont {Q.}~\bibnamefont {Zhao}},\ }\href {\doibase 10.1103/PhysRevD.89.034001} {\bibfield  {journal} {\bibinfo  {journal} {Phys. Rev. D}\ }\textbf {\bibinfo {volume} {89}},\ \bibinfo {pages} {034001} (\bibinfo {year} {2014})},\ \Eprint {http://arxiv.org/abs/1309.4303} {arXiv:1309.4303 [hep-ph]} \BibitemShut {NoStop}%
\bibitem [{\citenamefont {Li}\ and\ \citenamefont {Voloshin}(2014)}]{Li:2013ssa}%
  \BibitemOpen
  \bibfield  {author} {\bibinfo {author} {\bibfnamefont {X.}~\bibnamefont {Li}}\ and\ \bibinfo {author} {\bibfnamefont {M.~B.}\ \bibnamefont {Voloshin}},\ }\href {\doibase 10.1142/S0217732314500606} {\bibfield  {journal} {\bibinfo  {journal} {Mod. Phys. Lett. A}\ }\textbf {\bibinfo {volume} {29}},\ \bibinfo {pages} {1450060} (\bibinfo {year} {2014})},\ \Eprint {http://arxiv.org/abs/1309.1681} {arXiv:1309.1681 [hep-ph]} \BibitemShut {NoStop}%
\bibitem [{\citenamefont {Cleven}\ \emph {et~al.}(2015)\citenamefont {Cleven}, \citenamefont {Guo}, \citenamefont {Hanhart}, \citenamefont {Wang},\ and\ \citenamefont {Zhao}}]{Cleven:2015era}%
  \BibitemOpen
  \bibfield  {author} {\bibinfo {author} {\bibfnamefont {M.}~\bibnamefont {Cleven}}, \bibinfo {author} {\bibfnamefont {F.-K.}\ \bibnamefont {Guo}}, \bibinfo {author} {\bibfnamefont {C.}~\bibnamefont {Hanhart}}, \bibinfo {author} {\bibfnamefont {Q.}~\bibnamefont {Wang}}, \ and\ \bibinfo {author} {\bibfnamefont {Q.}~\bibnamefont {Zhao}},\ }\href {\doibase 10.1103/PhysRevD.92.014005} {\bibfield  {journal} {\bibinfo  {journal} {Phys. Rev. D}\ }\textbf {\bibinfo {volume} {92}},\ \bibinfo {pages} {014005} (\bibinfo {year} {2015})},\ \Eprint {http://arxiv.org/abs/1505.01771} {arXiv:1505.01771 [hep-ph]} \BibitemShut {NoStop}%
\bibitem [{\citenamefont {Close}\ and\ \citenamefont {Page}(2005)}]{Close:2005iz}%
  \BibitemOpen
  \bibfield  {author} {\bibinfo {author} {\bibfnamefont {F.~E.}\ \bibnamefont {Close}}\ and\ \bibinfo {author} {\bibfnamefont {P.~R.}\ \bibnamefont {Page}},\ }\href {\doibase 10.1016/j.physletb.2005.09.016} {\bibfield  {journal} {\bibinfo  {journal} {Phys. Lett. B}\ }\textbf {\bibinfo {volume} {628}},\ \bibinfo {pages} {215} (\bibinfo {year} {2005})},\ \Eprint {http://arxiv.org/abs/hep-ph/0507199} {arXiv:hep-ph/0507199} \BibitemShut {NoStop}%
\bibitem [{\citenamefont {Kou}\ and\ \citenamefont {Pene}(2005)}]{Kou:2005gt}%
  \BibitemOpen
  \bibfield  {author} {\bibinfo {author} {\bibfnamefont {E.}~\bibnamefont {Kou}}\ and\ \bibinfo {author} {\bibfnamefont {O.}~\bibnamefont {Pene}},\ }\href {\doibase 10.1016/j.physletb.2005.09.013} {\bibfield  {journal} {\bibinfo  {journal} {Phys. Lett. B}\ }\textbf {\bibinfo {volume} {631}},\ \bibinfo {pages} {164} (\bibinfo {year} {2005})},\ \Eprint {http://arxiv.org/abs/hep-ph/0507119} {arXiv:hep-ph/0507119} \BibitemShut {NoStop}%
\bibitem [{\citenamefont {Kalashnikova}\ and\ \citenamefont {Nefediev}(2008)}]{Kalashnikova:2008qr}%
  \BibitemOpen
  \bibfield  {author} {\bibinfo {author} {\bibfnamefont {Y.~S.}\ \bibnamefont {Kalashnikova}}\ and\ \bibinfo {author} {\bibfnamefont {A.~V.}\ \bibnamefont {Nefediev}},\ }\href {\doibase 10.1103/PhysRevD.77.054025} {\bibfield  {journal} {\bibinfo  {journal} {Phys. Rev. D}\ }\textbf {\bibinfo {volume} {77}},\ \bibinfo {pages} {054025} (\bibinfo {year} {2008})},\ \Eprint {http://arxiv.org/abs/0801.2036} {arXiv:0801.2036 [hep-ph]} \BibitemShut {NoStop}%
\bibitem [{\citenamefont {Berwein}\ \emph {et~al.}(2015)\citenamefont {Berwein}, \citenamefont {Brambilla}, \citenamefont {Tarr\'us~Castell\`a},\ and\ \citenamefont {Vairo}}]{Berwein:2015vca}%
  \BibitemOpen
  \bibfield  {author} {\bibinfo {author} {\bibfnamefont {M.}~\bibnamefont {Berwein}}, \bibinfo {author} {\bibfnamefont {N.}~\bibnamefont {Brambilla}}, \bibinfo {author} {\bibfnamefont {J.}~\bibnamefont {Tarr\'us~Castell\`a}}, \ and\ \bibinfo {author} {\bibfnamefont {A.}~\bibnamefont {Vairo}},\ }\href {\doibase 10.1103/PhysRevD.92.114019} {\bibfield  {journal} {\bibinfo  {journal} {Phys. Rev. D}\ }\textbf {\bibinfo {volume} {92}},\ \bibinfo {pages} {114019} (\bibinfo {year} {2015})},\ \Eprint {http://arxiv.org/abs/1510.04299} {arXiv:1510.04299 [hep-ph]} \BibitemShut {NoStop}%
\bibitem [{\citenamefont {Brambilla}\ \emph {et~al.}(2023)\citenamefont {Brambilla}, \citenamefont {Lai}, \citenamefont {Mohapatra},\ and\ \citenamefont {Vairo}}]{Brambilla:2022hhi}%
  \BibitemOpen
  \bibfield  {author} {\bibinfo {author} {\bibfnamefont {N.}~\bibnamefont {Brambilla}}, \bibinfo {author} {\bibfnamefont {W.~K.}\ \bibnamefont {Lai}}, \bibinfo {author} {\bibfnamefont {A.}~\bibnamefont {Mohapatra}}, \ and\ \bibinfo {author} {\bibfnamefont {A.}~\bibnamefont {Vairo}},\ }\href {\doibase 10.1103/PhysRevD.107.054034} {\bibfield  {journal} {\bibinfo  {journal} {Phys. Rev. D}\ }\textbf {\bibinfo {volume} {107}},\ \bibinfo {pages} {054034} (\bibinfo {year} {2023})},\ \Eprint {http://arxiv.org/abs/2212.09187} {arXiv:2212.09187 [hep-ph]} \BibitemShut {NoStop}%
\bibitem [{\citenamefont {Ali}\ \emph {et~al.}(2018)\citenamefont {Ali}, \citenamefont {Maiani}, \citenamefont {Borisov}, \citenamefont {Ahmed}, \citenamefont {Jamil~Aslam}, \citenamefont {Parkhomenko}, \citenamefont {Polosa},\ and\ \citenamefont {Rehman}}]{Ali:2017wsf}%
  \BibitemOpen
  \bibfield  {author} {\bibinfo {author} {\bibfnamefont {A.}~\bibnamefont {Ali}}, \bibinfo {author} {\bibfnamefont {L.}~\bibnamefont {Maiani}}, \bibinfo {author} {\bibfnamefont {A.~V.}\ \bibnamefont {Borisov}}, \bibinfo {author} {\bibfnamefont {I.}~\bibnamefont {Ahmed}}, \bibinfo {author} {\bibfnamefont {M.}~\bibnamefont {Jamil~Aslam}}, \bibinfo {author} {\bibfnamefont {A.~Y.}\ \bibnamefont {Parkhomenko}}, \bibinfo {author} {\bibfnamefont {A.~D.}\ \bibnamefont {Polosa}}, \ and\ \bibinfo {author} {\bibfnamefont {A.}~\bibnamefont {Rehman}},\ }\href {\doibase 10.1140/epjc/s10052-017-5501-6} {\bibfield  {journal} {\bibinfo  {journal} {Eur. Phys. J. C}\ }\textbf {\bibinfo {volume} {78}},\ \bibinfo {pages} {29} (\bibinfo {year} {2018})},\ \Eprint {http://arxiv.org/abs/1708.04650} {arXiv:1708.04650 [hep-ph]} \BibitemShut {NoStop}%
\bibitem [{\citenamefont {Bhavsar}\ \emph {et~al.}(2020)\citenamefont {Bhavsar}, \citenamefont {Shah}, \citenamefont {Patel},\ and\ \citenamefont {Vinodkumar}}]{Bhavsar:2020pog}%
  \BibitemOpen
  \bibfield  {author} {\bibinfo {author} {\bibfnamefont {T.}~\bibnamefont {Bhavsar}}, \bibinfo {author} {\bibfnamefont {M.}~\bibnamefont {Shah}}, \bibinfo {author} {\bibfnamefont {S.}~\bibnamefont {Patel}}, \ and\ \bibinfo {author} {\bibfnamefont {P.~C.}\ \bibnamefont {Vinodkumar}},\ }\href {\doibase 10.1016/j.nuclphysa.2020.121856} {\bibfield  {journal} {\bibinfo  {journal} {Nucl. Phys. A}\ }\textbf {\bibinfo {volume} {1000}},\ \bibinfo {pages} {121856} (\bibinfo {year} {2020})},\ \Eprint {http://arxiv.org/abs/2002.06363} {arXiv:2002.06363 [hep-ph]} \BibitemShut {NoStop}%
\bibitem [{\citenamefont {Lebed}(2017)}]{Lebed:2017min}%
  \BibitemOpen
  \bibfield  {author} {\bibinfo {author} {\bibfnamefont {R.~F.}\ \bibnamefont {Lebed}},\ }\href {\doibase 10.1103/PhysRevD.96.116003} {\bibfield  {journal} {\bibinfo  {journal} {Phys. Rev. D}\ }\textbf {\bibinfo {volume} {96}},\ \bibinfo {pages} {116003} (\bibinfo {year} {2017})},\ \Eprint {http://arxiv.org/abs/1709.06097} {arXiv:1709.06097 [hep-ph]} \BibitemShut {NoStop}%
\bibitem [{\citenamefont {Ablikim}\ \emph {et~al.}(2023)\citenamefont {Ablikim} \emph {et~al.}}]{BESIII:2023cmv}%
  \BibitemOpen
  \bibfield  {author} {\bibinfo {author} {\bibfnamefont {M.}~\bibnamefont {Ablikim}} \emph {et~al.} (\bibinfo {collaboration} {BESIII}),\ }\href {\doibase 10.1103/PhysRevLett.130.121901} {\bibfield  {journal} {\bibinfo  {journal} {Phys. Rev. Lett.}\ }\textbf {\bibinfo {volume} {130}},\ \bibinfo {pages} {121901} (\bibinfo {year} {2023})},\ \Eprint {http://arxiv.org/abs/2301.07321} {arXiv:2301.07321 [hep-ex]} \BibitemShut {NoStop}%
\bibitem [{\citenamefont {Cleven}\ \emph {et~al.}(2014)\citenamefont {Cleven}, \citenamefont {Wang}, \citenamefont {Guo}, \citenamefont {Hanhart}, \citenamefont {Mei\ss{}ner},\ and\ \citenamefont {Zhao}}]{Cleven:2013mka}%
  \BibitemOpen
  \bibfield  {author} {\bibinfo {author} {\bibfnamefont {M.}~\bibnamefont {Cleven}}, \bibinfo {author} {\bibfnamefont {Q.}~\bibnamefont {Wang}}, \bibinfo {author} {\bibfnamefont {F.-K.}\ \bibnamefont {Guo}}, \bibinfo {author} {\bibfnamefont {C.}~\bibnamefont {Hanhart}}, \bibinfo {author} {\bibfnamefont {U.-G.}\ \bibnamefont {Mei\ss{}ner}}, \ and\ \bibinfo {author} {\bibfnamefont {Q.}~\bibnamefont {Zhao}},\ }\href {\doibase 10.1103/PhysRevD.90.074039} {\bibfield  {journal} {\bibinfo  {journal} {Phys. Rev. D}\ }\textbf {\bibinfo {volume} {90}},\ \bibinfo {pages} {074039} (\bibinfo {year} {2014})},\ \Eprint {http://arxiv.org/abs/1310.2190} {arXiv:1310.2190 [hep-ph]} \BibitemShut {NoStop}%
\bibitem [{\citenamefont {Qin}\ \emph {et~al.}(2016)\citenamefont {Qin}, \citenamefont {Xue},\ and\ \citenamefont {Zhao}}]{Qin:2016spb}%
  \BibitemOpen
  \bibfield  {author} {\bibinfo {author} {\bibfnamefont {W.}~\bibnamefont {Qin}}, \bibinfo {author} {\bibfnamefont {S.-R.}\ \bibnamefont {Xue}}, \ and\ \bibinfo {author} {\bibfnamefont {Q.}~\bibnamefont {Zhao}},\ }\href {\doibase 10.1103/PhysRevD.94.054035} {\bibfield  {journal} {\bibinfo  {journal} {Phys. Rev. D}\ }\textbf {\bibinfo {volume} {94}},\ \bibinfo {pages} {054035} (\bibinfo {year} {2016})},\ \Eprint {http://arxiv.org/abs/1605.02407} {arXiv:1605.02407 [hep-ph]} \BibitemShut {NoStop}%
\bibitem [{\citenamefont {Zhou}\ \emph {et~al.}(2023)\citenamefont {Zhou}, \citenamefont {Li},\ and\ \citenamefont {Xiao}}]{Zhou:2023yjv}%
  \BibitemOpen
  \bibfield  {author} {\bibinfo {author} {\bibfnamefont {Z.-Y.}\ \bibnamefont {Zhou}}, \bibinfo {author} {\bibfnamefont {C.-Y.}\ \bibnamefont {Li}}, \ and\ \bibinfo {author} {\bibfnamefont {Z.}~\bibnamefont {Xiao}},\ }\href@noop {} {\  (\bibinfo {year} {2023})},\ \Eprint {http://arxiv.org/abs/2304.07052} {arXiv:2304.07052 [hep-ph]} \BibitemShut {NoStop}%
\bibitem [{\citenamefont {Chen}\ \emph {et~al.}(2016{\natexlab{b}})\citenamefont {Chen}, \citenamefont {Daub}, \citenamefont {Guo}, \citenamefont {Kubis}, \citenamefont {Mei\ss{}ner},\ and\ \citenamefont {Zou}}]{Chen:2015jgl}%
  \BibitemOpen
  \bibfield  {author} {\bibinfo {author} {\bibfnamefont {Y.-H.}\ \bibnamefont {Chen}}, \bibinfo {author} {\bibfnamefont {J.~T.}\ \bibnamefont {Daub}}, \bibinfo {author} {\bibfnamefont {F.-K.}\ \bibnamefont {Guo}}, \bibinfo {author} {\bibfnamefont {B.}~\bibnamefont {Kubis}}, \bibinfo {author} {\bibfnamefont {U.-G.}\ \bibnamefont {Mei\ss{}ner}}, \ and\ \bibinfo {author} {\bibfnamefont {B.-S.}\ \bibnamefont {Zou}},\ }\href {\doibase 10.1103/PhysRevD.93.034030} {\bibfield  {journal} {\bibinfo  {journal} {Phys. Rev. D}\ }\textbf {\bibinfo {volume} {93}},\ \bibinfo {pages} {034030} (\bibinfo {year} {2016}{\natexlab{b}})},\ \Eprint {http://arxiv.org/abs/1512.03583} {arXiv:1512.03583 [hep-ph]} \BibitemShut {NoStop}%
\bibitem [{\citenamefont {Chen}\ \emph {et~al.}(2017)\citenamefont {Chen}, \citenamefont {Cleven}, \citenamefont {Daub}, \citenamefont {Guo}, \citenamefont {Hanhart}, \citenamefont {Kubis}, \citenamefont {Mei\ss{}ner},\ and\ \citenamefont {Zou}}]{Chen:2016mjn}%
  \BibitemOpen
  \bibfield  {author} {\bibinfo {author} {\bibfnamefont {Y.-H.}\ \bibnamefont {Chen}}, \bibinfo {author} {\bibfnamefont {M.}~\bibnamefont {Cleven}}, \bibinfo {author} {\bibfnamefont {J.~T.}\ \bibnamefont {Daub}}, \bibinfo {author} {\bibfnamefont {F.-K.}\ \bibnamefont {Guo}}, \bibinfo {author} {\bibfnamefont {C.}~\bibnamefont {Hanhart}}, \bibinfo {author} {\bibfnamefont {B.}~\bibnamefont {Kubis}}, \bibinfo {author} {\bibfnamefont {U.-G.}\ \bibnamefont {Mei\ss{}ner}}, \ and\ \bibinfo {author} {\bibfnamefont {B.-S.}\ \bibnamefont {Zou}},\ }\href {\doibase 10.1103/PhysRevD.95.034022} {\bibfield  {journal} {\bibinfo  {journal} {Phys. Rev. D}\ }\textbf {\bibinfo {volume} {95}},\ \bibinfo {pages} {034022} (\bibinfo {year} {2017})},\ \Eprint {http://arxiv.org/abs/1611.00913} {arXiv:1611.00913 [hep-ph]} \BibitemShut {NoStop}%
\bibitem [{\citenamefont {Baru}\ \emph {et~al.}(2021)\citenamefont {Baru}, \citenamefont {Epelbaum}, \citenamefont {Filin}, \citenamefont {Hanhart}, \citenamefont {Mizuk}, \citenamefont {Nefediev},\ and\ \citenamefont {Ropertz}}]{Baru:2020ywb}%
  \BibitemOpen
  \bibfield  {author} {\bibinfo {author} {\bibfnamefont {V.}~\bibnamefont {Baru}}, \bibinfo {author} {\bibfnamefont {E.}~\bibnamefont {Epelbaum}}, \bibinfo {author} {\bibfnamefont {A.~A.}\ \bibnamefont {Filin}}, \bibinfo {author} {\bibfnamefont {C.}~\bibnamefont {Hanhart}}, \bibinfo {author} {\bibfnamefont {R.~V.}\ \bibnamefont {Mizuk}}, \bibinfo {author} {\bibfnamefont {A.~V.}\ \bibnamefont {Nefediev}}, \ and\ \bibinfo {author} {\bibfnamefont {S.}~\bibnamefont {Ropertz}},\ }\href {\doibase 10.1103/PhysRevD.103.034016} {\bibfield  {journal} {\bibinfo  {journal} {Phys. Rev. D}\ }\textbf {\bibinfo {volume} {103}},\ \bibinfo {pages} {034016} (\bibinfo {year} {2021})},\ \Eprint {http://arxiv.org/abs/2012.05034} {arXiv:2012.05034 [hep-ph]} \BibitemShut {NoStop}%
\bibitem [{\citenamefont {Molnar}\ \emph {et~al.}(2019)\citenamefont {Molnar}, \citenamefont {Danilkin},\ and\ \citenamefont {Vanderhaeghen}}]{Molnar:2019uos}%
  \BibitemOpen
  \bibfield  {author} {\bibinfo {author} {\bibfnamefont {D.~A.~S.}\ \bibnamefont {Molnar}}, \bibinfo {author} {\bibfnamefont {I.}~\bibnamefont {Danilkin}}, \ and\ \bibinfo {author} {\bibfnamefont {M.}~\bibnamefont {Vanderhaeghen}},\ }\href {\doibase 10.1016/j.physletb.2019.134851} {\bibfield  {journal} {\bibinfo  {journal} {Phys. Lett. B}\ }\textbf {\bibinfo {volume} {797}},\ \bibinfo {pages} {134851} (\bibinfo {year} {2019})},\ \Eprint {http://arxiv.org/abs/1903.08458} {arXiv:1903.08458 [hep-ph]} \BibitemShut {NoStop}%
\bibitem [{\citenamefont {Danilkin}\ \emph {et~al.}(2020)\citenamefont {Danilkin}, \citenamefont {Molnar},\ and\ \citenamefont {Vanderhaeghen}}]{Danilkin:2020kce}%
  \BibitemOpen
  \bibfield  {author} {\bibinfo {author} {\bibfnamefont {I.}~\bibnamefont {Danilkin}}, \bibinfo {author} {\bibfnamefont {D.~A.~S.}\ \bibnamefont {Molnar}}, \ and\ \bibinfo {author} {\bibfnamefont {M.}~\bibnamefont {Vanderhaeghen}},\ }\href {\doibase 10.1103/PhysRevD.102.016019} {\bibfield  {journal} {\bibinfo  {journal} {Phys. Rev. D}\ }\textbf {\bibinfo {volume} {102}},\ \bibinfo {pages} {016019} (\bibinfo {year} {2020})},\ \Eprint {http://arxiv.org/abs/2004.13499} {arXiv:2004.13499 [hep-ph]} \BibitemShut {NoStop}%
\bibitem [{\citenamefont {Xue}\ \emph {et~al.}(2018)\citenamefont {Xue}, \citenamefont {Jing}, \citenamefont {Guo},\ and\ \citenamefont {Zhao}}]{Xue:2017xpu}%
  \BibitemOpen
  \bibfield  {author} {\bibinfo {author} {\bibfnamefont {S.-R.}\ \bibnamefont {Xue}}, \bibinfo {author} {\bibfnamefont {H.-J.}\ \bibnamefont {Jing}}, \bibinfo {author} {\bibfnamefont {F.-K.}\ \bibnamefont {Guo}}, \ and\ \bibinfo {author} {\bibfnamefont {Q.}~\bibnamefont {Zhao}},\ }\href {\doibase 10.1016/j.physletb.2018.02.027} {\bibfield  {journal} {\bibinfo  {journal} {Phys. Lett. B}\ }\textbf {\bibinfo {volume} {779}},\ \bibinfo {pages} {402} (\bibinfo {year} {2018})},\ \Eprint {http://arxiv.org/abs/1708.06961} {arXiv:1708.06961 [hep-ph]} \BibitemShut {NoStop}%
\bibitem [{\citenamefont {Peng}\ \emph {et~al.}(2023)\citenamefont {Peng}, \citenamefont {Yan}, \citenamefont {S\'anchez~S\'anchez},\ and\ \citenamefont {Pavon~Valderrama}}]{Peng:2022nrj}%
  \BibitemOpen
  \bibfield  {author} {\bibinfo {author} {\bibfnamefont {F.-Z.}\ \bibnamefont {Peng}}, \bibinfo {author} {\bibfnamefont {M.-J.}\ \bibnamefont {Yan}}, \bibinfo {author} {\bibfnamefont {M.}~\bibnamefont {S\'anchez~S\'anchez}}, \ and\ \bibinfo {author} {\bibfnamefont {M.}~\bibnamefont {Pavon~Valderrama}},\ }\href {\doibase 10.1103/PhysRevD.107.016001} {\bibfield  {journal} {\bibinfo  {journal} {Phys. Rev. D}\ }\textbf {\bibinfo {volume} {107}},\ \bibinfo {pages} {016001} (\bibinfo {year} {2023})},\ \Eprint {http://arxiv.org/abs/2205.13590} {arXiv:2205.13590 [hep-ph]} \BibitemShut {NoStop}%
\bibitem [{\citenamefont {Ji}\ \emph {et~al.}(2022)\citenamefont {Ji}, \citenamefont {Dong}, \citenamefont {Guo},\ and\ \citenamefont {Zou}}]{Ji:2022blw}%
  \BibitemOpen
  \bibfield  {author} {\bibinfo {author} {\bibfnamefont {T.}~\bibnamefont {Ji}}, \bibinfo {author} {\bibfnamefont {X.-K.}\ \bibnamefont {Dong}}, \bibinfo {author} {\bibfnamefont {F.-K.}\ \bibnamefont {Guo}}, \ and\ \bibinfo {author} {\bibfnamefont {B.-S.}\ \bibnamefont {Zou}},\ }\href {\doibase 10.1103/PhysRevLett.129.102002} {\bibfield  {journal} {\bibinfo  {journal} {Phys. Rev. Lett.}\ }\textbf {\bibinfo {volume} {129}},\ \bibinfo {pages} {102002} (\bibinfo {year} {2022})},\ \Eprint {http://arxiv.org/abs/2205.10994} {arXiv:2205.10994 [hep-ph]} \BibitemShut {NoStop}%
\bibitem [{\citenamefont {von Detten}\ \emph {et~al.}(2023)\citenamefont {von Detten}, \citenamefont {Hanhart},\ and\ \citenamefont {Baru}}]{vonDetten:2023uja}%
  \BibitemOpen
  \bibfield  {author} {\bibinfo {author} {\bibfnamefont {L.}~\bibnamefont {von Detten}}, \bibinfo {author} {\bibfnamefont {C.}~\bibnamefont {Hanhart}}, \ and\ \bibinfo {author} {\bibfnamefont {V.}~\bibnamefont {Baru}},\ }in\ \href@noop {} {\emph {\bibinfo {booktitle} {{17th International Workshop on Meson Physics}}}}\ (\bibinfo {year} {2023})\ \Eprint {http://arxiv.org/abs/2309.11970} {arXiv:2309.11970 [hep-ph]} \BibitemShut {NoStop}%
\bibitem [{\citenamefont {Nakamura}\ \emph {et~al.}(2023)\citenamefont {Nakamura}, \citenamefont {Li}, \citenamefont {Peng}, \citenamefont {Sun},\ and\ \citenamefont {Zhou}}]{Nakamura:2023obk}%
  \BibitemOpen
  \bibfield  {author} {\bibinfo {author} {\bibfnamefont {S.~X.}\ \bibnamefont {Nakamura}}, \bibinfo {author} {\bibfnamefont {X.~H.}\ \bibnamefont {Li}}, \bibinfo {author} {\bibfnamefont {H.~P.}\ \bibnamefont {Peng}}, \bibinfo {author} {\bibfnamefont {Z.~T.}\ \bibnamefont {Sun}}, \ and\ \bibinfo {author} {\bibfnamefont {X.~R.}\ \bibnamefont {Zhou}},\ }\href@noop {} {\  (\bibinfo {year} {2023})},\ \Eprint {http://arxiv.org/abs/2312.17658} {arXiv:2312.17658 [hep-ph]} \BibitemShut {NoStop}%
\bibitem [{\citenamefont {Chen}\ \emph {et~al.}(2018)\citenamefont {Chen}, \citenamefont {Liu},\ and\ \citenamefont {Matsuki}}]{Chen:2017uof}%
  \BibitemOpen
  \bibfield  {author} {\bibinfo {author} {\bibfnamefont {D.-Y.}\ \bibnamefont {Chen}}, \bibinfo {author} {\bibfnamefont {X.}~\bibnamefont {Liu}}, \ and\ \bibinfo {author} {\bibfnamefont {T.}~\bibnamefont {Matsuki}},\ }\href {\doibase 10.1140/epjc/s10052-018-5635-1} {\bibfield  {journal} {\bibinfo  {journal} {Eur. Phys. J. C}\ }\textbf {\bibinfo {volume} {78}},\ \bibinfo {pages} {136} (\bibinfo {year} {2018})},\ \Eprint {http://arxiv.org/abs/1708.01954} {arXiv:1708.01954 [hep-ph]} \BibitemShut {NoStop}%
\bibitem [{\citenamefont {Weinberg}(1965)}]{Weinberg:1965zz}%
  \BibitemOpen
  \bibfield  {author} {\bibinfo {author} {\bibfnamefont {S.}~\bibnamefont {Weinberg}},\ }\href {\doibase 10.1103/PhysRev.137.B672} {\bibfield  {journal} {\bibinfo  {journal} {Phys. Rev.}\ }\textbf {\bibinfo {volume} {137}},\ \bibinfo {pages} {B672} (\bibinfo {year} {1965})}\BibitemShut {NoStop}%
\bibitem [{\citenamefont {Guo}\ \emph {et~al.}(2009)\citenamefont {Guo}, \citenamefont {Hanhart},\ and\ \citenamefont {Meissner}}]{Guo:2009wr}%
  \BibitemOpen
  \bibfield  {author} {\bibinfo {author} {\bibfnamefont {F.-K.}\ \bibnamefont {Guo}}, \bibinfo {author} {\bibfnamefont {C.}~\bibnamefont {Hanhart}}, \ and\ \bibinfo {author} {\bibfnamefont {U.-G.}\ \bibnamefont {Meissner}},\ }\href {\doibase 10.1103/PhysRevLett.103.082003} {\bibfield  {journal} {\bibinfo  {journal} {Phys. Rev. Lett.}\ }\textbf {\bibinfo {volume} {103}},\ \bibinfo {pages} {082003} (\bibinfo {year} {2009})},\ \bibinfo {note} {[Erratum: Phys.Rev.Lett. 104, 109901 (2010)]},\ \Eprint {http://arxiv.org/abs/0907.0521} {arXiv:0907.0521 [hep-ph]} \BibitemShut {NoStop}%
\bibitem [{\citenamefont {Guo}\ \emph {et~al.}(2011)\citenamefont {Guo}, \citenamefont {Hanhart}, \citenamefont {Li}, \citenamefont {Meissner},\ and\ \citenamefont {Zhao}}]{Guo:2010ak}%
  \BibitemOpen
  \bibfield  {author} {\bibinfo {author} {\bibfnamefont {F.-K.}\ \bibnamefont {Guo}}, \bibinfo {author} {\bibfnamefont {C.}~\bibnamefont {Hanhart}}, \bibinfo {author} {\bibfnamefont {G.}~\bibnamefont {Li}}, \bibinfo {author} {\bibfnamefont {U.-G.}\ \bibnamefont {Meissner}}, \ and\ \bibinfo {author} {\bibfnamefont {Q.}~\bibnamefont {Zhao}},\ }\href {\doibase 10.1103/PhysRevD.83.034013} {\bibfield  {journal} {\bibinfo  {journal} {Phys. Rev. D}\ }\textbf {\bibinfo {volume} {83}},\ \bibinfo {pages} {034013} (\bibinfo {year} {2011})},\ \Eprint {http://arxiv.org/abs/1008.3632} {arXiv:1008.3632 [hep-ph]} \BibitemShut {NoStop}%
\bibitem [{\citenamefont {Wang}\ \emph {et~al.}(2013{\natexlab{b}})\citenamefont {Wang}, \citenamefont {Hanhart},\ and\ \citenamefont {Zhao}}]{Wang:2013hga}%
  \BibitemOpen
  \bibfield  {author} {\bibinfo {author} {\bibfnamefont {Q.}~\bibnamefont {Wang}}, \bibinfo {author} {\bibfnamefont {C.}~\bibnamefont {Hanhart}}, \ and\ \bibinfo {author} {\bibfnamefont {Q.}~\bibnamefont {Zhao}},\ }\href {\doibase 10.1016/j.physletb.2013.06.049} {\bibfield  {journal} {\bibinfo  {journal} {Phys. Lett. B}\ }\textbf {\bibinfo {volume} {725}},\ \bibinfo {pages} {106} (\bibinfo {year} {2013}{\natexlab{b}})},\ \Eprint {http://arxiv.org/abs/1305.1997} {arXiv:1305.1997 [hep-ph]} \BibitemShut {NoStop}%
\bibitem [{\citenamefont {Matuschek}\ \emph {et~al.}(2021)\citenamefont {Matuschek}, \citenamefont {Baru}, \citenamefont {Guo},\ and\ \citenamefont {Hanhart}}]{Matuschek:2020gqe}%
  \BibitemOpen
  \bibfield  {author} {\bibinfo {author} {\bibfnamefont {I.}~\bibnamefont {Matuschek}}, \bibinfo {author} {\bibfnamefont {V.}~\bibnamefont {Baru}}, \bibinfo {author} {\bibfnamefont {F.-K.}\ \bibnamefont {Guo}}, \ and\ \bibinfo {author} {\bibfnamefont {C.}~\bibnamefont {Hanhart}},\ }\href {\doibase 10.1140/epja/s10050-021-00413-y} {\bibfield  {journal} {\bibinfo  {journal} {Eur. Phys. J. A}\ }\textbf {\bibinfo {volume} {57}},\ \bibinfo {pages} {101} (\bibinfo {year} {2021})},\ \Eprint {http://arxiv.org/abs/2007.05329} {arXiv:2007.05329 [hep-ph]} \BibitemShut {NoStop}%
\bibitem [{\citenamefont {Braaten}\ and\ \citenamefont {Lu}(2007)}]{Braaten:2007dw}%
  \BibitemOpen
  \bibfield  {author} {\bibinfo {author} {\bibfnamefont {E.}~\bibnamefont {Braaten}}\ and\ \bibinfo {author} {\bibfnamefont {M.}~\bibnamefont {Lu}},\ }\href {\doibase 10.1103/PhysRevD.76.094028} {\bibfield  {journal} {\bibinfo  {journal} {Phys. Rev. D}\ }\textbf {\bibinfo {volume} {76}},\ \bibinfo {pages} {094028} (\bibinfo {year} {2007})},\ \Eprint {http://arxiv.org/abs/0709.2697} {arXiv:0709.2697 [hep-ph]} \BibitemShut {NoStop}%
\bibitem [{\citenamefont {Hanhart}\ \emph {et~al.}(2010)\citenamefont {Hanhart}, \citenamefont {Kalashnikova},\ and\ \citenamefont {Nefediev}}]{Hanhart:2010wh}%
  \BibitemOpen
  \bibfield  {author} {\bibinfo {author} {\bibfnamefont {C.}~\bibnamefont {Hanhart}}, \bibinfo {author} {\bibfnamefont {Y.~S.}\ \bibnamefont {Kalashnikova}}, \ and\ \bibinfo {author} {\bibfnamefont {A.~V.}\ \bibnamefont {Nefediev}},\ }\href {\doibase 10.1103/PhysRevD.81.094028} {\bibfield  {journal} {\bibinfo  {journal} {Phys. Rev. D}\ }\textbf {\bibinfo {volume} {81}},\ \bibinfo {pages} {094028} (\bibinfo {year} {2010})},\ \Eprint {http://arxiv.org/abs/1002.4097} {arXiv:1002.4097 [hep-ph]} \BibitemShut {NoStop}%
\bibitem [{\citenamefont {Guo}(2020)}]{Guo:2020oqk}%
  \BibitemOpen
  \bibfield  {author} {\bibinfo {author} {\bibfnamefont {F.-K.}\ \bibnamefont {Guo}},\ }\href {\doibase 10.11804/NuclPhysRev.37.2019CNPC52} {\bibfield  {journal} {\bibinfo  {journal} {Nucl. Phys. Rev.}\ }\textbf {\bibinfo {volume} {37}},\ \bibinfo {pages} {406} (\bibinfo {year} {2020})},\ \Eprint {http://arxiv.org/abs/2001.05884} {arXiv:2001.05884 [hep-ph]} \BibitemShut {NoStop}%
\bibitem [{\citenamefont {Watson}(1952)}]{Watson:1952ji}%
  \BibitemOpen
  \bibfield  {author} {\bibinfo {author} {\bibfnamefont {K.~M.}\ \bibnamefont {Watson}},\ }\href {\doibase 10.1103/PhysRev.88.1163} {\bibfield  {journal} {\bibinfo  {journal} {Phys. Rev.}\ }\textbf {\bibinfo {volume} {88}},\ \bibinfo {pages} {1163} (\bibinfo {year} {1952})}\BibitemShut {NoStop}%
\bibitem [{\citenamefont {Baru}\ \emph {et~al.}(2023)\citenamefont {Baru}, \citenamefont {Epelbaum}, \citenamefont {Filin}, \citenamefont {Hanhart},\ and\ \citenamefont {Nefediev}}]{Baru:2022xne}%
  \BibitemOpen
  \bibfield  {author} {\bibinfo {author} {\bibfnamefont {V.}~\bibnamefont {Baru}}, \bibinfo {author} {\bibfnamefont {E.}~\bibnamefont {Epelbaum}}, \bibinfo {author} {\bibfnamefont {A.~A.}\ \bibnamefont {Filin}}, \bibinfo {author} {\bibfnamefont {C.}~\bibnamefont {Hanhart}}, \ and\ \bibinfo {author} {\bibfnamefont {A.~V.}\ \bibnamefont {Nefediev}},\ }\href {\doibase 10.1103/PhysRevD.107.014027} {\bibfield  {journal} {\bibinfo  {journal} {Phys. Rev. D}\ }\textbf {\bibinfo {volume} {107}},\ \bibinfo {pages} {014027} (\bibinfo {year} {2023})},\ \Eprint {http://arxiv.org/abs/2211.08038} {arXiv:2211.08038 [hep-ph]} \BibitemShut {NoStop}%
\bibitem [{\citenamefont {Guo}\ \emph {et~al.}(2013)\citenamefont {Guo}, \citenamefont {Hanhart}, \citenamefont {Mei\ss{}ner}, \citenamefont {Wang},\ and\ \citenamefont {Zhao}}]{Guo:2013zbw}%
  \BibitemOpen
  \bibfield  {author} {\bibinfo {author} {\bibfnamefont {F.-K.}\ \bibnamefont {Guo}}, \bibinfo {author} {\bibfnamefont {C.}~\bibnamefont {Hanhart}}, \bibinfo {author} {\bibfnamefont {U.-G.}\ \bibnamefont {Mei\ss{}ner}}, \bibinfo {author} {\bibfnamefont {Q.}~\bibnamefont {Wang}}, \ and\ \bibinfo {author} {\bibfnamefont {Q.}~\bibnamefont {Zhao}},\ }\href {\doibase 10.1016/j.physletb.2013.06.053} {\bibfield  {journal} {\bibinfo  {journal} {Phys. Lett. B}\ }\textbf {\bibinfo {volume} {725}},\ \bibinfo {pages} {127} (\bibinfo {year} {2013})},\ \Eprint {http://arxiv.org/abs/1306.3096} {arXiv:1306.3096 [hep-ph]} \BibitemShut {NoStop}%
\bibitem [{\citenamefont {Ablikim}\ \emph {et~al.}(2017{\natexlab{c}})\citenamefont {Ablikim} \emph {et~al.}}]{BESIII:2017tqk}%
  \BibitemOpen
  \bibfield  {author} {\bibinfo {author} {\bibfnamefont {M.}~\bibnamefont {Ablikim}} \emph {et~al.} (\bibinfo {collaboration} {BESIII}),\ }\href {\doibase 10.1103/PhysRevD.96.032004} {\bibfield  {journal} {\bibinfo  {journal} {Phys. Rev. D}\ }\textbf {\bibinfo {volume} {96}},\ \bibinfo {pages} {032004} (\bibinfo {year} {2017}{\natexlab{c}})},\ \bibinfo {note} {[Erratum: Phys.Rev.D 99, 019903 (2019)]},\ \Eprint {http://arxiv.org/abs/1703.08787} {arXiv:1703.08787 [hep-ex]} \BibitemShut {NoStop}%
\bibitem [{\citenamefont {Baru}\ \emph {et~al.}(2010)\citenamefont {Baru}, \citenamefont {Hanhart}, \citenamefont {Kalashnikova}, \citenamefont {Kudryavtsev},\ and\ \citenamefont {Nefediev}}]{Baru:2010ww}%
  \BibitemOpen
  \bibfield  {author} {\bibinfo {author} {\bibfnamefont {V.}~\bibnamefont {Baru}}, \bibinfo {author} {\bibfnamefont {C.}~\bibnamefont {Hanhart}}, \bibinfo {author} {\bibfnamefont {Y.~S.}\ \bibnamefont {Kalashnikova}}, \bibinfo {author} {\bibfnamefont {A.~E.}\ \bibnamefont {Kudryavtsev}}, \ and\ \bibinfo {author} {\bibfnamefont {A.~V.}\ \bibnamefont {Nefediev}},\ }\href {\doibase 10.1140/epja/i2010-10929-7} {\bibfield  {journal} {\bibinfo  {journal} {Eur. Phys. J. A}\ }\textbf {\bibinfo {volume} {44}},\ \bibinfo {pages} {93} (\bibinfo {year} {2010})},\ \Eprint {http://arxiv.org/abs/1001.0369} {arXiv:1001.0369 [hep-ph]} \BibitemShut {NoStop}%
\bibitem [{\citenamefont {Baru}\ \emph {et~al.}(2022{\natexlab{a}})\citenamefont {Baru}, \citenamefont {Dong}, \citenamefont {Du}, \citenamefont {Filin}, \citenamefont {Guo}, \citenamefont {Hanhart}, \citenamefont {Nefediev}, \citenamefont {Nieves},\ and\ \citenamefont {Wang}}]{Baru:2021ldu}%
  \BibitemOpen
  \bibfield  {author} {\bibinfo {author} {\bibfnamefont {V.}~\bibnamefont {Baru}}, \bibinfo {author} {\bibfnamefont {X.-K.}\ \bibnamefont {Dong}}, \bibinfo {author} {\bibfnamefont {M.-L.}\ \bibnamefont {Du}}, \bibinfo {author} {\bibfnamefont {A.}~\bibnamefont {Filin}}, \bibinfo {author} {\bibfnamefont {F.-K.}\ \bibnamefont {Guo}}, \bibinfo {author} {\bibfnamefont {C.}~\bibnamefont {Hanhart}}, \bibinfo {author} {\bibfnamefont {A.}~\bibnamefont {Nefediev}}, \bibinfo {author} {\bibfnamefont {J.}~\bibnamefont {Nieves}}, \ and\ \bibinfo {author} {\bibfnamefont {Q.}~\bibnamefont {Wang}},\ }\href {\doibase 10.1016/j.physletb.2022.137290} {\bibfield  {journal} {\bibinfo  {journal} {Phys. Lett. B}\ }\textbf {\bibinfo {volume} {833}},\ \bibinfo {pages} {137290} (\bibinfo {year} {2022}{\natexlab{a}})},\ \Eprint {http://arxiv.org/abs/2110.07484} {arXiv:2110.07484 [hep-ph]} \BibitemShut {NoStop}%
\bibitem [{\citenamefont {Workman}\ \emph {et~al.}(2022)\citenamefont {Workman} \emph {et~al.}}]{ParticleDataGroup:2022pth}%
  \BibitemOpen
  \bibfield  {author} {\bibinfo {author} {\bibfnamefont {R.~L.}\ \bibnamefont {Workman}} \emph {et~al.} (\bibinfo {collaboration} {Particle Data Group}),\ }\href {\doibase 10.1093/ptep/ptac097} {\bibfield  {journal} {\bibinfo  {journal} {PTEP}\ }\textbf {\bibinfo {volume} {2022}},\ \bibinfo {pages} {083C01} (\bibinfo {year} {2022})}\BibitemShut {NoStop}%
\bibitem [{\citenamefont {Zhang}\ \emph {et~al.}(2022)\citenamefont {Zhang}, \citenamefont {Hanhart}, \citenamefont {Mei\ss{}ner},\ and\ \citenamefont {Xie}}]{Zhang:2021hcl}%
  \BibitemOpen
  \bibfield  {author} {\bibinfo {author} {\bibfnamefont {X.}~\bibnamefont {Zhang}}, \bibinfo {author} {\bibfnamefont {C.}~\bibnamefont {Hanhart}}, \bibinfo {author} {\bibfnamefont {U.-G.}\ \bibnamefont {Mei\ss{}ner}}, \ and\ \bibinfo {author} {\bibfnamefont {J.-J.}\ \bibnamefont {Xie}},\ }\href {\doibase 10.1140/epja/s10050-021-00661-y} {\bibfield  {journal} {\bibinfo  {journal} {Eur. Phys. J. A}\ }\textbf {\bibinfo {volume} {58}},\ \bibinfo {pages} {20} (\bibinfo {year} {2022})},\ \Eprint {http://arxiv.org/abs/2107.03168} {arXiv:2107.03168 [hep-ph]} \BibitemShut {NoStop}%
\bibitem [{\citenamefont {Ablikim}\ \emph {et~al.}(2013)\citenamefont {Ablikim} \emph {et~al.}}]{BESIII:2013ouc}%
  \BibitemOpen
  \bibfield  {author} {\bibinfo {author} {\bibfnamefont {M.}~\bibnamefont {Ablikim}} \emph {et~al.} (\bibinfo {collaboration} {BESIII}),\ }\href {\doibase 10.1103/PhysRevLett.111.242001} {\bibfield  {journal} {\bibinfo  {journal} {Phys. Rev. Lett.}\ }\textbf {\bibinfo {volume} {111}},\ \bibinfo {pages} {242001} (\bibinfo {year} {2013})},\ \Eprint {http://arxiv.org/abs/1309.1896} {arXiv:1309.1896 [hep-ex]} \BibitemShut {NoStop}%
\bibitem [{\citenamefont {Hanhart}\ \emph {et~al.}(2001)\citenamefont {Hanhart}, \citenamefont {Sibirtsev},\ and\ \citenamefont {Speth}}]{Hanhart:2001ft}%
  \BibitemOpen
  \bibfield  {author} {\bibinfo {author} {\bibfnamefont {C.}~\bibnamefont {Hanhart}}, \bibinfo {author} {\bibfnamefont {A.}~\bibnamefont {Sibirtsev}}, \ and\ \bibinfo {author} {\bibfnamefont {J.}~\bibnamefont {Speth}},\ }\href@noop {} {\  (\bibinfo {year} {2001})},\ \Eprint {http://arxiv.org/abs/hep-ph/0107245} {arXiv:hep-ph/0107245} \BibitemShut {NoStop}%
\bibitem [{\citenamefont {Ablikim}\ \emph {et~al.}(2015)\citenamefont {Ablikim} \emph {et~al.}}]{BESIII:2015pqw}%
  \BibitemOpen
  \bibfield  {author} {\bibinfo {author} {\bibfnamefont {M.}~\bibnamefont {Ablikim}} \emph {et~al.} (\bibinfo {collaboration} {BESIII}),\ }\href {\doibase 10.1103/PhysRevD.92.092006} {\bibfield  {journal} {\bibinfo  {journal} {Phys. Rev. D}\ }\textbf {\bibinfo {volume} {92}},\ \bibinfo {pages} {092006} (\bibinfo {year} {2015})},\ \Eprint {http://arxiv.org/abs/1509.01398} {arXiv:1509.01398 [hep-ex]} \BibitemShut {NoStop}%
\bibitem [{\citenamefont {Ablikim}\ \emph {et~al.}(2017{\natexlab{d}})\citenamefont {Ablikim} \emph {et~al.}}]{BESIII:2017bua}%
  \BibitemOpen
  \bibfield  {author} {\bibinfo {author} {\bibfnamefont {M.}~\bibnamefont {Ablikim}} \emph {et~al.} (\bibinfo {collaboration} {BESIII}),\ }\href {\doibase 10.1103/PhysRevLett.119.072001} {\bibfield  {journal} {\bibinfo  {journal} {Phys. Rev. Lett.}\ }\textbf {\bibinfo {volume} {119}},\ \bibinfo {pages} {072001} (\bibinfo {year} {2017}{\natexlab{d}})},\ \Eprint {http://arxiv.org/abs/1706.04100} {arXiv:1706.04100 [hep-ex]} \BibitemShut {NoStop}%
\bibitem [{\citenamefont {Li}\ and\ \citenamefont {Voloshin}(2013)}]{Li:2013yka}%
  \BibitemOpen
  \bibfield  {author} {\bibinfo {author} {\bibfnamefont {X.}~\bibnamefont {Li}}\ and\ \bibinfo {author} {\bibfnamefont {M.~B.}\ \bibnamefont {Voloshin}},\ }\href {\doibase 10.1103/PhysRevD.88.034012} {\bibfield  {journal} {\bibinfo  {journal} {Phys. Rev. D}\ }\textbf {\bibinfo {volume} {88}},\ \bibinfo {pages} {034012} (\bibinfo {year} {2013})},\ \Eprint {http://arxiv.org/abs/1307.1072} {arXiv:1307.1072 [hep-ph]} \BibitemShut {NoStop}%
\bibitem [{\citenamefont {Chen}\ \emph {et~al.}(2023{\natexlab{b}})\citenamefont {Chen}, \citenamefont {Du},\ and\ \citenamefont {Guo}}]{Chen:2023def}%
  \BibitemOpen
  \bibfield  {author} {\bibinfo {author} {\bibfnamefont {Y.-H.}\ \bibnamefont {Chen}}, \bibinfo {author} {\bibfnamefont {M.-L.}\ \bibnamefont {Du}}, \ and\ \bibinfo {author} {\bibfnamefont {F.-K.}\ \bibnamefont {Guo}},\ }\href@noop {} {\  (\bibinfo {year} {2023}{\natexlab{b}})},\ \Eprint {http://arxiv.org/abs/2310.15965} {arXiv:2310.15965 [hep-ph]} \BibitemShut {NoStop}%
\bibitem [{\citenamefont {Guo}\ \emph {et~al.}(2015)\citenamefont {Guo}, \citenamefont {Hanhart}, \citenamefont {Wang},\ and\ \citenamefont {Zhao}}]{Guo:2014iya}%
  \BibitemOpen
  \bibfield  {author} {\bibinfo {author} {\bibfnamefont {F.-K.}\ \bibnamefont {Guo}}, \bibinfo {author} {\bibfnamefont {C.}~\bibnamefont {Hanhart}}, \bibinfo {author} {\bibfnamefont {Q.}~\bibnamefont {Wang}}, \ and\ \bibinfo {author} {\bibfnamefont {Q.}~\bibnamefont {Zhao}},\ }\href {\doibase 10.1103/PhysRevD.91.051504} {\bibfield  {journal} {\bibinfo  {journal} {Phys. Rev. D}\ }\textbf {\bibinfo {volume} {91}},\ \bibinfo {pages} {051504} (\bibinfo {year} {2015})},\ \Eprint {http://arxiv.org/abs/1411.5584} {arXiv:1411.5584 [hep-ph]} \BibitemShut {NoStop}%
\bibitem [{\citenamefont {Cleven}\ and\ \citenamefont {Zhao}(2017)}]{Cleven:2016qbn}%
  \BibitemOpen
  \bibfield  {author} {\bibinfo {author} {\bibfnamefont {M.}~\bibnamefont {Cleven}}\ and\ \bibinfo {author} {\bibfnamefont {Q.}~\bibnamefont {Zhao}},\ }\href {\doibase 10.1016/j.physletb.2017.02.041} {\bibfield  {journal} {\bibinfo  {journal} {Phys. Lett. B}\ }\textbf {\bibinfo {volume} {768}},\ \bibinfo {pages} {52} (\bibinfo {year} {2017})},\ \Eprint {http://arxiv.org/abs/1611.04408} {arXiv:1611.04408 [hep-ph]} \BibitemShut {NoStop}%
\bibitem [{\citenamefont {Casalbuoni}\ \emph {et~al.}(1993)\citenamefont {Casalbuoni}, \citenamefont {Deandrea}, \citenamefont {Di~Bartolomeo}, \citenamefont {Gatto}, \citenamefont {Feruglio},\ and\ \citenamefont {Nardulli}}]{Casalbuoni:1992fd}%
  \BibitemOpen
  \bibfield  {author} {\bibinfo {author} {\bibfnamefont {R.}~\bibnamefont {Casalbuoni}}, \bibinfo {author} {\bibfnamefont {A.}~\bibnamefont {Deandrea}}, \bibinfo {author} {\bibfnamefont {N.}~\bibnamefont {Di~Bartolomeo}}, \bibinfo {author} {\bibfnamefont {R.}~\bibnamefont {Gatto}}, \bibinfo {author} {\bibfnamefont {F.}~\bibnamefont {Feruglio}}, \ and\ \bibinfo {author} {\bibfnamefont {G.}~\bibnamefont {Nardulli}},\ }\href {\doibase 10.1016/0370-2693(93)91521-N} {\bibfield  {journal} {\bibinfo  {journal} {Phys. Lett. B}\ }\textbf {\bibinfo {volume} {309}},\ \bibinfo {pages} {163} (\bibinfo {year} {1993})},\ \Eprint {http://arxiv.org/abs/hep-ph/9304280} {arXiv:hep-ph/9304280} \BibitemShut {NoStop}%
\bibitem [{\citenamefont {Casalbuoni}\ \emph {et~al.}(1997)\citenamefont {Casalbuoni}, \citenamefont {Deandrea}, \citenamefont {Di~Bartolomeo}, \citenamefont {Gatto}, \citenamefont {Feruglio},\ and\ \citenamefont {Nardulli}}]{Casalbuoni:1996pg}%
  \BibitemOpen
  \bibfield  {author} {\bibinfo {author} {\bibfnamefont {R.}~\bibnamefont {Casalbuoni}}, \bibinfo {author} {\bibfnamefont {A.}~\bibnamefont {Deandrea}}, \bibinfo {author} {\bibfnamefont {N.}~\bibnamefont {Di~Bartolomeo}}, \bibinfo {author} {\bibfnamefont {R.}~\bibnamefont {Gatto}}, \bibinfo {author} {\bibfnamefont {F.}~\bibnamefont {Feruglio}}, \ and\ \bibinfo {author} {\bibfnamefont {G.}~\bibnamefont {Nardulli}},\ }\href {\doibase 10.1016/S0370-1573(96)00027-0} {\bibfield  {journal} {\bibinfo  {journal} {Phys. Rept.}\ }\textbf {\bibinfo {volume} {281}},\ \bibinfo {pages} {145} (\bibinfo {year} {1997})},\ \Eprint {http://arxiv.org/abs/hep-ph/9605342} {arXiv:hep-ph/9605342} \BibitemShut {NoStop}%
\bibitem [{\citenamefont {Colangelo}\ \emph {et~al.}(2006)\citenamefont {Colangelo}, \citenamefont {De~Fazio},\ and\ \citenamefont {Ferrandes}}]{Colangelo:2005gb}%
  \BibitemOpen
  \bibfield  {author} {\bibinfo {author} {\bibfnamefont {P.}~\bibnamefont {Colangelo}}, \bibinfo {author} {\bibfnamefont {F.}~\bibnamefont {De~Fazio}}, \ and\ \bibinfo {author} {\bibfnamefont {R.}~\bibnamefont {Ferrandes}},\ }\href {\doibase 10.1016/j.physletb.2006.01.021} {\bibfield  {journal} {\bibinfo  {journal} {Phys. Lett. B}\ }\textbf {\bibinfo {volume} {634}},\ \bibinfo {pages} {235} (\bibinfo {year} {2006})},\ \Eprint {http://arxiv.org/abs/hep-ph/0511317} {arXiv:hep-ph/0511317} \BibitemShut {NoStop}%
\bibitem [{\citenamefont {Dong}\ \emph {et~al.}(2021)\citenamefont {Dong}, \citenamefont {Guo},\ and\ \citenamefont {Zou}}]{Dong:2021juy}%
  \BibitemOpen
  \bibfield  {author} {\bibinfo {author} {\bibfnamefont {X.-K.}\ \bibnamefont {Dong}}, \bibinfo {author} {\bibfnamefont {F.-K.}\ \bibnamefont {Guo}}, \ and\ \bibinfo {author} {\bibfnamefont {B.-S.}\ \bibnamefont {Zou}},\ }\href {\doibase 10.13725/j.cnki.pip.2021.02.001} {\bibfield  {journal} {\bibinfo  {journal} {Progr. Phys.}\ }\textbf {\bibinfo {volume} {41}},\ \bibinfo {pages} {65} (\bibinfo {year} {2021})},\ \Eprint {http://arxiv.org/abs/2101.01021} {arXiv:2101.01021 [hep-ph]} \BibitemShut {NoStop}%
\bibitem [{\citenamefont {Baru}\ \emph {et~al.}(2022{\natexlab{b}})\citenamefont {Baru}, \citenamefont {Dong}, \citenamefont {Guo}, \citenamefont {Hanhart}, \citenamefont {Nefediev},\ and\ \citenamefont {Zou}}]{Baru:2022vmi}%
  \BibitemOpen
  \bibfield  {author} {\bibinfo {author} {\bibfnamefont {V.}~\bibnamefont {Baru}}, \bibinfo {author} {\bibfnamefont {X.~K.}\ \bibnamefont {Dong}}, \bibinfo {author} {\bibfnamefont {F.~K.}\ \bibnamefont {Guo}}, \bibinfo {author} {\bibfnamefont {C.}~\bibnamefont {Hanhart}}, \bibinfo {author} {\bibfnamefont {A.}~\bibnamefont {Nefediev}}, \ and\ \bibinfo {author} {\bibfnamefont {B.~S.}\ \bibnamefont {Zou}},\ }\href {\doibase 10.21468/SciPostPhysProc.6.007} {\bibfield  {journal} {\bibinfo  {journal} {SciPost Phys. Proc.}\ }\textbf {\bibinfo {volume} {6}},\ \bibinfo {pages} {007} (\bibinfo {year} {2022}{\natexlab{b}})}\BibitemShut {NoStop}%
\bibitem [{\citenamefont {Colangelo}\ \emph {et~al.}(2004)\citenamefont {Colangelo}, \citenamefont {De~Fazio},\ and\ \citenamefont {Pham}}]{Colangelo:2003sa}%
  \BibitemOpen
  \bibfield  {author} {\bibinfo {author} {\bibfnamefont {P.}~\bibnamefont {Colangelo}}, \bibinfo {author} {\bibfnamefont {F.}~\bibnamefont {De~Fazio}}, \ and\ \bibinfo {author} {\bibfnamefont {T.~N.}\ \bibnamefont {Pham}},\ }\href {\doibase 10.1103/PhysRevD.69.054023} {\bibfield  {journal} {\bibinfo  {journal} {Phys. Rev. D}\ }\textbf {\bibinfo {volume} {69}},\ \bibinfo {pages} {054023} (\bibinfo {year} {2004})},\ \Eprint {http://arxiv.org/abs/hep-ph/0310084} {arXiv:hep-ph/0310084} \BibitemShut {NoStop}%
\bibitem [{\citenamefont {Colangelo}\ \emph {et~al.}(2002)\citenamefont {Colangelo}, \citenamefont {De~Fazio},\ and\ \citenamefont {Pham}}]{Colangelo:2002mj}%
  \BibitemOpen
  \bibfield  {author} {\bibinfo {author} {\bibfnamefont {P.}~\bibnamefont {Colangelo}}, \bibinfo {author} {\bibfnamefont {F.}~\bibnamefont {De~Fazio}}, \ and\ \bibinfo {author} {\bibfnamefont {T.~N.}\ \bibnamefont {Pham}},\ }\href {\doibase 10.1016/S0370-2693(02)02306-7} {\bibfield  {journal} {\bibinfo  {journal} {Phys. Lett. B}\ }\textbf {\bibinfo {volume} {542}},\ \bibinfo {pages} {71} (\bibinfo {year} {2002})},\ \Eprint {http://arxiv.org/abs/hep-ph/0207061} {arXiv:hep-ph/0207061} \BibitemShut {NoStop}%
\bibitem [{\citenamefont {Daub}\ \emph {et~al.}(2016)\citenamefont {Daub}, \citenamefont {Hanhart},\ and\ \citenamefont {Kubis}}]{Daub:2015xja}%
  \BibitemOpen
  \bibfield  {author} {\bibinfo {author} {\bibfnamefont {J.~T.}\ \bibnamefont {Daub}}, \bibinfo {author} {\bibfnamefont {C.}~\bibnamefont {Hanhart}}, \ and\ \bibinfo {author} {\bibfnamefont {B.}~\bibnamefont {Kubis}},\ }\href {\doibase 10.1007/JHEP02(2016)009} {\bibfield  {journal} {\bibinfo  {journal} {JHEP}\ }\textbf {\bibinfo {volume} {02}},\ \bibinfo {pages} {009} (\bibinfo {year} {2016})},\ \Eprint {http://arxiv.org/abs/1508.06841} {arXiv:1508.06841 [hep-ph]} \BibitemShut {NoStop}%
\bibitem [{\citenamefont {Ropertz}\ \emph {et~al.}(2018)\citenamefont {Ropertz}, \citenamefont {Hanhart},\ and\ \citenamefont {Kubis}}]{Ropertz:2018stk}%
  \BibitemOpen
  \bibfield  {author} {\bibinfo {author} {\bibfnamefont {S.}~\bibnamefont {Ropertz}}, \bibinfo {author} {\bibfnamefont {C.}~\bibnamefont {Hanhart}}, \ and\ \bibinfo {author} {\bibfnamefont {B.}~\bibnamefont {Kubis}},\ }\href {\doibase 10.1140/epjc/s10052-018-6416-6} {\bibfield  {journal} {\bibinfo  {journal} {Eur. Phys. J. C}\ }\textbf {\bibinfo {volume} {78}},\ \bibinfo {pages} {1000} (\bibinfo {year} {2018})},\ \Eprint {http://arxiv.org/abs/1809.06867} {arXiv:1809.06867 [hep-ph]} \BibitemShut {NoStop}%
\bibitem [{\citenamefont {Chen}\ \emph {et~al.}(2019)\citenamefont {Chen}, \citenamefont {Dai}, \citenamefont {Guo},\ and\ \citenamefont {Kubis}}]{Chen:2019mgp}%
  \BibitemOpen
  \bibfield  {author} {\bibinfo {author} {\bibfnamefont {Y.-H.}\ \bibnamefont {Chen}}, \bibinfo {author} {\bibfnamefont {L.-Y.}\ \bibnamefont {Dai}}, \bibinfo {author} {\bibfnamefont {F.-K.}\ \bibnamefont {Guo}}, \ and\ \bibinfo {author} {\bibfnamefont {B.}~\bibnamefont {Kubis}},\ }\href {\doibase 10.1103/PhysRevD.99.074016} {\bibfield  {journal} {\bibinfo  {journal} {Phys. Rev. D}\ }\textbf {\bibinfo {volume} {99}},\ \bibinfo {pages} {074016} (\bibinfo {year} {2019})},\ \Eprint {http://arxiv.org/abs/1902.10957} {arXiv:1902.10957 [hep-ph]} \BibitemShut {NoStop}%
\bibitem [{\citenamefont {Mannel}\ and\ \citenamefont {Urech}(1997)}]{Mannel:1995jt}%
  \BibitemOpen
  \bibfield  {author} {\bibinfo {author} {\bibfnamefont {T.}~\bibnamefont {Mannel}}\ and\ \bibinfo {author} {\bibfnamefont {R.}~\bibnamefont {Urech}},\ }\href {\doibase 10.1007/s002880050344} {\bibfield  {journal} {\bibinfo  {journal} {Z. Phys. C}\ }\textbf {\bibinfo {volume} {73}},\ \bibinfo {pages} {541} (\bibinfo {year} {1997})},\ \Eprint {http://arxiv.org/abs/hep-ph/9510406} {arXiv:hep-ph/9510406} \BibitemShut {NoStop}%
\end{thebibliography}%

\vspace{7cm}
\appendix
\section{Lagrangian and parameter determination}
\label{Sec:AppA}
To construct the Lagrangian we define superfields representing the different light-quark spin multiples \cite{Casalbuoni:1992fd}. The ground states of heavy mesons with light quark quantum numbers $j^P=\tfrac{1}{2}^-$ will be denoted by $H_a$. In the presence of one unit of angular momentum there are 2 spin 
multiplets, one with $j^P=\tfrac{3}{2}^+$ and one with $j^P=\tfrac{1}{2}^+$. In the following
the former is of relevance, which is
denoted by $T_a$. The states in the latter have widths of the order
of 300 MeV and are only included implicitly.
All together
we may thus write

\begin{equation}
\begin{aligned}
H^{(Q)}_a&=\frac{1+\s{v}}{2} \left[ D_a^{* \mu} \gamma_\mu-D_a \gamma_5 \right] \ , \\
T^{(Q)^\mu}_a&=\frac{1+\slashed{v}}{2} \left[ D_{2a}^{\mu \nu} \gamma_\nu {\color{white}\frac{1}{1}}  \right.\\
& \qquad \left. -\sqrt{\tfrac{3}{2}} D_{1 a \nu} \gamma_5 (g^{\mu \nu} - \tfrac{1}{3}\gamma^\nu (\gamma^\mu-v^\mu)) \right] \ , 
\end{aligned}
\end{equation}

where $a$ is the SU(3) flavor index. We have, e.g., for $j^P=\tfrac{1}{2}^-$

\begin{equation}
\begin{aligned}
D_a&=(D^0, D^+, D_s^+) \\
D^{*}_{a \mu}&=(D^{*0}_\mu,D^{*+}_\mu,D^{*+}_{s \mu})\,.
\end{aligned}
\end{equation}

The superfields creating heavy mesons are given by 

\begin{equation}
\begin{aligned}
\bar{H}^{(Q)}_a =& \gamma_0 H_a^{(Q) \dagger} \gamma_0 \,=\left[ D_a^{* \dagger \mu} \gamma_\mu+D_a^\dagger \gamma_5 \right] \frac{1+\slashed{v}}{2}\\
\bar{T}^{(Q)^\mu}_a =&\gamma_0 T_a^{(Q)\mu \dagger} \gamma_0=\left[ D_{2a}^{\mu \nu \,\dagger} \gamma_\nu \right.\\
+ \sqrt{\tfrac{3}{2}}&\left. D_{1 a \nu}^\dagger  (g^{\mu \nu} - \tfrac{1}{3} (\gamma^\mu-v^\mu)\gamma^\nu)  \gamma_5 \right] \frac{1+\slashed{v}}{2} \,.
\end{aligned}
\end{equation}

The corresponding superfields containing an anti-heavy quark $\bar{Q}$ can be constructed using the charge conjugation operator $\mathcal{C}=i\gamma^2 \gamma^0$, where we are following the convention ${\mathcal{C} D_a^{(Q)} \mathcal{C}^{-1}= D_a^{(\bar Q)}}$ and ${\mathcal{C} D_a^{* (Q)} \mathcal{C}^{-1}=-D_a^{* (\bar Q)}}$.

\begin{widetext}
    
\begin{equation}
\begin{aligned}
H_a^{(\bar{Q})}&= \left[ D^{* (\bar{Q})}_{a\mu} \gamma^\mu- D_a^{(\bar{Q})} \gamma_5 \right] \frac{1-\slashed{v}}{2}\\
\bar{H}_a^{(\bar{Q})}&=\gamma_0 H_a^{(\bar{Q})\dagger} \gamma_0=\frac{1-\slashed{v}}{2}  \left[ D_a^{* (\bar{Q}) \, \mu \, \dagger} \gamma_\mu+D_a^{ (\bar{Q}) \, \dagger} \gamma_5 \right]\\[10pt]
T_a^{(\bar{Q}) \mu}&=\left[ D_{2a}^{(\bar{Q}) \mu \nu} \gamma_\nu - \sqrt{\tfrac{3}{2}} D_{1a \nu}^{(\bar{Q})} \gamma_5 (g^{\mu \nu} -\tfrac{1}{3}(\gamma^\mu-v^\mu) \gamma^\nu) \right] \frac{1-\s{v}}{2}\\
\bar{T}_{a}^{(\bar{Q})\,\mu}&=\gamma_0 T_{a}^{(\bar{Q})\,\mu\,\dagger} \gamma_0=\frac{1-\slashed{v}}{2}  \left[ D_{2a}^{(\bar{Q}) \, \mu \nu \,\dagger} \gamma_\nu +\sqrt{\tfrac{3}{2}} D_{1 a \nu}^{(\bar{Q}) \ \dagger}  (g^{\mu \nu} - \tfrac{1}{3} \gamma^\nu   (\gamma^\mu-v^\mu)\gamma_5 )\right]\,.
\end{aligned}
\end{equation}

\end{widetext}

The heavy field operators contain a factor $\sqrt{M_\text{H}}$ and therefore have dimension 3/2.\\[.3cm]
Pseudoscalar mesons couple  through the vector $\mathcal{V}_\mu$ and axialvector $\mathcal{A}_\mu$ current containing an even and odd number of boson fields, respectively, 

\begin{equation}
\begin{aligned}
\mathcal{V}_\mu &=\frac{1}{2}(u^\dagger \partial_\mu u + u \partial_\mu u^\dagger)\\
\mathcal{A}_\mu &=\frac{i}{2}(u^\dagger \partial_\mu u - u \partial_\mu u^\dagger)\, ,
\end{aligned}
\end{equation}

conserving chiral symmetry. Chiral symmetry violation is introduced via constructions
of the kind

\begin{equation}
    \chi_\pm = u^\dagger \chi u^\dagger \pm u \chi^\dagger u \ ,
\end{equation}

with $\chi=2B{\mathcal M}$, where $\mathcal M$ is the quark mass matrix and $B$
is a scale parameter related to the chiral condensate.
Here the  exponential parameterization is employed for the light Goldstone boson fields:

\begin{equation}
\begin{aligned}
u=&\exp\left(  i \frac{\Phi}{\sqrt{2} \fpi} \right) \\
\Phi=&
\begin{pmatrix}
\frac{\pi^0}{\sqrt{2}}+\frac{\eta_8}{\sqrt{6}} & \pi^+ & K^+\\
\pi^- & -\frac{\pi^0}{\sqrt{2}}+\frac{\eta_8}{\sqrt{6}} & K^0\\
K^- &\bar{K}^0 & -\frac{2\eta_8}{\sqrt{6}}
\end{pmatrix},
\label{eq:goldstone_fields}
\end{aligned}
\end{equation}

with $\fpi=92\mev$ denoting the pion decay constant in the chiral limit. The Lagrangian is constructed by imposing invariance under heavy-quark spin and chiral transformation \cite{Casalbuoni:1996pg,Colangelo:2005gb,Dong:2021juy}. The kinetic terms are

\begin{equation}
\begin{aligned}
	\mathcal{L}_\text{kin}=&i \langle \bar{H}_b v_\mu D^{\mu}_{ba}  H_a \rangle  \\
 +&\frac{\fpi^2}{4} \left(\langle \partial^\mu u \partial_\mu u^\dagger \rangle + \langle \chi_+ \rangle \right)
 \\
 +&\langle \bar{T}^\mu_b ( i v_\nu D^\nu_{ba}- \delta_{ba} \Delta_T) T_{a \mu} \rangle 
\end{aligned}
\end{equation}

and the relevant terms for the interaction are given by

\begin{equation}
\begin{aligned}
\mathcal{L}_\text{int}&=g \langle H_b^{(Q)} {\mathcal{A}}\sss _{ba} \gamma_5 \bar{H}^{(Q)}_a \rangle +  k \langle T_b^{{(Q)}\mu} {\mathcal{A}}\sss _{ba} \gamma_5 \bar{T}_b^{(Q)} \rangle\\
&+  \frac{h_1}{\Lambda_\chi} \langle T_b^{(Q) \mu} (D_\mu {\mathcal{A}}\sss) _{ba} \gamma_5 \bar{H}_a^{(Q)}	\rangle \\
&+\frac{h_2}{\Lambda_\chi} \langle T_b^{(Q)\mu} (\s{D} \mathcal{A}_\mu)_{ba} \gamma_5 \bar{H}_a^{(Q)} \rangle\\
&+ g \langle \bar H_a^{(\bar{Q})} {\mathcal{A}}\sss _{ab} \gamma_5 H^{(\bar{Q})}_b \rangle + k \langle \bar T_a^{{(\bar{Q})}\mu} {\mathcal{A}}\sss _{ab}  \gamma_5 T_b^{(\bar{Q})} \rangle\\
&+  \frac{h_1}{\Lambda_\chi} \langle \bar T_a^{(\bar{Q}) \mu} ( {\mathcal{A}}\sss \overleftarrow{D}_\mu)_{ab} \gamma_5 H_b^{(\bar{Q})}	\rangle \\
& + \frac{h_2}{\Lambda_\chi} \langle \bar T_a^{(\bar{Q})\mu} ( \mathcal{A}_\mu \overleftarrow{\s{D}})_{ab} \gamma_5 H_b^{(\bar{Q})} \rangle + \text{h.c.} \,\, .
\label{lagrangian}
\end{aligned}
\end{equation}

The relation between the decay width and  the effective coupling of a resonance $R$
with total angular momentum $J_R$ decaying into the two-body final state $a$ 
in the narrow width approximation is given by~\cite{ParticleDataGroup:2022pth}

\begin{equation}
\Gamma_{R \rightarrow a}=\frac1{m_R} \rho_a(m_R^2)
\left(\frac{1}{2J_R+1}\right)\sum_{\rm pol.}\left|{\mathcal M}_{R\to a}\right|^2\, ,
\label{eq:coupling}
\end{equation}

where $m_R$ denotes the resonance mass, the phase space factor is $\rho_a(m_R^2)=2p_a(m_R)/(16\pi m_R)$
and $p_a$ denotes the relative momentum of the decay particles in the rest frame
of the resonance, 

\begin{equation}
\begin{aligned}
&p_a(m_R)=\\
&\frac{\sqrt{(m_R^2-(m_{a,1}+m_{a,2})^2)(m_R^2-(m_{a,1}-m_{a,2})^2)}}{(2m_R)}\ ,
\end{aligned}
\end{equation}

with $m_{a,i}$ for the masses of the particles in channel $a$. 
The summation runs over the polarizations of the final and initial state, respectively,
if necessary. The pertinent matrix elements can be read off the Lagrangian
given in Eq.~(\ref{lagrangian}) straightforwardly allowing one
 to determine the couplings from the experimentally measured decay widths.
 
The squared matrix element for the transition of $D^{* a} \to D^b \phi_{ab} $, summed over the 
$D^*$ polarizations, is given by

\begin{equation}
\begin{aligned}
\sum_\text{pol.} |\mathcal{M}_{D^* D \pi}|^2=\frac{2 g^2 c_{ab}^2 p_{\pi D}(m_{D^*})^2}{\fpi^2} m_{D^*} m_D \, ,
\end{aligned}
\end{equation}

where the coefficient $c_{ab}$ can be read off from the Goldstone boson matrix provided
in Eq.~(\ref{eq:goldstone_fields}): $c_{+0}=1$ and $c_{++}=1/\sqrt{2}$.
Using eq. \eqref{eq:coupling} we extract for $D^{* +} \rightarrow D^0 \pi^+$

\begin{equation}
\begin{aligned}
&|g(D^{* +} \rightarrow D^0 \pi^+)|=\\
&\sqrt{\frac{12 \fpi^2 \pi \Gamma(D^{* +} \rightarrow D^0 \pi^+)}{p_{\pi D}(m_{D^*})^3} \frac{m_{D^{* +}}}{m_{D^0}}}\approx 0.57 \, ,
\end{aligned}
\end{equation}

where the central values listed in the Review of Particle Physics by the Particle Data Group~\cite{ParticleDataGroup:2022pth} were used 

\begin{equation}
\begin{aligned}
    \Gamma(D^{* +} \rightarrow D^0 \pi^+)&=\text{BR}(D^{* +} \rightarrow D^0 \pi^+) \cdot \Gamma^{D^{* +}}_\text{full} \\
    &=0.677 \cdot 83.4\,\text{keV} = 56.4\,\text{keV} \,.
\end{aligned}
\end{equation}

Analogously from ${D^{* +} \rightarrow D^+ \pi^0}$ we find

\begin{equation}
\begin{aligned}
&|g(D^{* +} \rightarrow D^+ \pi^0)|=\\
&\sqrt{\frac{24 \fpi^2 \pi \Gamma(D^{* +} \rightarrow D^+ \pi^0)}{p_{\pi D}(m_{D^*})^3} \frac{m_{D^{* +}}}{m_{D^+}}}\approx 0.57 \, ,
\end{aligned}
\end{equation}

with $\Gamma(D^{* +} \rightarrow D^+ \pi^0)=25.6\,\text{keV}$. It is this value that we use in our calculations in line with Ref.~\cite{Baru:2022vmi}. Since in this work we do not aim at a calculation with controlled
uncertainties but more at demonstrating the consistency of the existing data with just
a single molecular state in the mass range studied, we feel   safe to not keep track with
the individual uncertainties of the parameters employed.
The interaction of the $j_l^P=\tfrac{3}{2}^+$ doublet $\{D_2,D_1\}$ with $\{D^*,D\}$ and the Goldstone bosons $\Phi$ 
given in Eq.~(\ref{lagrangian}) can be re-expressed as
\begin{equation}
\mathcal{L}_{TH\pi}
=-\frac{h^\prime}{\sqrt{2} \fpi} \langle T^{(Q) \mu} \gamma^\nu (\partial_\mu \partial_\nu \Phi) \gamma_5 \bar{H}^{(Q)} \rangle+ \text{h.c.} \, ,
\label{hpdef}
\end{equation}
where $h^\prime=(h_1+h_2)/\Lambda_\chi$.\\
The decay of the narrow $D_1$ into $D^* \pi$ is predominately in a $D$-wave, since the $S$-wave is suppressed by heavy-quark spin symmetry, which calls for the conservation of the light quark total angular momentum
in the decay. However, violations of this symmetry in the charm sector can be sizable. 
To get an estimate for the $S$-wave strength in the $D_1$ decay, we can use the fact that the spin partner of the $D_1$, the $D_2$, can only decay into $D^* \pi$ and $D \pi$ in a pure $D$-wave due to the total angular momentum 
conservation~\cite{Guo:2020oqk}.
Adding the partial widths, according to Eq.~(\ref{eq:coupling}), the total width of the $D_2$ is given by
\begin{equation}
\begin{aligned}
    \Gamma_{D_2}&= \frac{1}{5} 
    \frac{\rho_{\pi D^*}(m_{D_2}^2)}{m_{D_2}}\sum_\text{pol.} |\mathcal{M}_{D_2 \rightarrow D^* \pi}|^2\\
    &+  \frac{1}{5} 
    \frac{\rho_{\pi D}(m_{D_2}^2)}{m_{D_2}}\sum_\text{pol.} |\mathcal{M}_{D_2 \rightarrow D \pi}|^2 
    ,
\end{aligned}
\end{equation}
with
\begin{equation}
\begin{aligned}
    \sum_\text{pol.} |\mathcal{M}_{D_2 \rightarrow D^* \pi}|^2 &=\frac{3}{2} \frac{2 h^{\prime 2}}{\fpi^2} p_{\pi D^*}(m_{D_2})^4 m_{D_2} m_{D^*}\\
    \sum_\text{pol.} |\mathcal{M}_{D_2 \rightarrow D \pi}|^2 &= \frac{3}{2}\frac{4 h^{\prime 2}}{3 \fpi^2} p_{\pi D}(m_{D_2})^4 m_{D_2} m_D \, ,
\end{aligned}
\end{equation}
where the factor $3/2$ in front of each term results from adding
the partial widths of the $D^{(*)+}\pi^0$ and $D^{(*)0}\pi^+$ in line
with what was done for the decay of the $D^*$.
Using $\Gamma_{D_2}=47.3\,\mev$~\cite{ParticleDataGroup:2022pth},
one can extract $h^\prime=0.82\,\gev^{-1}$.
 Our calculation is not sensitive to the sign
of this coupling which we chose positive.

Allowing for a $D^*\pi$ $S$-wave, the expression for the total width of the
$D_1$ reads
\begin{equation}
\begin{aligned}
    \Gamma_{D_1}&= \frac{1}{3} 
    \frac{\rho_{\pi D^*}(m_{D_1}^2)}{m_{D_1}}\sum_\text{pol.} |\mathcal{M}_{D_1 \rightarrow D^* \pi}^{s-{\rm wave}}|^2\\
    &+ \frac{1}{3} 
    \frac{\rho_{\pi D^*}(m_{D_1}^2)}{m_{D_1}}\sum_\text{pol.} |\mathcal{M}_{D_1 \rightarrow D^* \pi}^{d-{\rm wave}}|^2
 ,
\end{aligned}
\end{equation}

From Eq.~(\ref{hpdef}) one finds
\begin{equation}
\sum_\text{pol.}  |\mathcal{M}_{D_1 \rightarrow D^* \pi}^{d-{\rm wave}}|^2   =\frac{h^{\prime  2}}{\fpi^2} p_{D^*\pi}(m_{D_1})^4  m_{D_1} m_{D^*} \ ,
\end{equation}
where again a factor 3/2 was included to account for the two possible
final states. With $h^{\prime}$ fixed above, one 
finds $\Gamma_{D_1}^{d\text{-wave}}=15\mev$, in agreement with Ref. \cite{Guo:2020oqk}.
Since the total width of the $D_1(2420)$ is 31 MeV~\cite{ParticleDataGroup:2022pth},
the remainder must be generated by the $S$-wave decay.
Using
\begin{equation}
\mathcal{L}_{D_1 D^* \pi}^\text{s-wave}=i \frac{h^\prime_s}{\sqrt{6} \fpi} \left( D_{1b} \cdot D^{* \dagger}_a\right) \partial_0 \phi_{ba}
\end{equation}
one gets
\begin{equation}
   \sum_\text{pol.} |\mathcal{M}_{D_1 \rightarrow D^* \pi}^{s-{\rm wave}}|^2=\frac{h^{\prime 2}_s \omega_\pi^2}{6 \fpi^2} \frac{3}{2} m_{D_1} m_{D^*} \ , 
\end{equation}
where $\omega_\pi$ denotes the energy of the pion and again the factor $3/2$ 
accounts for the two decay channels
$D_1^+ \rightarrow D^{* 0} \pi^+$ and $D^{* +} \pi^0$. This
 leads to $h^\prime_s=0.57$.
Below we study a pion angular distribution, which is sensitive to the relative
sign of $h^\prime$ and $h^\prime_s$. We here already account for the observation
that the data call for equal signs of the two. 
\\
Photons couple via the field-strength tensor $F^{\mu \nu}=\partial_\mu A_\nu-\partial_\nu A_\mu$, where $A_\mu$ denotes the photon field. In this way  gauge invariance is
preserved automatically. 
The production of a vector resonance  from a photon is thus described by
\begin{equation}
\mathcal{L}_{V \gamma}=\frac{e}{2f_V} V_{\mu\nu} F^{\mu\nu} \approx\frac{e m_V^2}{f_V} V_\mu A^\mu \, ,
\label{eq:photon_VMD}
\end{equation}
where $V_{\mu\nu}=\partial_\mu V_\nu-\partial_\nu V_\mu$ and $V$ denotes either
the field for the $Y(4230)$ or the $\psi(4160)$.
The implications of heavy quark spin symmetry on charmonium production from photons are discussed later in the chapter, but as the production of the $Y(4230)$ must go via the broad $D_1(2430)$, thus we may allow for an additional phase in case of a point-like production.  For the decay of $Y(4230) \rightarrow X(3872) \gamma$ we can describe the E1 transition of $D_1$ going to $D^* \gamma$ with the following Lagrangian
\begin{equation}
\mathcal{L}_{TH\gamma}=\frac{c_a}{2} \langle T^i_a \bar{H}_a \rangle E^i \,,
\end{equation}
where $E^i$ denotes the electric component of the photon field.

We now come to the description of the ground state doubly heavy vector fields
of relevance to this study.
Heavy quark spin symmetry allows us to write $Q \bar Q$ superfields~\cite{Colangelo:2003sa}.
The $\ell=0$ superfield $R^{(Q\bar{Q})}$ contains the \{${\jpsi}$,$\eta_c$\} doublet
\be
R^{(Q\bar{Q})}=\frac{1+\slashed{v}}{2}\left[ \jpsi^\mu \gamma_\mu - \eta_c \gamma_5 \right] \frac{1-\slashed{v}}{2}\, ,
\ee
where the interaction with $D/D^*$ is given by the Lagrangian
\be
\mathcal{L}_{HHR}=\frac{g_{HHR}}{2} \langle R^{(Q\bar{Q})} \bar{H}_{2a}  {\overset{\leftrightarrow}{\partial}}\ssss \bar{H}_{1a} \rangle \,,
\ee
with $A\overset{\leftrightarrow}{\partial_\mu} B=A (\partial_\mu B)-(\partial_\mu A) B$. The resulting vertex factors are
\begin{equation}
\begin{aligned}
V_{\jpsi D D} {=}& g_{\jpsi D \bar{D}} (\epsilon_{\jpsi} \cdot q)\\
V_{\jpsi D^* D} {=}& g_{\jpsi D^* \bar{D}} i \epsilon_{ijk} \epsilon_{\jpsi}^i \epsilon_{D^*}^j q^k\\
V_{\jpsi D^* D^*}{=}& -g_{\jpsi D^* \bar{D}^*} \left[ (\epsilon_{\jpsi} \cdot \epsilon_2) (\epsilon_1 \cdot q) \right.\\
- (\epsilon_{\jpsi} &\left. {} \cdot q) (\epsilon_1 \cdot \epsilon_2) + (\epsilon_{\jpsi} \cdot \epsilon_1) (\epsilon_2 \cdot q) \right]\, ,
\end{aligned}
\end{equation}
with $q=k_1^{(Q)}-k_2^{(\bar{Q})}=$ denoting the relative residual momentum between the $D$ mesons. At leading order the masses of the multiples are degenerate ${m_{D^*}=m_D=m_H}$ and $q$ simplifies to ${q=p_1-m_H v -p_2+m_H v=p_1-p_2}$. The coupling is traditionally
parameterized as
\be
g_{\jpsi A B}=\frac{m_{\jpsi} \sqrt{m_A m_B}}{m_D f_{\jpsi}}\, ,
\ee
which includes the leading spin-symmetry violating effects 
via the mass factors.
The $\ell=1$ superfield $P^{(Q\bar{Q})}$ contains the spin triplet $\chi_{c0},\chi_{c1},\chi_{c2}$ and the singlet $h_c$
\begin{equation}
\begin{aligned}
{P^{(Q\bar{Q})}}^\mu&=\frac{1+\s{v}}{2} \left( \chi_2^{\mu \alpha} \gamma_\alpha + \frac{1}{\sqrt{s}} \epsilon^{\mu \alpha \beta \gamma} v_\alpha \gamma_\beta \chi_{1\gamma} \right.\\
& \quad \left. + \frac{1}{\sqrt{3}} (\gamma^\mu-v^\mu) \chi_0 + h_1^\mu \gamma_5 \right) \frac{1-\s{v}}{2}\, .
\end{aligned}
\end{equation}
Due to the Proca-constraint, $P^\mu v_\mu$, the leading order Lagrangian
for the interaction of the $\ell=1$ spin-multiplet with $D$ and $D^*$ contains
only a single term:
\begin{equation}
\mathcal{L}_{HHP} = i\frac{g_{HHP}}{2} \left\langle {P^{(Q\bar{Q})}}^\mu \bar{H}_{2a}  \gamma_\mu \bar{H}_{1a} \right\rangle \ .
\end{equation}
From this the vertex factors evaluate to
\begin{equation}
\begin{aligned}
V_{\hc \Ds D} &= -2 g^{\hc} (\epsilon^*_{\Ds} \cdot \epsilon_{\hc}) \sqrt{m_\hc m_{D^*} m_D}\\
V_{\hc \Ds \Ds} &=2 i g^{\hc} \epsilon_{\alpha \beta \tau \sigma} p^\alpha \epsilon_{\hc}^\beta \epsilon^{* \tau}_1 \epsilon^{* \sigma}_2 \sqrt{\frac{m_{D^*}^2}{m_\hc}}\\
&=2 i g^{\hc} \epsilon^{ijk} \epsilon_{\hc}^i \epsilon^{* j}_1 \epsilon^{* k}_2 \sqrt{m_{D^*}^2 m_\hc } \, ,
\end{aligned}
\end{equation}
where we fixed $\alpha=0$ such that $p^\alpha\approx m_\hc$. The coupling is 
parameterized as
\begin{equation}
g^\hc=-\frac{m_{\chi_{c0}}}{3} \frac{1}{f_{\chi_{c0}}} \,,
\end{equation}
where $f_\jpsi=416 \mev$ was determined using vector-meson dominance (VMD) \cite{Guo:2010ak} and $f_{\chi_{c0}}=510\pm40$ from numerical results of QCD sum rules \cite{Colangelo:2002mj}.
Those parameters carry a systematic uncertainty which is
difficult to quantify. In
the fits we allow $f_\jpsi$ to vary within $10\%$ of its value, while we fix $f_{\chi_{c0}}$ to its central value, since our fit is not sensitive to this quantity.

\section{Inclusion of $\pi\pi-\bar KK$ final state interaction}
\label{sec:pipi_fsi}
An amplitude ${\cal M}$ corresponding to the given  isospin $I$ (the isospin index is omitted in what follows) can be projected to a partial wave ${\cal M}^l$ with definite angular momentum $l$
\be
{\cal M}^l(s)=\frac{1}{2 \sqrt{2}^\alpha} \int_{-1}^{1} \text{d} z \,\, {\cal M}(s,z) P^l(z)\, ,
\ee
where $P^l$ denotes the Legendre polynomial of degree $l$,  $z$ the scattering angle and $\alpha=0,1,2$ is a symmetry factor for identical particles in initial and final states (e.g. $\alpha=2$ for $AA \to BB$, $\alpha=1$ for $AA \to C \bar{C}$ and $\alpha=0$ for $AB \to AB$). The full amplitude can be reconstructed using the orthogonality relation of the Legendre polynomials
\be
{\cal M}(s,z)=\sqrt{2}^\alpha \sum_l (2l+1) {\cal M}^l(s) P^l(z)\, .
\ee
On the other hand,
from the unitarity of the $S$-matrix the discontinuity of the production amplitude 
$\mathcal{M}^l$ is given by
\be
\text{disc} \mathcal{M}^l_j(s)=2 i \sum_k T_{jk}^*(s) \sigma_k(s)  \mathcal{M}^l_k(s) \,,
\ee
where $\sigma_k=\sqrt{1-(4m_k^2/s)}\Theta(s-4 m_k^2)$ and the subscript indices $j, k,...$ refer to the coupled channels, which in our case correspond to $\pi\pi$ and $K\bar K$. Furthermore, $T_{jk}$
corresponds to the meson-meson coupled-channel amplitude.
 The solution is given by Muskhelishvili Omn\'es function encoding the $\pi\pi/K\bar K$ final state interaction. 
Therefore, the full amplitude is given by the sum of the  amplitudes  ${\cal M}_j$  and $\Gamma$,  involving the left-hand and right-hand (unitarity) cuts only, respectively, which reads    
\begin{widetext}
    
\begin{equation}
\begin{aligned}
\mathcal{M}_j^\text{full}(s)=\mathcal{M}_j+\Gamma_j&=\mathcal{M}_j+\sum_k \Omega_{jk} \left[  \left(\mathcal{P}_{n-1}\right)_k+\sum_{lm}\frac{s^n}{\pi} \int \frac{\D z}{z^n} \frac{\Omega^{-1}_{kl}(z) T_{lm}(z) \sigma_m(z) \mathcal{M}^0_{m}(z)}{z-s}  \right]\\
&=\left[ \mathcal{M}^{l>0}_j+\mathcal{M}^0_j\right]+\sum_k \left[ \Omega_{j k} \left( \left(\mathcal{P}_{n-1}\right)_k+\frac{s^n}{\pi}\,\,\text{P.V.} \int \left[...\right] \right)+i T_{jk} \sigma_k \mathcal{M}^0_{k} \right],
\label{ampwithFSI}
\end{aligned}
\end{equation}

\end{widetext}
where $\mathcal{P}_{n-1}$ is a polynomial of the order $n-1$, which is discussed below, and $\Omega$ is the S-wave Omn\'es matrix. The amplitudes   ${\cal M}_j$ correspond to the diagrams discussed in Secs.~\ref{sec:general_Jpsipipi} and \ref{sec:general_hcpipi}, while the right-hand cut in $\Gamma$ emerges from the $\pi\pi/K\bar K$ FSI  in an $S$ wave. 
The necessary input for the amplitude $T$ and the Omn\'es matrix is taken  from Refs.~\cite{Daub:2015xja,Ropertz:2018stk}.
We now have a closer look at the principal value integral.
Using the short hand notation
$\sum_{kl}\Omega^{-1}_{jk} T_{kl} \sigma_l \mathcal{M}_{l}=f_j(z)$ we get

\begin{eqnarray}
  \text{P.V.}& & \hspace{-0.3cm}\int \frac{dz}{z^{n}} \frac{f_k(z)}{z-s} \nonumber \\ 
=\text{P.V.}& & \hspace{-0.3cm}\int \text{d}z \left[ \frac{f_k(z)}{z^{n-1}} - \frac{f_k(s)}{s^{n-1}} + \frac{f_k(s)}{s^{n-1}}\right] 
\nonumber \\
& & \hspace{3cm}\times\frac{1}{z (z-s)}
\nonumber \\ \nonumber
& &\hspace{-1.5cm}=\int \text{d}z \left[ \frac{f_k(z)}{z^{n-1}}-\frac{f_k(s)}{s^{n-1}} \right] \\ \nonumber
& & \quad +  \frac{f_k(s)}{s^{n-1}}  {\,\,\, \text{P.V.}\int} \text{d}z \frac{1}{z (z-s)} \ .
\end{eqnarray}

The integral in the last line can be evaluated 
analytically

\be \nonumber
{\text{P.V.}\int_{s_\text{th}}^\infty} \text{d}z \frac{1}{z (z-s)}
=\frac{1}{s}\ln \left( \frac{1}{s/s_\text{th}-1}\right)
\ee

and that in the second 
needs to be done
numerically.
However, in the exploratory study
we aim at here, we neglect this part which we expect to play a role of a  correction. 
The quality of the fits we achieve might be taken as justification of this treatment a posteriori,
although some shift in importance of different contributions in a complete analysis cannot be
excluded.
Thus,
the modified $\pi \pi$ and $\bar K K$ s-wave is given by  %

\begin{widetext}
    
\be\label{Eq:M0mod}
(\mathcal{M}_{j}^0)_\text{mod}= \mathcal{M}^0_j + \sum_{k} 
 \Omega_{j k}  \left(\mathcal{P}_{n-1}\right)_k+ 
\left(i T_{jk} \sigma_k+\frac{1}{\pi} \ln \left( \frac{1}{s/s_\text{th}-1}\right) T_{jk} \sigma_k \right)  \mathcal{M}^0_k \, ,
\ee

\end{widetext}

such that the full amplitude  \eqref{ampwithFSI} is approximated  as

\be\label{eq:Mjmod}
\mathcal{M}_{j\text{ mod}}= \mathcal{M}^{l>0}_j+(\mathcal{M}_{j}^0)_\text{mod}\, .
\ee

This also enables us to project the FSI onto the $K\bar{K}$-channel, allowing us to also determine $Y(4230) \rightarrow \jpsi \pi \pi \rightarrow \jpsi \bar{K} K$.

The subtraction polynomial $\left( \mathcal{P}_{n-1}\right)_j$ in Eqs.~(\ref{ampwithFSI}) and \eqref{Eq:M0mod} is matched to the $Y(4230) \rightarrow \jpsi \phi \phi$ chiral contact term. In Ref. \cite{Chen:2019mgp} it was found that both SU(3) singlet and octet components of the light quarks contribute in the $Y(4230)$, which can be decomposed as

\begin{equation}
\begin{aligned}
        | Y(4230) \rangle &=\left(c_1|V_1^\text{light}  \rangle + c_8 | V_8^\text{light} \rangle\,\right)\\
  &\hspace{3.5cm}  {\otimes} |V^\text{heavy}\rangle,
    \end{aligned}
\end{equation}

where 

\begin{equation}
\begin{aligned}
  V_1^\text{light}&{=} \tfrac{1}{\sqrt{3}}(u \bar u {+} d \bar d {+} s \bar s) \\
  V_8^\text{light}& {=} \tfrac{1}{\sqrt{6}}(u \bar u {+} d \bar d {-} 2 s \bar s) \ .  
\end{aligned}
\end{equation}

The Lagrangian $\mathcal{L}_{Y \psi \phi \phi}$ at leading order in chiral expansion is given by 
\cite{Mannel:1995jt}

\begin{equation}
\begin{aligned}
\mathcal{L}_{Y \psi \phi \phi}&{=} g_1 \langle V_1^\alpha J_\alpha^\dagger \rangle \langle u_\mu u^\mu \rangle {+} h_1 \langle V_1^\alpha J_\alpha^\dagger \rangle \langle u_\mu u _\nu \rangle v^\mu v^\nu \\
&{+} g_8 \langle J_\alpha^\dagger \rangle  \langle V_8^\alpha u_\mu u^\mu \rangle {+} h_8 \langle J_\alpha^\dagger \rangle  \langle V_8^\alpha u_\mu u _\nu \rangle v^\mu v^\nu \\
&{+} \text{h.c.} \, ,
\end{aligned}
\end{equation}

with $J=(\psi/\sqrt{3}) \mathds{1}$. The resulting $s$-wave projected chiral contact terms 
relevant for the $J/\psi\pi\pi$ $J/\psi K\bar K$ final states are given by

\be
\begin{aligned}
M^{0}_{\pi \pi}=&-\frac{2}{f^2} \sqrt{m_Y m_\jpsi} \left(g_1 + \frac{g_8}{\sqrt{2}}\right) (s-2 m_\pi^2)\\
&+\frac{h_1 + \frac{h_8}{\sqrt{2}}}{2} \left[ s+q^2\left( 1-\frac{\sigma_\pi}{3} \right) \right]\\
M^{0}_{K K}=&-\frac{2}{f^2} \sqrt{m_Y m_\jpsi} \left(g_1 - \frac{g_8}{2 \sqrt{2}}\right) (s-2 m_K^2)\\
&+\frac{h_1 - \frac{h_8}{2 \sqrt{2}}}{2} \left[ s+q^2\left( 1-\frac{\sigma_K}{3} \right) \right]\, ,
\end{aligned}
\ee

with $q^2=\lambda^{1/2}(m_Y^2,m_\jpsi^2,s)/(2 m_Y)$,  resulting in

\begin{equation}
\begin{aligned}
    \left[\Omega \mathcal{P}_{n-1}\right]_{\pi\pi}&=\Omega_{11} M^{0}_{\pi \pi} +\frac{2}{\sqrt{3}} \Omega_{12}  M^{0}_{K K}\\
    \left[\Omega \mathcal{P}_{n-1}\right]_{K K}&=\Omega_{21} M^{0}_{\pi \pi} +\frac{2}{\sqrt{3}} \Omega_{22}  M^{0}_{K K}\, .
    \label{eq:R2_ct_pipi}
\end{aligned}
\end{equation}

To summarize, the amplitudes, incorporating the $\pi\pi/K\bar K$ FSI, used in our calculations, are provided by Eqs.~\eqref{Eq:M0mod},
\eqref{eq:Mjmod} and \eqref{eq:R2_ct_pipi}.

\section{Pole uncertainty}
\label{sec:pole_uncertainty}

Within our calculation the
pole position is fixed by three parameters
(see Eq.~(\ref{Eq:G})): $g_{Y0}$, $m_0$ and $\Gamma_{\rm in}$, with  $m_0$ ($\Gamma_{\rm in}$) influencing
only the real (imaginary) part of the pole location
and  $g_{Y0}$ influencing both. To estimate the uncertainty of the pole position we performed a $\chi^2$ fit, however, with two approximations. First, we only allow the three resonance parameters $m_0,g_{Y0}$ and $\Gamma_{\rm in}$ to vary. Second, as $\jpsi \pi \pi$ has the best statistics of all the available data and the fit
suggests a negligible contribution of the $\psi(4160)$
to this channel, it is
by far the most restrictive final state for the $Y(4230)$ pole location. Therefore,  to estimate the uncertainty of the $Y(4230)$ pole location, we focus solely on the  $\jpsi \pi \pi$ channel. 
We checked that the inclusion of $D \bar D^* \pi$ yields no significant change to the uncertainty of the pole, supporting that the main influence on the pole position is driven by the data on the $\jpsi \pi \pi$ channel.
In the analysis, we allow $m_0$ and $\Gamma_\text{in}$ to vary within $\pm10 \mev$ and $g_{y0}$ by $\pm0.2 \mev$ around their best fit values.
These parameter ranges allow the pole to vary over a sufficiently wide range, including all values within the $1\sigma$ range around the best fit pole position.
Within these ranges random 
combinations of the three parameters are picked
under the requirement that
for each  parameter-set the change in the $\chi^2$ value must lie within 
\begin{equation}
    \chi^2_\text{best fit} - \chi^2_\text{random parameters} \leq \Delta_{\chi^2}(p,3) \, ,
    \label{eq:chi2_test}
\end{equation}
where the three in $\Delta_{\chi^2}(p,3)$ indicates that
three parameters are varied, and 
$p\approx0.683$ corresponds to
the  $1\sigma$ band.
To evaluate $\Delta_{\chi^2}(p,k)$ 
the $\chi^2$ cumulative-distribution function needs to equate $p$
\begin{equation}
    \frac{1}{\Gamma(k/2)} \gamma\left(\frac{k}{2},\frac{\Delta_{\chi^2}}{2}\right)=p \, ,
\end{equation}
where $\Gamma$ and $\gamma$ denote the regular and lower incomplete Gamma functions, respectively
\begin{equation}
    \gamma(x,a)=\int_0^a \text{d}t \,\, t^{x-1} e^{-t}, \quad \Re(a)>0 \, .
\end{equation}
Solving for $k=3$ degrees of freedom, 
the $1\sigma$ deviation in the $\chi^2$ value reads $$\Delta_{\chi^2}(p=1\sigma,k=3)\approx3.525\, .$$ 

\begin{figure}[t]
    \centering
    \includegraphics[width=\linewidth,trim={0pt 0pt 11pt 80pt}]{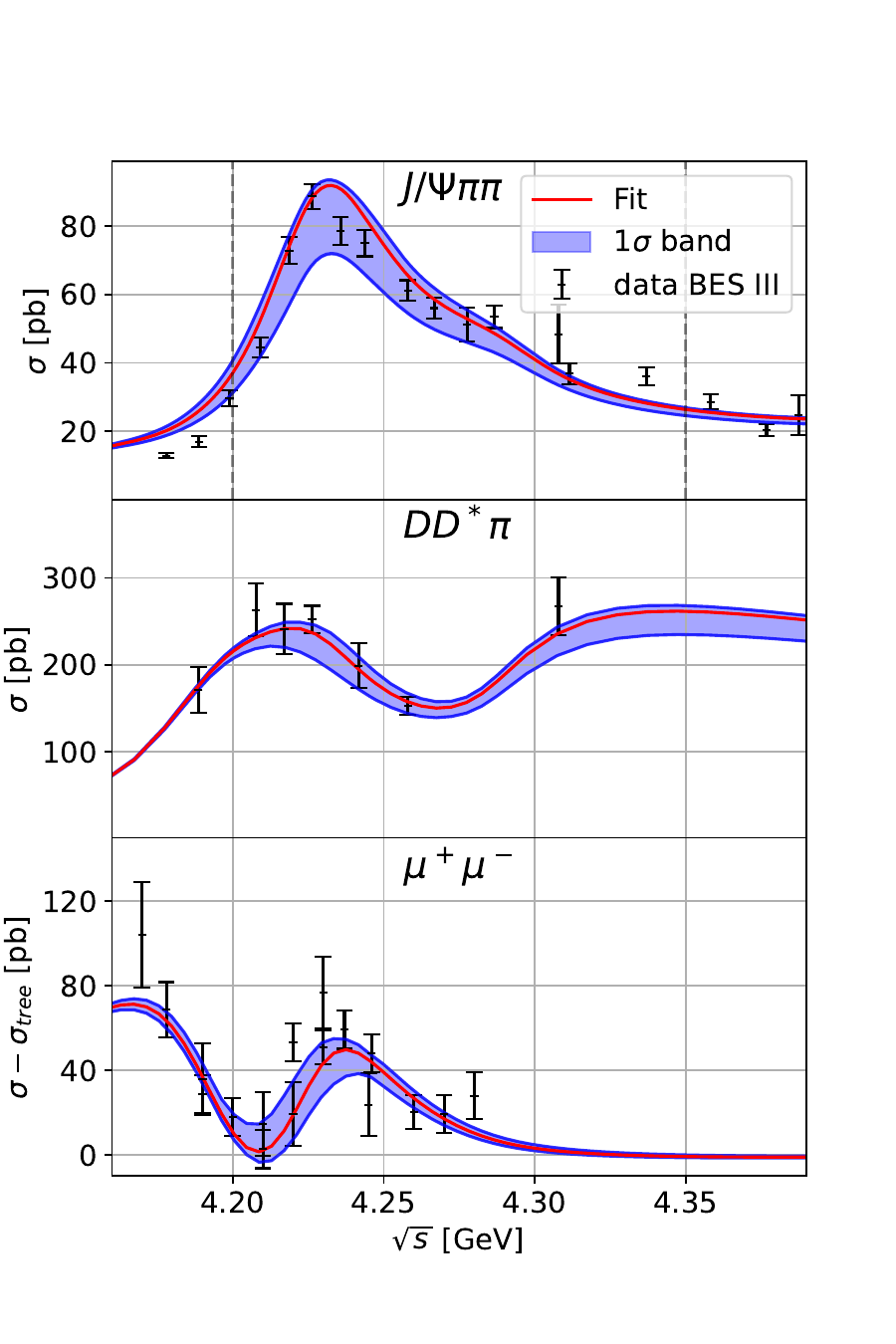}
    \includegraphics[width=\linewidth,trim={14pt 25pt 0pt 40pt}]{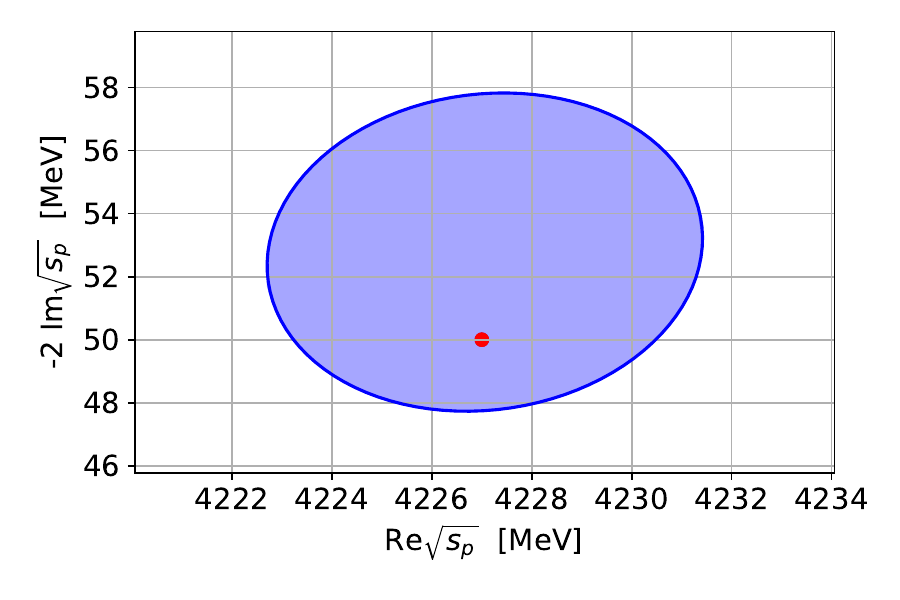}
    \caption{
     Uncertainty estimate of the $Y(4230)$ pole position. Upper three panels: 
Total cross sections for $\jpsi \pi^+ \pi^-,D^0 D^{*-} \pi^+$ and $\mu^+\mu^-$ final states with the $1 \sigma$ uncertainty band extracted from fits to the $\jpsi \pi \pi$ data and propagated to the other channels, as described in the text.  Red line in all plots corresponds to the best global fit to all data considered in this study (see the main text for details). 
  Lower panel: The best pole position  of the $Y(4230)$ (red dot) and the $1 \sigma$ uncertainty (blue ellipse) extracted from fits to the $\jpsi \pi \pi$ data. 
\label{fig:Y_uncertainty}}
    \end{figure}
The fits only included data from 4.2 to 4.35 GeV because this energy interval is expected to be under control due to the theoretical mechanisms considered in this work. As mentioned in the main text, deviations of our results from the data beyond this energy range are expected due to, in particular,  the absence of contributions from the $D_1\bar D^*$ and $D_2\bar D^*$ channels.
The results for $N\approx300$ random generated parameter sets that fulfill the condition of Eq.~\eqref{eq:chi2_test} are shown in Fig.~\ref{fig:Y_uncertainty}.
The upper panel shows the obtained $1\sigma$ uncertainty band for the $J/\psi \pi\pi$ total cross section,  while the bottom plot shows the corresponding spread of the pole position, which results in 
\begin{equation}\label{Eq:Ypole}
\sqrt{s_\text{pole}^{Y(4230)}}=\left( 4227{\pm} 4 {-} \frac{i}{2}\left(50^{+8}_{-2}\right) \right) \mev \, .
\end{equation}
In addition, the two plots  in the middle of Fig.~\ref{fig:Y_uncertainty} demonstrate the effect of the $1\sigma$ uncertainty extracted 
from fits to the $J/\psi \pi \pi$ channel on the total cross sections in the $D \bar D^* \pi$ and $\mu^+ \mu^-$ channels. As pointed out earlier, the variations in these channels are considerably less pronounced compared to those in the $J/\psi \pi \pi$ channel.

Our error estimate in Eq.~\eqref{Eq:Ypole} is supported by the observation that the uncertainty we find using this method is of the same order of magnitude as that  provided by the BESIII collaboration
extracted from the $J/\psi \pi\pi$ channel.

\section{Loop calculation}
\label{sec:loop_calc}
\subsection{Triangle}

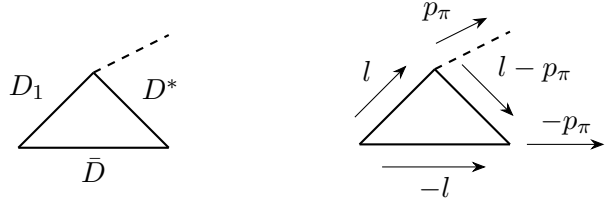
\begin{figure*}
\begin{tikzpicture}
		\begin{feynman}[large]
	\vertex [](a) at (0,0) ;
	\vertex [] (a2) at (0,-0.3);
	\vertex [](b) at (1,1) ;
	\vertex [] (out1) at (2,1.5);
	\vertex [] (c) at (2,0) ;
	\vertex [] (c2) at (2,-0.3) ;
	\vertex [] (i) at (-1.5,-0.3) {};
    \vertex [] (c2) at (2,-.6) {{\color{white}a}};
	
	\diagram*{
		(a) -- [edge label= {\(D_1\)}] (b),
		(b) -- [edge label= {\(D^*\)}] (c),
		(a) -- [edge label'= {\(\bar{D}\)}] (c),
		(a) -- [opacity=0] (c),
        (b) -- [scalar] (out1),
	};
	\end{feynman}
	\end{tikzpicture}
	$\quad$
\begin{tikzpicture}
\begin{feynman}[large]
\vertex [](a) at (0,0) ;
\vertex [] (a2) at (0,-0.3);
\vertex [](b) at (1,1) ;
\vertex [] (out1) at (2,1.5);
\vertex [] (c) at (2,0) ;
\vertex [] (c2) at (2,-0.3) ;
\vertex [] (out2) at (3.5,-0.3);
\vertex [] (i) at (-1.5,-0.3) {};

\diagram*{
    (a) -- [momentum=$l$] (b),
    (b) -- [momentum=$l-p_\pi$] (c),
    (a) -- [momentum'=$-l$] (c),
    (b) -- [opacity=0, momentum=$p_\pi$] (out1),
    (c2) -- [opacity=0, momentum=$-p_\pi$] (out2),
    (b) -- [scalar] (out1),
};
\end{feynman}
\end{tikzpicture}
\centering
\label{fig:Triangle_D1DDs}
\caption{Momentum assignment of the triangle loop.  }
\end{figure*}

The triangle diagram shown in figure \ref{fig:Triangle_D1DDs} only has one time ordering and is given by

\begin{widetext}

\begin{equation}
\mathcal{T}_{D_1 D D^*}=\int \frac{ \text{d}^3 l}{(2\pi)^3} \frac{1}{8 \omega_D \omega_{D_1} \omega_{D^*}} \frac{1}{E-\omega_{D_1}-\omega_D} \frac{1}{E-\omega_\pi-\omega_{D^*}-\omega_D}\, ,
\label{eq:Triangle_D1DDs_int}
\end{equation}

\end{widetext}

with

\begin{equation}
\begin{aligned}
	\omega_D=&\sqrt{m_D^2+l^2}\\\
	\omega_{D^*}=&\sqrt{m_\Ds^2+ l^2+p_\pi^2-2 l p_\pi z }\\
	\omega_{D_1}=&\sqrt{(m_{D_1}-i \Gamma_{D_1}/2)^2+l^2}\\
    \omega_\pi=&\sqrt{m_\pi^2+p_\pi^2}\, .
\end{aligned}
\end{equation}

In the center of mass frame of the $Y(4230)$ one can choose the momenta in a way that only the $D^*$ has an angular dependence, where $\Vec{p_\pi}=p_\pi (0,0,1)^\text{T}$, such that $z=\cos{\theta}$ denotes the cosine of the polar angle of the loop momentum $\vec{l}$. Due to the width of the $D_1$ only the last propagator in eq. \eqref{eq:Triangle_D1DDs_int} has poles on the real axis, as in comparison the $D^*$ width is negligible small. We define

\begin{equation}
	f(E,l)=\frac{l^2}{8 \omega_D \omega_{D_1}} \frac{1}{E-\omega_{D_1}-\omega_D}\, ,
\end{equation}

such that f is regular in $l$. The integral can now be rewritten as

\begin{equation}
\begin{aligned}
	&\mathcal{T}_{D_1 D D^*}=\frac{1}{(2 \pi)^2}\int_{0}^{\Lambda} \text{d} l f(E,l) \\
 &\quad \times \int_{-1}^{1} \text{d} z \frac{1}{\omega_{D^*}} \frac{1}{E-\omega_\pi-\omega_{D^*}-\omega_D}\, ,
 \label{eq:Tri_int}
\end{aligned}
\end{equation}

where the trivial integration over the loop momentums azimuthal angle is performed. Doing a variable transformation

\begin{equation}
	\text{d} \omega_{D^*}=\frac{l p_\pi}{\omega_{D^*}} \text{d} z \, ,
\end{equation}

the angular integration becomes

\begin{equation}
	\int_{-1}^{1} \text{d} \omega_{D^*} \frac{1}{p_\pi l} \frac{1}{E-\omega_\pi-\omega_D^{*}-\omega_D} \, .
    \label{eq:int_wd}
\end{equation}

The inverse factor of $l$ is canceled by $f(E,l)$, while $p_\pi$ is canceled by the phase space integration.

With the relation

\begin{equation}
	\frac{1}{x-x_0 \pm i \epsilon}=\mathcal{P}\left( \frac{1}{x-x_0} \right) \mp i \pi \delta(x-x_0) \, ,
\end{equation}

where $\mathcal{P}$ denotes the Cauchy principal value, Eq.~\eqref{eq:int_wd} becomes

\begin{equation}
\begin{aligned}
	&\frac{1}{E-\omega_\pi-\omega_{D^*}-\omega_D+i \epsilon}=\\
 &\quad -\mathcal{P}\left( \frac{1}{\omega_{D^*}-(E-\omega_D-\omega_\pi)} \right)\\
 &\quad -i \pi \delta(\omega_{D^*}-(E-\omega_\pi-\omega_D)) \, .
 \end{aligned}
\end{equation}

The $\delta$ function can be rewritten as

\begin{equation}
	\delta(\omega_{D^*}-(E-\omega_\pi-\omega_D))=\frac{\omega_{D^*}}{l p_\pi} \delta(z-z_0) \, ,
\end{equation}

with

\begin{equation}
	z_0=-\frac{l^2+p_\pi^2+m_{D^*}^2-(E-\omega_\pi-\omega_D)^2}{2 l p_\pi} \, .
\end{equation}

Now Eq.~\eqref{eq:Tri_int} takes the form

\begin{widetext}

\begin{equation}
	I=\frac{1}{(2\pi)^2} \int_{0}^{\infty} \text{d} l \frac{f(E,l)}{l p_\pi} \int_{-1}^{1} \text{d} z \left[ -\mathcal{P}\left( \frac{1}{\omega_{D^*}-(E-\omega_\pi-\omega_D)} \right) - i \pi \delta(z-z_0) \right]\, ,
\end{equation}

with

\begin{equation}
	\mathcal{P}\left( \int_{\omega_{D^*}({z=-1})}^{\omega_{D^*}({z=1})} \text{d} \omega_{D^*} \frac{1}{\omega_{D^*}-(E-\omega_\pi-\omega_D)} \right)=\log \left(  \left| \frac{E-\omega_{D^*}|_{z=1}-\omega_\pi-\omega_D}{E-\omega_{D^*}|_{z=-1}-\omega_\pi-\omega_D} \right| \right) \, .
\end{equation}

\end{widetext}

Finally we arrive at

\begin{equation}
\begin{aligned}
	I(E)=&\int_{0}^{\Lambda} \text{d} l \frac{\tilde{f}(E,l)}{p_\pi} \\
  &\quad \times \left[  \log \left( \left| \frac{E-\omega_{D^*}|_{z=1}-\omega_\pi-\omega_D}{E-\omega_{D^*}|_{z=-1}-\omega_\pi-\omega_D} \right| \right) \right.\\
	&\quad \left. + i \pi \Theta \left( \frac{E-\omega_{D^*}|_{z=1}-\omega_\pi-\omega_D}{E-\omega_{D^*}|_{z=-1}-\omega_\pi-\omega_D} \right)\right]\, ,
\end{aligned}
\end{equation}

where the remaining radial integration can be preformed numerically. In case of the $\jpsi D^{(*)} \bar{D}^{(*)}$ vertex, which scales with the loop-momentum itself, the integrand is modified accordingly 

\begin{equation}
    \mathcal{T}(C,\text{Num})=C \int \frac{\text{d}^3 l}{ 2 \pi^3} \frac{\text{Num}(l,p_1,p_2)}{8 \omega_1 \omega_2 \omega_3} G_1 G_2 
\end{equation}

with $\text{Num}$ denoting the momentum factors in the numerator and $C$ being a constant.\\
To decrease the number of sample points needed to achieve a stable result, it is useful to further split the $l$ integration at the poles of the propagator, as the distribution of Gauss-Legendre sample points is more dense at the integration borders

\begin{equation}
    \int_0^\Lambda \text{d} l = \int_0^{l^\text{P}_0} \int_{l^\text{P}_0}^{l^\text{P}_1} ... \int_{l^\text{P}_n}^{\Lambda} \text{d} l\, ,
    \label{eq:l_split}
\end{equation}
where the $l^\text{P}_i$ correspond to the poles of the propagator in $l$, which can be calculated analytically for each propagator.

\begin{figure*}[t]
\begin{tikzpicture}
	\begin{feynman}[small]
	\vertex [](a) at (0,0);
	\vertex [](b) at (2,-1.5);
	\vertex [](c) at (3,-0.5);
	\vertex [](d) at (1,1);
	\vertex [](p1) at (3,2);
	\vertex [](p2) at (3,-2.5);
		\diagram*{
		(a) -- [momentum=$l+p_1$] (d),
		(a) -- [momentum'=$-l-p_1$] (b),
		(d) -- [momentum=$l$] (c),
		(b) -- [momentum'=$-l-p_1-p_2$] (c),
		(d) -- [scalar, momentum=$p_1$] (p1),
		(b) -- [scalar, momentum'=$p_2$] (p2),
	};
	\end{feynman}
\end{tikzpicture}
$\qquad \qquad $
\begin{tikzpicture}
	\begin{feynman}[small]
	\vertex [](a) at (0,0);
	\vertex [](b) at (2,-1.5);
	\vertex [](c) at (3,-0.5);
	\vertex [](d) at (1,1);
	\vertex [](p1) at (3,2);
	\vertex [](p2) at (3,-2.5);
		\diagram*{
		(a) -- [edge label= {1},inner sep=0.1cm] (d),
		(a) -- [edge label'= {2},inner sep=0.1cm] (b),
		(d) -- [edge label= {3},inner sep=0.1cm] (c),
		(b) -- [edge label'= {4},inner sep=0.1cm] (c),
		(d) -- [scalar,edge label= {5},inner sep=0.1cm] (p1),
		(b) -- [scalar, edge label'= {6},inner sep=0.1cm] (p2),
	};
	\end{feynman}
\end{tikzpicture}
\centering
\caption{One time ordering of the first Feynman diagram shown in Fig.~\ref{fig:R2_box_decomp}.}
\label{fig: Box_int}
\end{figure*}
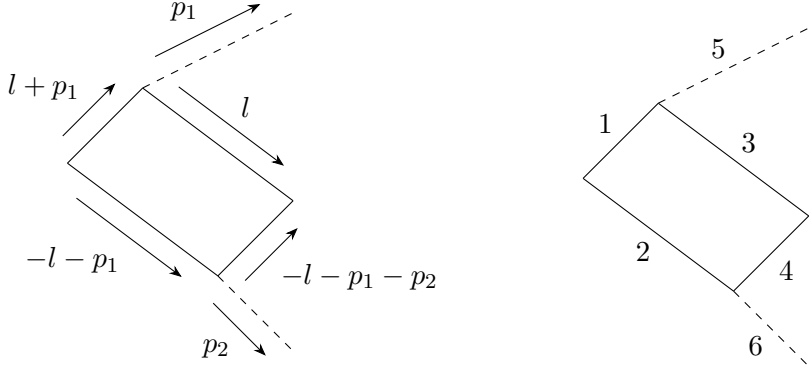

\subsection{Box}

The Integral for the scalar box shown in figure \ref{fig: Box_int} is given by

\begin{widetext}
    
\begin{equation}
    I=\int \frac{\text{d}^3l}{(2\pi)^3} \frac{\text{Num}(l,p_1,p_2)}{16 \omega_1 \omega_2 \omega_3 \omega_4}\frac{1}{E-\omega_{1}-\omega_2} \frac{1}{E-\omega_{5}-\omega_{3}-\omega_{2}}\frac{1}{E-\omega_{4}-\omega_{3}-\omega_{5}-\omega_{6}}\, .
\end{equation}

\end{widetext}

In this work the second cut corresponds for most box topologies to $\bar{D} D^* \pi$, which can go on-shell. Analogous to the triangle we isolate the singularity and define the remaining part in a function $f(E,l,\varphi,z,p_1,p_2)$ that is regular in $l$ and $z$. However, different to the triangle, it is not possible to to perform the integration of the polar angle analytically as $f$ is also dependent on $z$, such that we perform a numerical subtraction

\begin{widetext}
    
\begin{equation}
\begin{aligned}
    I=&\int \frac{\text{d}^3 l}{(2\pi)^3} f(E,l,\varphi,z,p_1,p_2) \frac{1}{E-\omega_{5}-\omega_{3}-\omega_{2}}\\
    =&\int_0^\Lambda \text{d}l \,\, l^2 \int_0^{2\pi} \text{d}\varphi \left[ \int_{-1}^{1} \text{d}z \, \frac{f(E,l,\varphi,z,p_1,p_2)-f(E,l,\varphi,z_0,p_1,p_2)}{E-\omega_{5}-\omega_{3}-\omega_{2}} \right.\\
    &\qquad \qquad \qquad \qquad  + \left. f(E,l,\varphi,z_0,p_1,p_2) \int_{-1}^{1} \text{d}z \, \frac{1}{E-\omega_{5}-\omega_{3}-\omega_{2}} \right] \, ,
\end{aligned}
\end{equation}

\end{widetext}

where $z_0$ is the pole of the propagator. The integration of the second summand con now be done analytically, resulting in

\begin{equation}
\begin{aligned}
    &\int_{-1}^{1} \text{d}z \, \frac{1}{E-\omega_{5}-\omega_{3}-\omega_{2}} =\\
    &\qquad \log \left[  \frac{(E-\omega_{5}-\omega_{3}-\omega_{2})|_{z=1}}{(E-\omega_{5}-\omega_{3}-\omega_{2})|_{z=-1}} \right] \, .
\end{aligned}
\end{equation}

The remaining $\varphi$ and $l$ integration are performed numerically using Gauss-Legendre integration. Analogously to the triangle the radial integration is split according to Eq.~\eqref{eq:l_split}. The general notation used in this work is

\begin{equation}
    \mathcal{B}(C,\text{Num}){=}{\sum_T} C {\int} \frac{\text{d}^3 l}{2 \pi^3} \frac{\text{Num}(l,p_1,p_2)}{16 \omega_1 \omega_2 \omega_3 \omega_4} G_1 G_2 G_3\, ,
    \label{eq:box_int}
\end{equation}

where $\sum_T$ stands for the sum over the different time orderings and $G_i$ denotes the propagators for the different cuts, e.g. $G_1=1/(E-w_{D_1}-\omega_{D})$. 

\end{document}